\documentclass[pre,showpacs,showkeys,superscriptaddress,amsmath,amssymb,color]{revtex4-1}
\usepackage{color}
\usepackage{amsmath}
\usepackage{graphicx}
\usepackage[latin1]{inputenc}
\usepackage{amsfonts}
\usepackage{amssymb}
\usepackage{float}
\usepackage{bm}
\usepackage{dcolumn}
\usepackage{hyperref}

\newcommand\etal{\textsl{et al.}}
\pdfoutput=1

\begin{document}

\title{Pattern-transition, microstructure and dynamics in two-dimensional vibrofluidized granular bed}

\author{Istafaul H. Ansari}
\author{Meheboob Alam}
\email{Corresponding author: meheboob@jncasr.ac.in}
\affiliation{Engineering Mechanics Unit, Jawaharlal Nehru Center for Advanced Scientific Research, Jakkur P.O., Bangalore 560064, India}
\date{\today}

\begin{abstract}
Experiments are conducted in a two-dimensional mono-layer vibrofluidized bed of glass beads,
with a goal to understand the transition scenario and the underlying microstructure and dynamics
in different patterned-states. At small shaking accelerations ($\Gamma=A\omega^2/g <1$, where $A$ and $\omega=2\pi f$ 
are the amplitude and angular frequency of shaking and $g$ is the gravitational acceleration), 
the particles remain attached with the base of the vibrating container -- this is known as the solid bed (SB).
With increasing $\Gamma$ (at large enough shaking amplitude $A/d$)
and/or with increasing $A/d$ (at large enough $\Gamma$),  the sequence of  transitions/bifurcations unfolds as follows: 
SB (``solid bed'') to BB (``bouncing bed'') to LS (``Leidenfrost state'') to ``2-roll Convection'' to ``1-roll Convection'' and finally to a gas-like state.
For a given length of the container,  the coarsening of multiple convection rolls leading to the genesis of a ``single-roll'' structure (dubbed the {\it multi-roll transition}),
and its subsequent transition to a granular-gas  are two novel findings of this work.
We show that the critical shaking intensity ($\Gamma_{BB}^{LS}$) for  ``$BB\to LS$''-transition 
has a power-law dependence on the particle loading ($F=h_0/d$, where $h_0$ is the number
of particle layers at rest and $d$ is the particle diameter) and the  shaking amplitude ($A/d$).
The characteristics of $BB$ and $LS$ states are studied  by calculating (i) the coarse-grained density and temperature profiles and (ii) the pair correlation function.
It is shown that while the contact network of particles in the BB-state represents a hexagonal-packed structure,
the contact network within the ``floating-cluster'' of the LS resembles a liquid-like state.
An unsteadiness of the Leidenfrost state has been uncovered wherein the interface (between the floating-cluster and the dilute collisional layer underneath)
and the top of the bed are found to  oscillate sinusoidally,  with its oscillation frequency closely matching with the frequency of external shaking. 
Therefore, the granular Leidenfrost state is a period-1  wave as is the case for BB-state.
\end{abstract}


\maketitle

\section{Introduction}

The earliest scientific work on pattern formation in granular media dates back to Chladni~\cite{Chladni1787} and  Faraday~\cite{Faraday1831}
who studied the now well-known heaping phenomenon in shallow layers of vibrating particles.
There have been a renewed interest to understand the pattern-formation scenario in a vibrated granular bed starting from early 1990's~\citep{DFL1989,CR1991,GHS1992,PB1992,Luding1994,LR1995,PH1995,BM1995,HYH1995,Warr1995,WB1996,Aoki1996,Knight1996,UMB1996,KWG1997,Bizon1998,RRC2000,YH2000,PSS2000,WHP2001,BRM2001,BKS2002,GMIZ2002,MPB2003,OO2003,KM2003,Huan2004,Sano2005,Eshuis2005,CPS2007}
and two comprehensive reviews on granular patterns till 2005 can be found in Refs.~\cite{Kudrolli2004,AT2006}.
Since then a vertically shaken box of particles has become a  canonical experimental setup to study granular patterns; more specifically,
under harmonic shaking via $y(t)=A\sin(2\pi ft)$, such a system is known to admit
a variety of interesting patterns:  the standing waves, `$f/n$' sub-harmonic waves via period-doubling bifurcations~\citep{DFL1989,WB1996,Sano2005},
surface waves and heaping~\citep{CR1991,PB1992},
Faraday waves and oscillons~\citep{UMB1996,Bizon1998}, convection~\citep{GHS1992,Luding1994,BM1995,HYH1995} and density-inverted state~\citep{MPB2003,Eshuis2005}. 
The primary control parameter for the onset and/or bifurcation of any of these patterns is the dimensionless shaking acceleration $\Gamma=A\omega^2/g$,
which is a relative measure of the driving acceleration with respect to the gravitational acceleration $g$.
For a review on various patterns in a quasi-2D vibrated bed as well as to know the  specific contributions of various research groups, 
we refer the readers to the introductory section of a recent work~\cite{Shukla2014}.  In the present experimental work,
we shall primarily  focus on unveiling the transition scenario among different patterns (the bouncing bed, the density-inverted state, the convection and the granular gas)
in a vibrofluidized bed with increasing $\Gamma$, and  this being a well-studied problem, here we provide only a brief review of the  recent  works
of Eshuis {\it et al}~\cite{Eshuis2005,Eshuis2007} and others which motivated the present study.

The above mentioned density-inverted state  was dubbed {\it granular} Leidenfrost state (LS) by Eshuis~\etal~\cite{Eshuis2005} who
established a possible connection  with the original Leidenfrost effect~\cite{Leidenfrost}:
a liquid drop  placed on a hot plate  can float over its own vapor layer if the temperature of the plate exceeds a minimum value (Leidenfrost temperature).
Akin to this, a dense, compact layer of particles can be supported 
by a dilute gaseous region of fast moving particles underneath it in vertically shaken granular bed beyond critical value of the shaking intensity.
More specifically, they found that a dense region of particles with crystalline-type structure 
can float over a granular gas at mild acceleration ($\Gamma\sim 10$) which bifurcates from a time-periodic bouncing bed state.
Previous molecular dynamics (MD) simulations had also predicted the possibility of such 
\textit{density inversion}~\cite{LR1995} or {\it floating-cluster}~\cite{MPB2003} in a similar setup.

The most recent and comprehensive experimental work of Eshuis~\etal~\cite{Eshuis2007},
on a shallow vertically-shaken granular materials held  in a ``quasi-2D'' box (of a few particle wide),
provides the complete phase diagram of various patterns consisting of
(i) solid bed, (ii) bouncing bed, (iii) sub-harmonic undulatory waves,  (iv) granular Leidenfrost state, (v) convection and (vi) granular gas.
They also carefully documented the related bifurcation scenario for each pattern as a function of $\Gamma$ and the particle loading.
They found that with increasing shaking intensity $\Gamma$ the convection-rolls
can appear either (i) from the bouncing-bed state at small particle loading, or (ii) from the Leidenfrost state at large particle loading. 
On the other hand, the primary onset of the Leidenfrost state was found to occur from the undulatory waves
with increasing $\Gamma$, however, the undulatory state can also exist over a range of 
$\Gamma$ between two Leidenfrost states (at lower and higher values of $\Gamma$).
The latter finding implies that two successive bifurcations from a ``sub-harmonic'' wavy-state 
to the Leidenfrost state resulted to the onset of convection at large particle loadings.
It must be noted that the temporal-order of the Leidenfrost state was never quantified~\cite{Eshuis2005,Eshuis2007,AAlam2012,AAlam2013} via experiments,
but it appears to be a quasi-steady or steady state from the related theoretical analyses~\cite{Shukla2014,Eshuis2010,Eshuis2013}.
In any case, the convection-rolls found at strong shaking (i.e.~at large values of $\Gamma>25$)
are the granular analog of Rayleigh-Benard convection, and they must be differentiated from
the so-called ``boundary-driven'' convection~\cite{GHS1992,Luding1994,BM1995,Knight1996} that
appear at milder shaking ($\Gamma\sim 5$).

In this work we carry out   experiments in a purely two-dimensional (2D) container that can accomodate only one layer of particles along the depth of the container,
and in doing so we uncover the complete sequence of bifurcations in a 2D-vibrofluidized bed  as well as provide quantitative results about different patterned states.
In a quasi-2D setup of a few-particle diameter depth, such as that employed by Eshuis~\etal~\cite{Eshuis2007},
it is difficult to obtain quantitative data on (i) the hydrodynamic fields (density, granular temperature
and velocity) as well as on (ii) particle-level quantities (microstructure, distribution function, etc.).
Such quantitative measurements are however possible in pure 2D-container that can accommodate
a mono-layer of particles along its depth as in the experimental work of Eshuis~\etal~\cite{Eshuis2005}.
Our primary focus is to (i) to quantify the onset of the transition from the bouncing bed to the Leidenfrost state in terms of different control parameters, 
(ii) to analyze the microstructure in the latter state and related signatures of transition,
(iii) to quantify the temporal-order (steady or time-periodic) of the Leidenfrost state,
and finally (iv) to uncover the possible routes to transition of the LS to convection and eventually to a gas. 
The characteristics and distinguishing features of different states have also been probed  with
the help of their mean-fields, namely, the density and temperature profiles along the vertical direction. 
It was speculated in Ref.~\cite{Eshuis2007} that the collective motion
of particles along the depth of the container may be a prerequisite for the onset of convection at strong shaking. 
On the contrary, we shall demonstrate that (v) the ``Rayleigh-Benard-type'' granular convection 
can be realized in experiments in a 2D mono-layer system and (vi) the route-to-transition to a gaseous state occurs via a novel ``single-roll''  convection.
In particular, the coarsening of a pair of convection rolls leading to the genesis of a ``single-roll'' structure
spanning the length of the container and its subsequent transition to a granular-gas were not reported in previous experiments.

\begin{figure}[!ht]
\centering
(a)
\includegraphics[width=2.6in,height=2in]{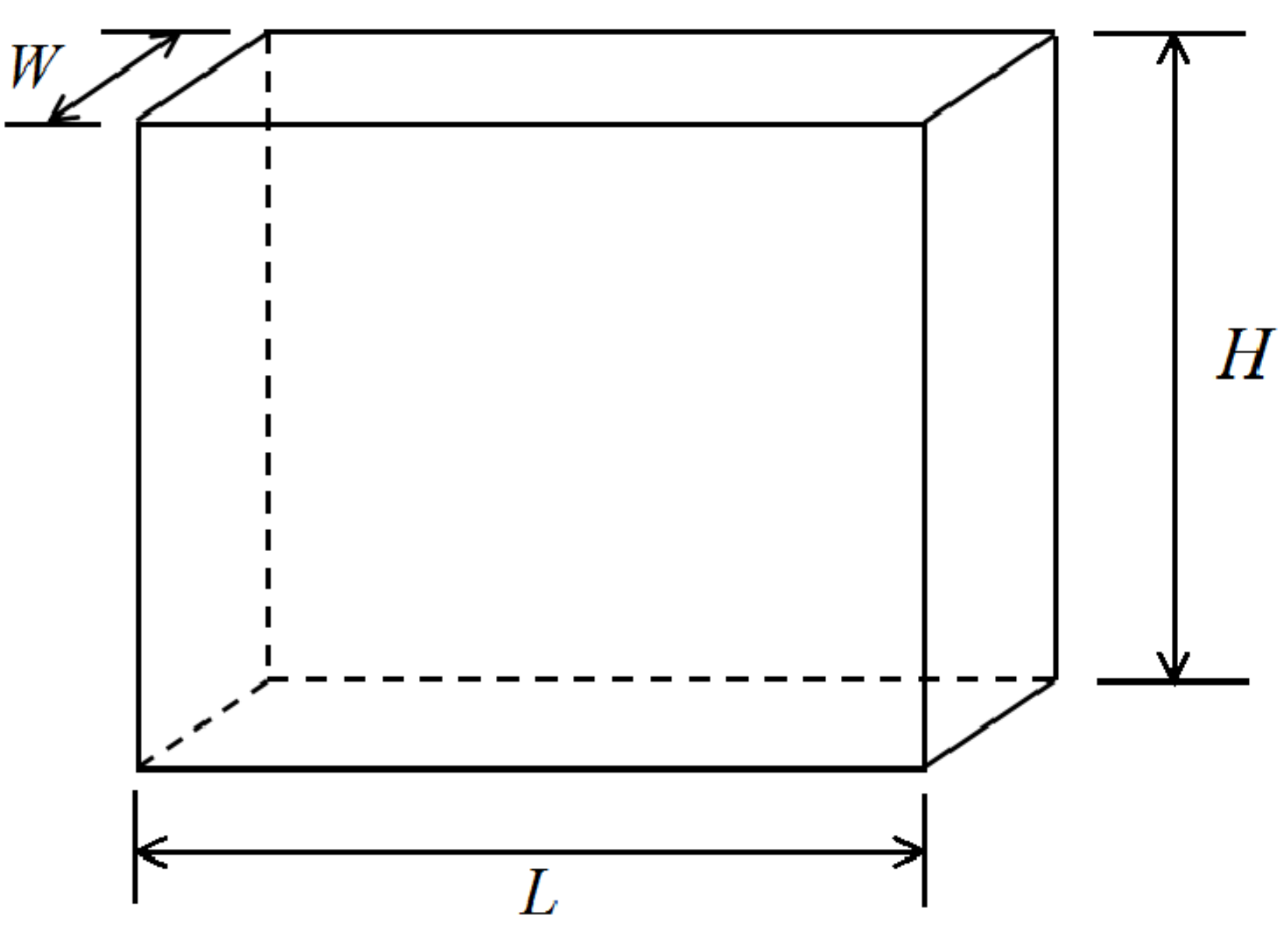}\\
(b)
\includegraphics[width=3.1in,height=2in]{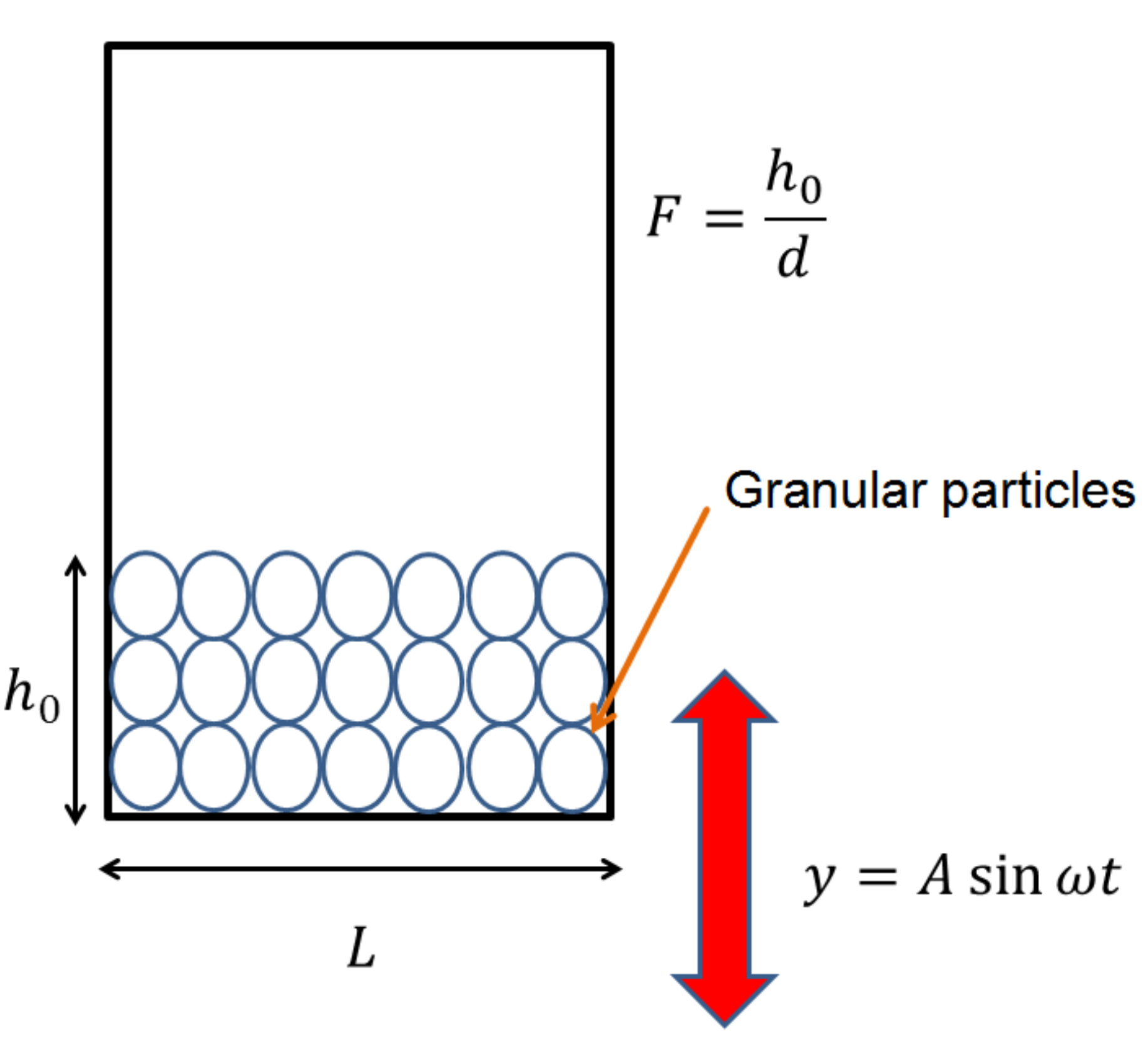}
\caption{
(Color Online)
(a) Schematic of the container with length $L$, width $W$ and height $H$.
(b) The container is partially-filled with a mono-layer (i.e.~$W/d\approx 1$)  of spherical glass-beads of diameter $d$,
and is vibrated harmonically, $y(t)=A\sin{\omega t} = A\sin{2\pi ft}$, via an electromagnetic shaker.
$F=h_0/d$ is the number of particle layers at rest.
}
\label{fig:fig1}
\end{figure}

\section{Experimental Setup and Methodology}

The experimental setup  consists of a quasi-two-dimensional rectangular Plexiglas 
container with length ($L$), width ($W$) and height ($H$) of $40$, $2.2$ and $100$ mm, respectively,
as depicted in Fig.~\ref{fig:fig1}(a), which is vibrated along the vertical direction ($y$) using an electromagnetic shaker. 
The width ($W$) of the cell has been chosen such that it fits a mono-layer of spherical glass-beads along its width, i.e. $W/d\approx 1$;
for example, with $d=2$ mm diameter glass beads, we have $W/d=1.1$ which was kept constant for experiments with even larger diameter particles. 
A similar Hele-Shaw container was used by Eshuis~\etal~\cite{Eshuis2005} whose work
motivated the present work to unveil the complete bifurcation scenario of patterns in a 2D `mono-layer' vibrofluidized bed.

\subsection{Experimental protocol and control parameters}

The container is partially filled with spherical glass balls (density $2500$kg/m$^3$)  of
specified height of $h_0=Fd$ (where $h_0$ is the number of particle layers, see Fig.~\ref{fig:fig1}$b$).
This particle-filled container  is mounted on an electromagnetic shaker (of Ling Dynamics System)
via a circular head expander~\cite{AAlam2012,AAlam2013}, and is vibrated vertically using a sine-wave of the form: 
\begin{equation}
  y=A\sin(\omega t)=A\sin(2\pi ft),
\end{equation}
where $A$ is the shaking amplitude and $f$ is the frequency of shaking. 
The shaker operates in a closed loop, controlled by a controller and an amplifier through a software interface. 
To generate a feedback signal of specified amplitude and frequency of the sinusoidal vibration, 
a piezoelectric accelerometer is mounted on the head expander.

There are four dimensionless control parameters in this system. The dimensionless shaking acceleration/intensity 
\begin{equation}
 \Gamma=\frac{A\omega^2}{g}=\frac{4\pi^2 A f^2}{g},
\end{equation} 
is the primary control parameter in this problem.
In addition, (ii) the dimensionless shaking amplitude $A/d$ and  (iii) the initial layer height,
\begin{equation}
  F=\frac{h_0}{d},
\end{equation} 
where $h_0$ is the number of particle layers at $t=0$, (iv) the length of the container ($L/d$),
which is of crucial importance for the observance of certain patterns (see Sec.III.B), constitute the other three control parameters.
In this paper, we have restricted to a relatively narrow container of length $L/d\sim (20,50)$.
The last control parameter is the coefficient of restitution ($e$) for particle collisions -- all results are presented for glass-beads ($e\approx 0.95$).
[Previous works~\cite{Eshuis2007} as well as the repetition of a few of the present experiments with steel-beads ($e\approx 0.9$)
suggest that our reported patterns are robust irrespective of the choice of specific particles.]

Most of the experiments are  carried out with $d=2.0\;mm$ diameter glass balls;
the filling height $F=h_0/d$ is  varied from $10$ to $50$. To check the robustness of
reported results, additional mono-layer experiments with $d=5.0\;mm$ diameter glass balls are also done
in a container with $L=100$, $W=5.5$ and $H=100$ mm, respectively, such that $W/d=1.1$. 
Once the container is filled with a specified number of layers of particles $F$,
the experiments are conducted at a specified shaking amplitude $A/d$, but by 
increasing the shaking frequency $f$ (and hence increasing the shaking acceleration $\Gamma$) linearly at a specified sweep/ramping rate.
(This type of frequency sweeping at a constant amplitude  is achieved by our closed-loop shaker system.)
We have used  a linear sweep rate of $0.01\ Hz/s$, unless stated otherwise,
and the results are found to be qualitatively similar for a sweep rate of $0.1\ Hz/s$. 
To obtain the phase-diagram of patterns to span the control parameter space in the ($\Gamma, A/d$)-plane,
the experiments are conducted for a range of $\Gamma$ and $A/d$.
The shaking intensity $\Gamma$ is varied from $0$ to $55$, with its upper limit being set by the maximum payload of our shaker;
the shaking amplitude $A/d$ is varied from $0.2$ to $4.0$, with its upper limit being set by the maximum permissible limit ($A=9mm$) of our shaker.

The granular particles held in the vibrating container are illuminated with two white  
LED light sources of power $25 W$, positioned at an oblique angle facing towards the container from both 
sides. Such type of lighting arrangement provided  uniform lighting over the region of interest~\cite{AAlam2012,AAlam2013}.
We employed a high-speed camera  (IDT MotionPro Y4S3), mostly at a frame rate of $1000$ frames per second,
to capture the temporal evolution of the collective motion of particles. 
The series of snapshots are grabbed in on-board camera memory and later on  
transferred to computer via USB for post-processing and data analysis.

\subsection{Image analysis for particle tracking}

The acquired images of the granular bed have been used to calculate 
(i) the coarse-grained density and the granular temperature fields (the results are discussed in Sec.~III.C),
(ii) the pair-correlation function (Sec.~IV.A)  and (iii) the oscillations within the Leidenfrost state (Sec.~IV.B).
For all cases, the particles detection and their position information, frame by frame,  are required to be extracted. 
Once the image analysis is completed, the particle coordinates are fed to 
particle-tracking routine to  calculate the velocity of each particle in all frames.

A number of particle tracking routines  have been developed by various  research groups and available as open-source.
In this study, we employed \textit{Particle Detector and Tracker} open-source distribution~\cite{CG1996} which consists of an \textit{ImageJ Plugin} for particles detection 
and tracking from digital  videos. This plugin implements the feature point detection and tracking algorithm as
described in Ref.~\cite{SK2005}. It presents an easy-to-use, computationally efficient, 
two-dimensional, feature point-tracking tool for the automated detection and analysis of  particle trajectories as recorded by high speed imaging.
The feature point tracking problem consists of detecting images of particles in a digital image sequence and linking these detections over 
time to follow the tracks of individual particles.

\begin{figure}[!ht]
\centering
\includegraphics[width=3.2in,height=3.2in]{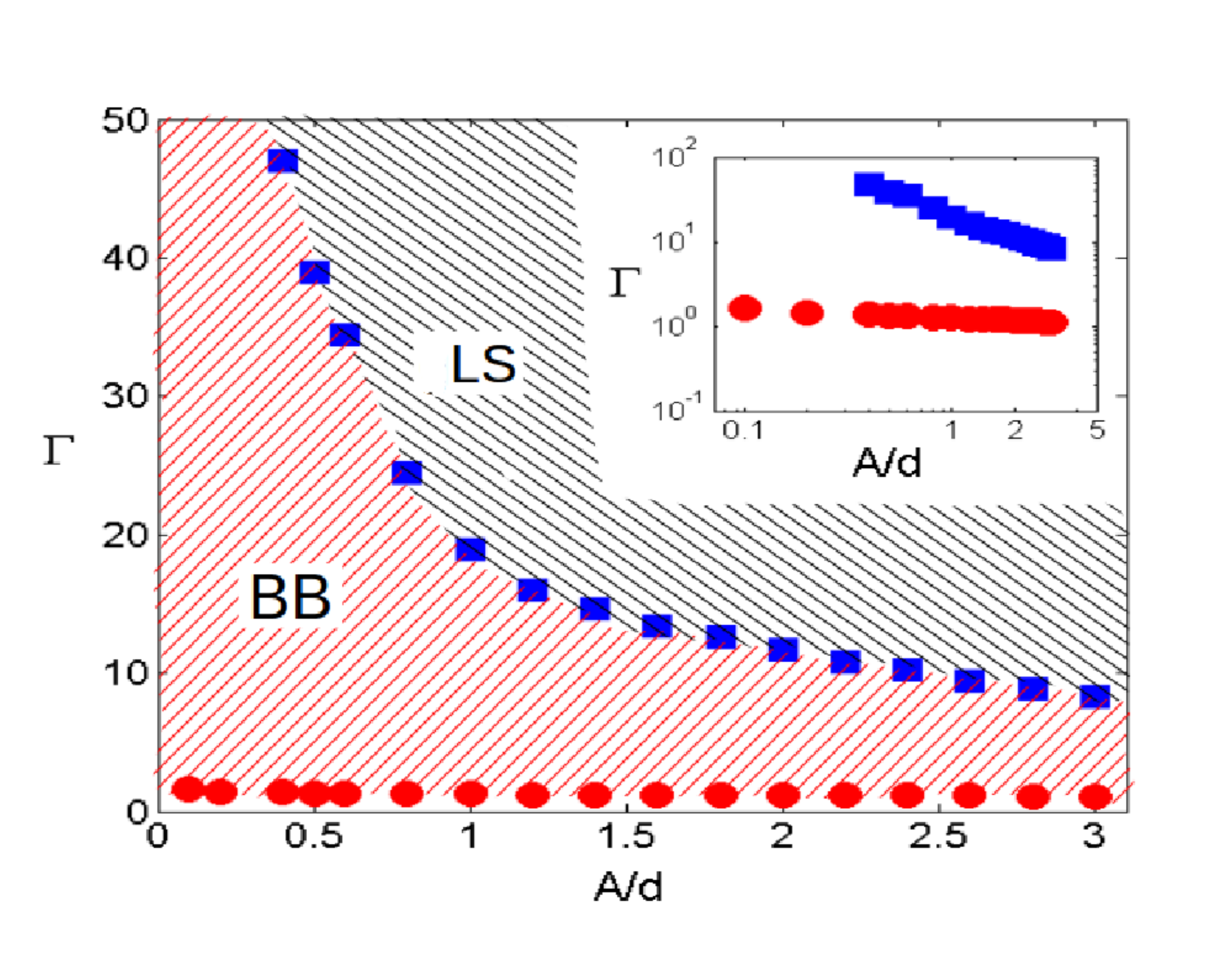}
\caption{
(Color online)
Phase diagram in the ($\Gamma, A/d$)-plane for $F=25$ layers of 
$d=2.0$ mm diameter glass balls confined in a Hele-Shaw container  of width $W/d=1.1$ and length $L/d=20$; 
inset shows the same phase-diagram in logarithmic scale.
Regions of bouncing bed (BB) and Leidenfrost state (LS) are hatched in the main panel; the region below red circles represents `solid bed' (SB).
The symbols represent approximate locations of the respective transition; they are obtained by running experiments at fixed values of $A/d$ by
 increasing frequency at a linear ramping-rate of $0.01$ Hz/s; the ``up-sweeping'' and ``down-sweeping'' experiments yielded almost the same result
for the onset of patterns, suggesting a ``supercritical'' transition between different states.
}
\label{fig:fig2}
\end{figure}

\section{Pattern Transition and Hydrodynamic Fields}

\subsection{Transition from bouncing-bed to Leidenfrost state: Phase diagram and scaling}

\begin{figure}[!ht]
\centering
\includegraphics[width=1.0in,height=1.8in]{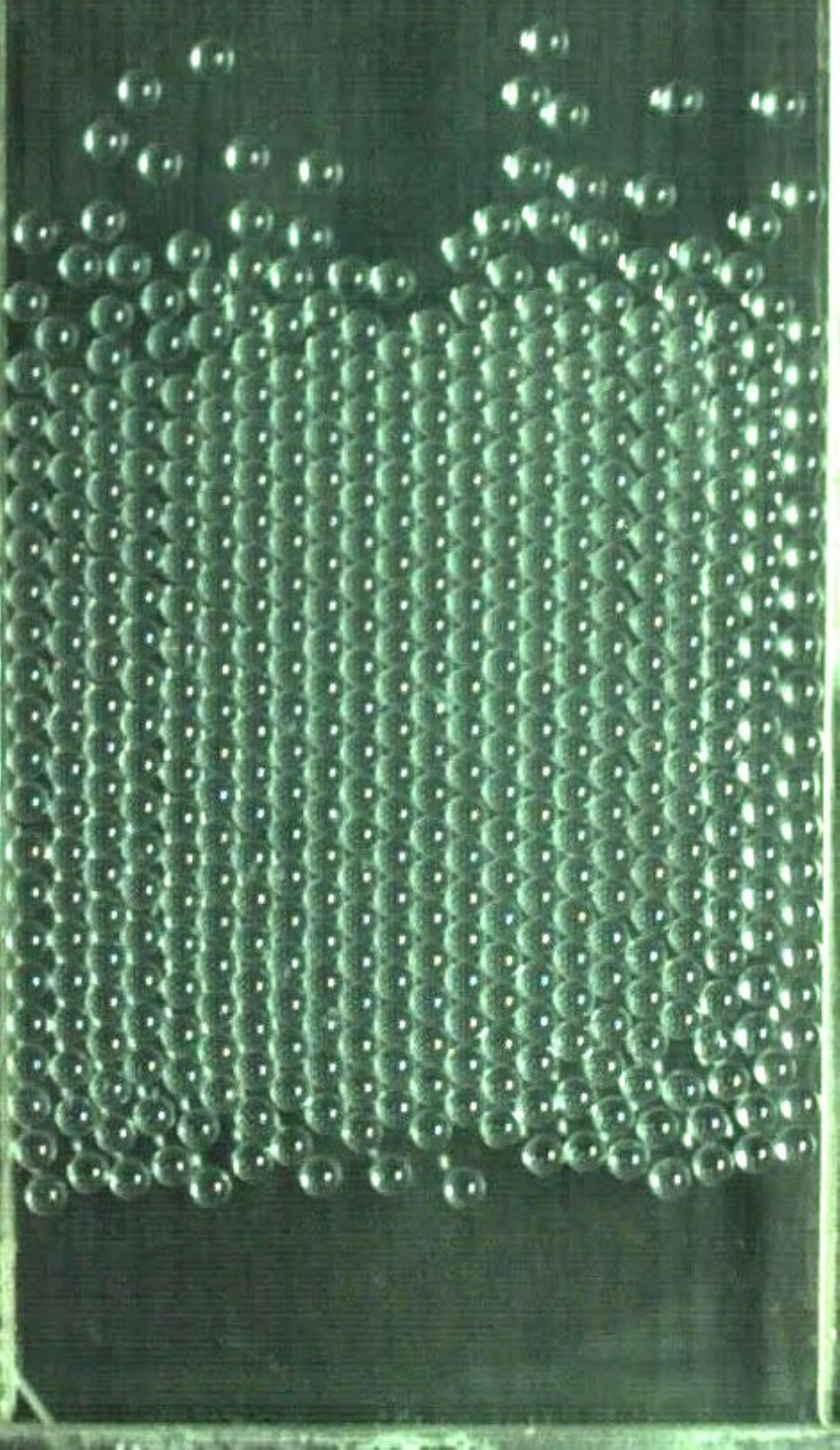}
\includegraphics[width=1.0in,height=1.8in]{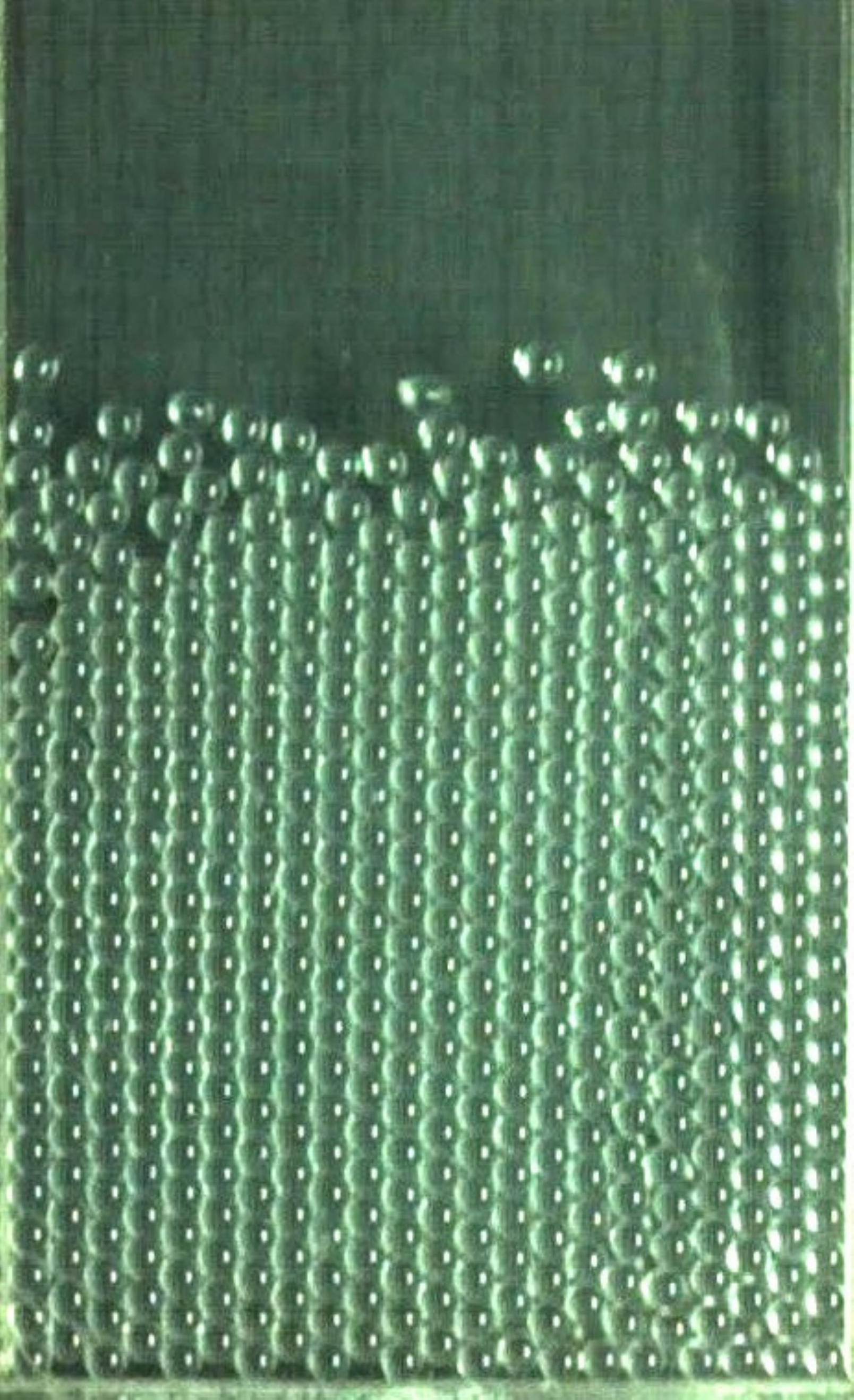}
\includegraphics[width=1.0in,height=1.8in]{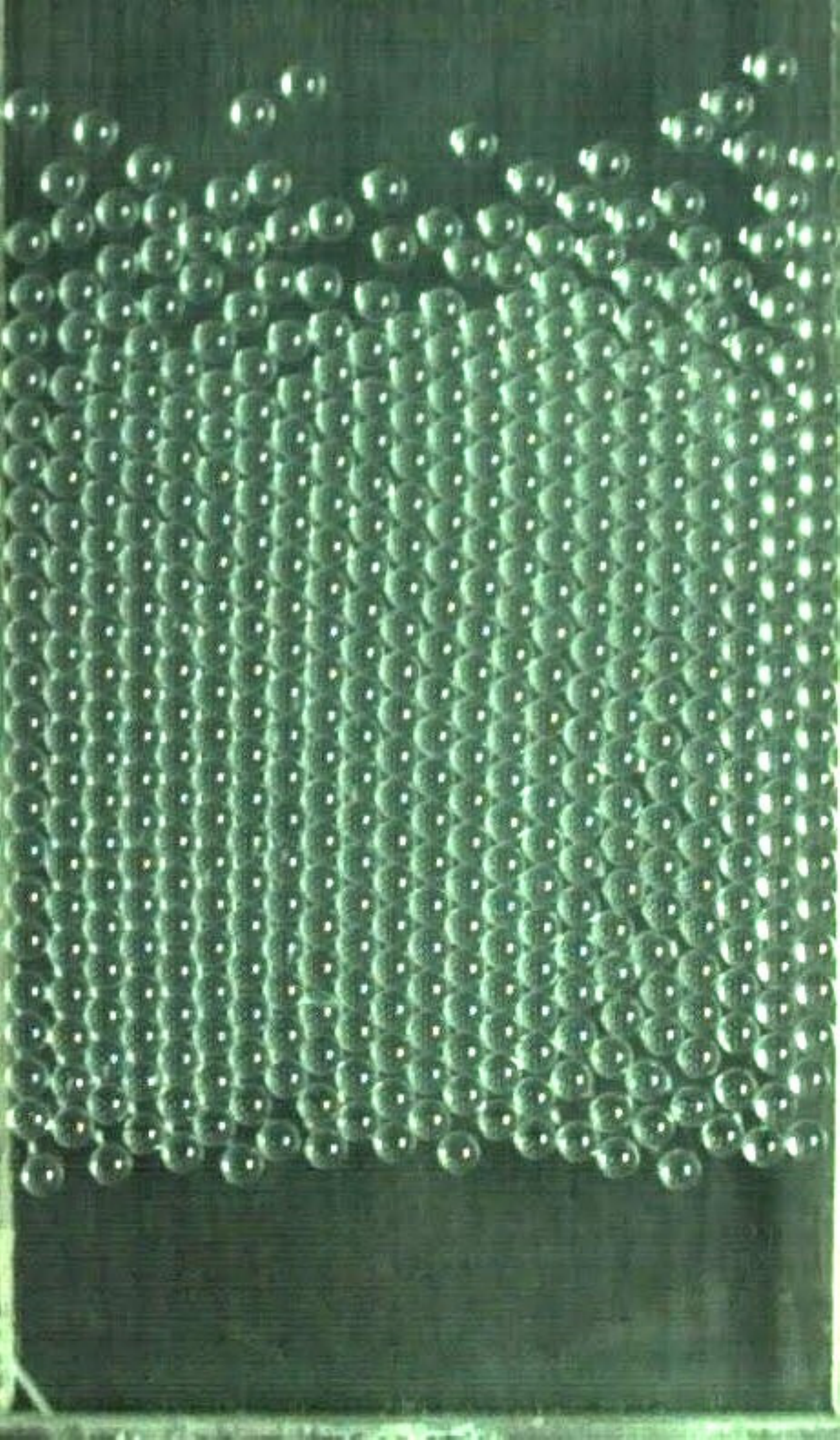}
\caption{
Snapshots of the bouncing-bed (\textit{BB})  at three successive instants of the oscillation cycle:
$t=0\tau$ (left), {$t=\tau/2$} (middle), and {$t=\tau$} (right). 
The shaking acceleration and amplitude are  $\Gamma=5$ ($f=16.08\;Hz$) and $A/d=2.4$, respectively,
with other parameters as in Fig.~\ref{fig:fig2}.
}
\label{fig:fig3}
\end{figure}

The phase diagram in ($\Gamma, A/d$)-plane for $F=25$ layers of $2.0\;mm$ diameter glass beads is displayed in Fig.~\ref{fig:fig2}. 
There are three regimes in Fig.~\ref{fig:fig2}: the Solid Bed (\textit{SB}), Bouncing Bed (\textit{BB}) and Leidenfrost state (\textit{LS}).
At any $A/d$ with $\Gamma \leq 1$, the granular bed moves synchronously with the shaker motion without getting detached from the container base-- this is the
regime of \textit{Solid Bed (SB)}. As the shaking intensity is increased beyond
some critical value ($\Gamma > 1$), the particles get detached from the base  of the container and starts
a collective motion resembling that of a single particle bouncing off a plate-- this is
dubbed \textit{Bouncing Bed (BB)} regime. Three successive snapshots of the \textit{BB} at
$t=0$, $\tau/2$, and $\tau$, where $\tau=1/f$ is the period of shaking, are displayed in Fig.~\ref{fig:fig3} for parameter values of $\Gamma=5$ and $A/d=2.4$.
The inset in Fig.~\ref{fig:fig2} indicates that the transition from the solid bed to the bouncing bed regimes occurs at a shaking intensity $\Gamma_{SB}^{BB}$
that remains relatively independent of the shaking amplitude $A/d$.
 Note that the data points in  Fig.~\ref{fig:fig2} represent boundaries between different states --
they  have been obtained  (i) via a visual inspection of running-images on the computer while carrying out experiments 
and (ii) later via a  `frame-by-frame' analysis of the acquired high-speed images ($1000$ frames/s); in addition, we have also
checked their accuracy by calculating the coarse-grained density and velocity fields (see Sec.~III.C) at two locations above and below
several transition points.

\begin{figure}[!ht]
\centering
\includegraphics[width=1.0in,height=1.8in]{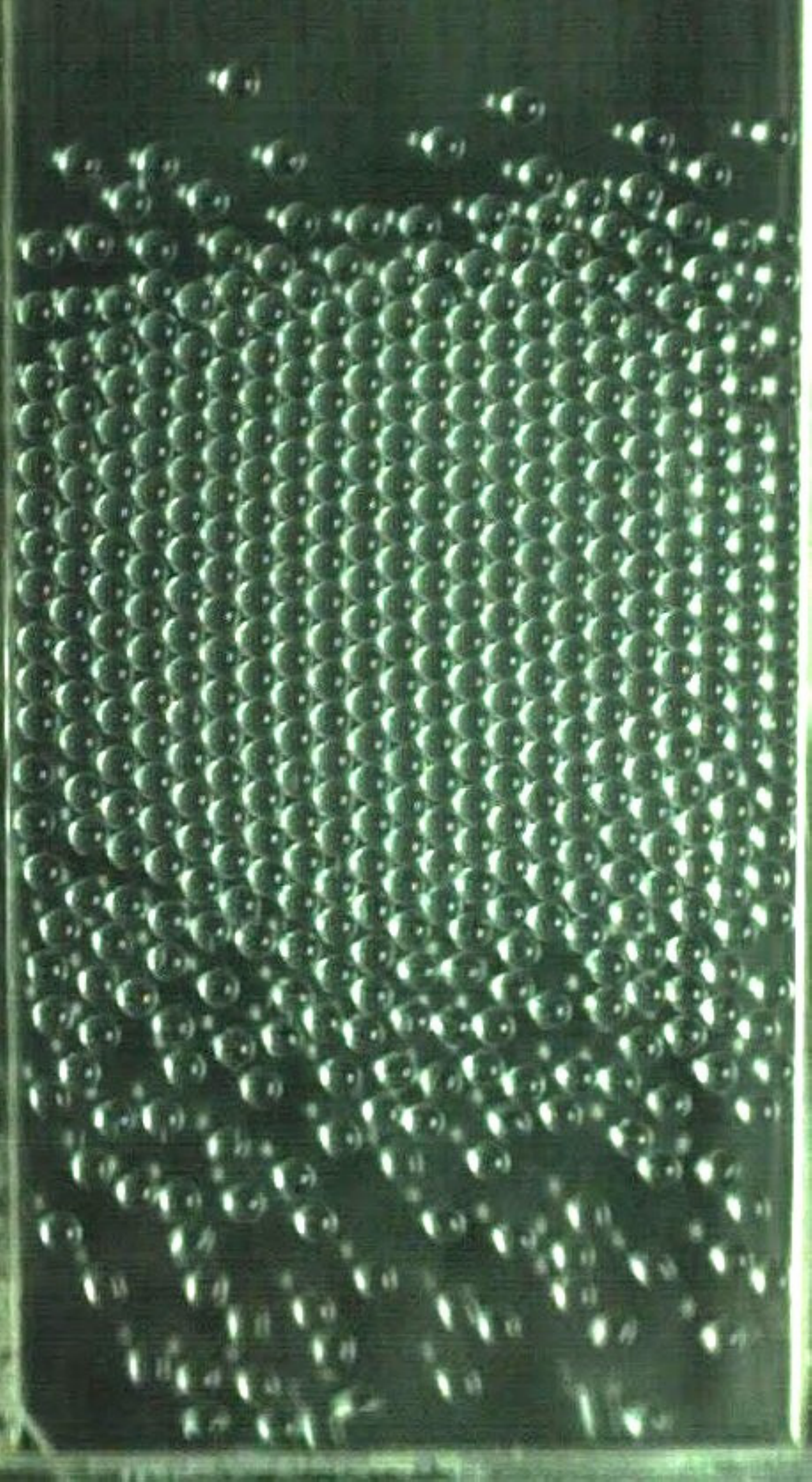}
\includegraphics[width=1.0in,height=1.8in]{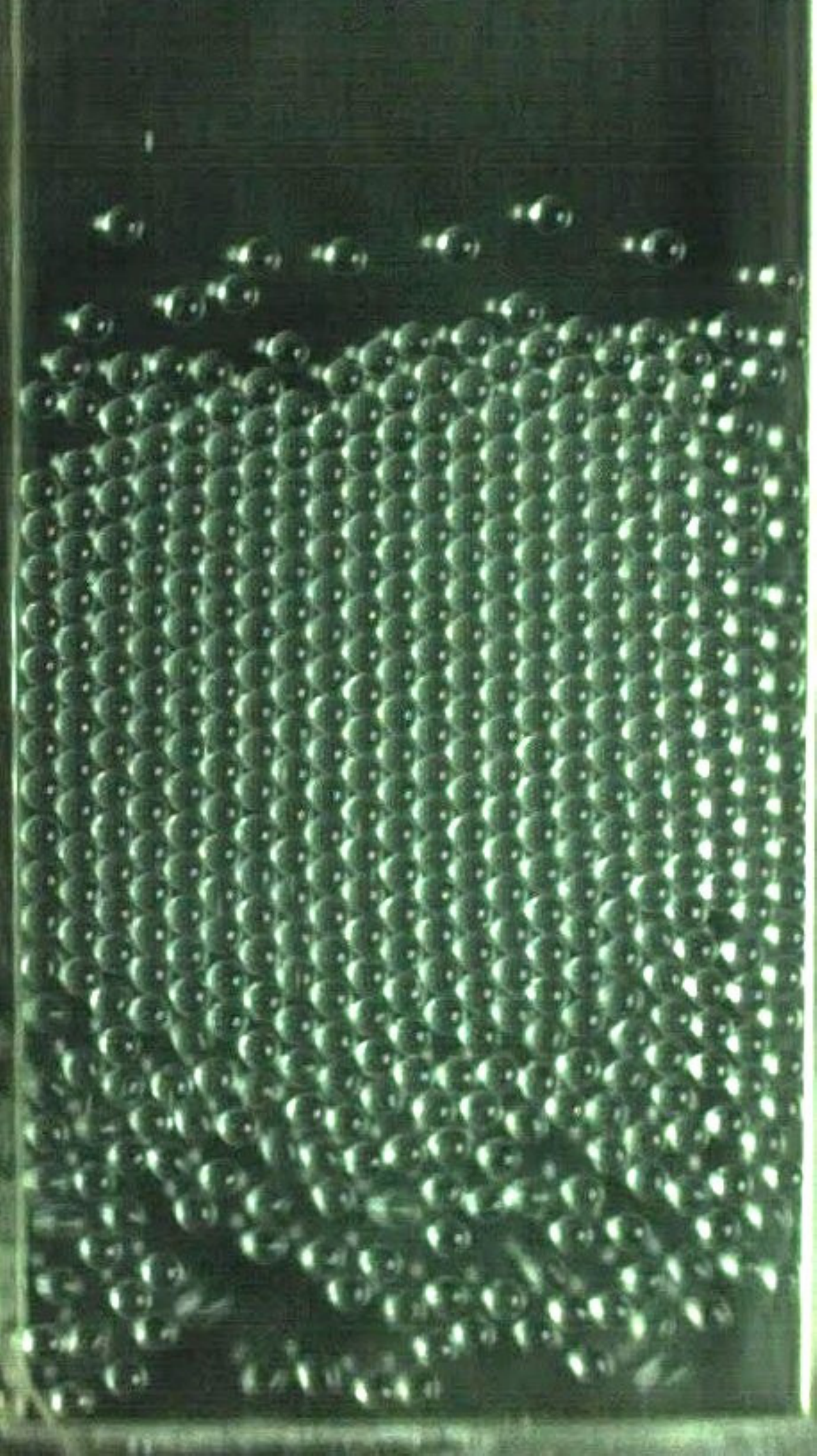}
\includegraphics[width=1.0in,height=1.8in]{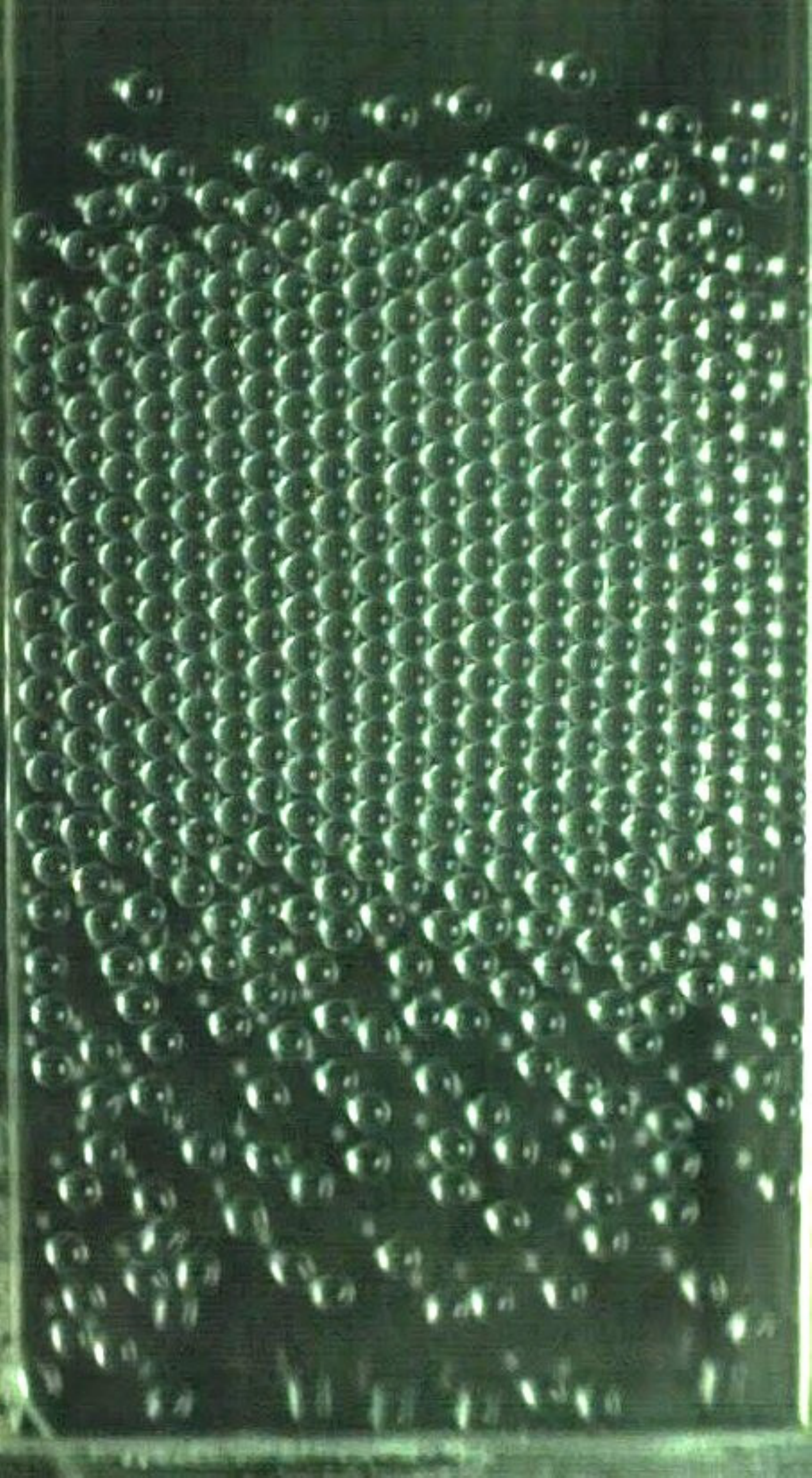}
\caption{
Snapshots of the granular Leidenfrost state (LS) at three successive time instants of the oscillation cycle: $t=0$ (left), $t=\tau/2$ (middle) and $t=\tau$ (right). 
The shaking acceleration is  $\Gamma=30$ ($f=39.4\;Hz$), with other parameters as in Fig.~\ref{fig:fig3}.
}
\label{fig:fig4}
\end{figure}

If one increases the shaking intensity ($\Gamma$) from the \textit{BB}-regime beyond some critical value
(see Fig.~\ref{fig:fig2}), the bouncing bed transits to a  ``density-inverted'' state~\cite{LR1995,MPB2003,Eshuis2005}.
The latter corresponds to a state in which a dense region of nearly crystal-packed particles floats
over a dilute region of fast moving particles, dubbed ``floating-cluster''~\cite{MPB2003} or granular ``Leidenfrost'' state (LS)~\cite{Eshuis2005}.
The characteristic features of the LS are evident from Fig.~\ref{fig:fig4} which shows
three successive snapshots at $t=0$, $\tau/2$ and $\tau$ of the \textit{LS} over an oscillation cycle  at a shaking acceleration of $\Gamma=30$.

The blue squares in Fig.~\ref{fig:fig2} (see also its inset) indicate that the transition from the bouncing bed  to the granular Leidenfrost state
depends strongly on the shaking amplitude $A/d$.  The inset on logarithmic-scale confirms that the corresponding critical shaking acceleration
$\Gamma_{BB}^{LS}$ at which this  transition occurs follows a power law:
\begin{equation}
  \Gamma_{BB}^{LS} \equiv \Gamma_c \approx  20.74\bigg(\frac{A}{d}\bigg)^{-\frac{7}{8}}.
\label{Eqn:GammaA/d1}
\end{equation}
At higher shaking amplitudes the onset of \textit{LS} is expected to occur at lower values of $\Gamma$
since the input energy (via shaker) to the granular materials is proportional to $\Gamma(A/d)$.

\begin{figure}[!ht]
\centering
\includegraphics[width=3.3in,height=3.3in]{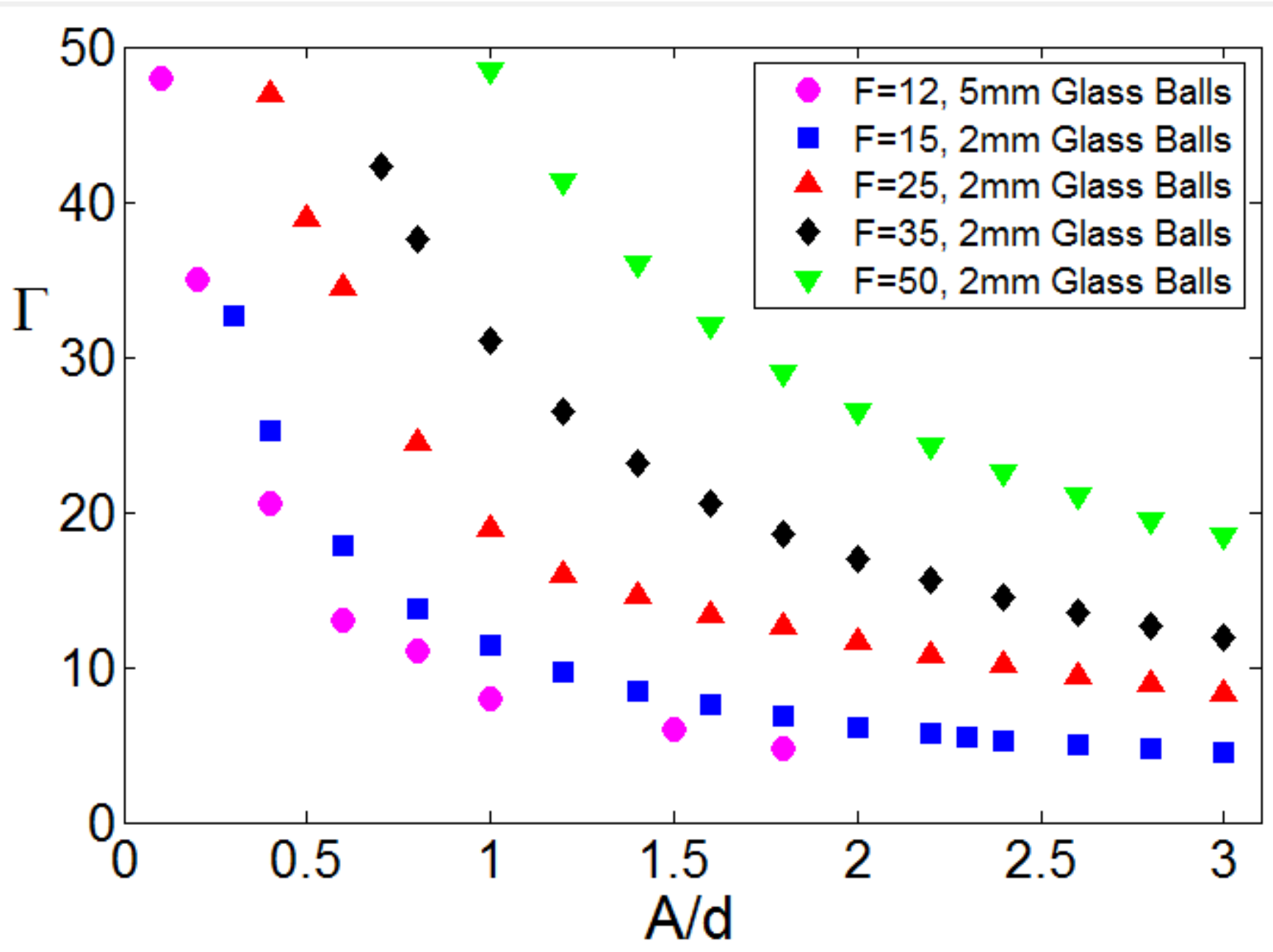}
\caption{
(Color online)
Effect of filling-height $F=h_0/d$ on the $BB\to LS$-transition, see the legend for $F$-values, with other parameters as in Fig.~\ref{fig:fig2}.
}
\label{fig:fig5}
\end{figure}

Similar to Fig.~\ref{fig:fig2}, we have performed a series of experiments by varying the initial filling-height $F$ of $2\ mm$ diameter glass-beads 
as well as by varying the diameter of beads. These data on the critical values of ($\Gamma, A/d$) for the onset of Leidenfrost state
are displayed in Fig.~\ref{fig:fig5} as denoted by different symbols.
It is seen that for a given $A/d$ the critical shaking acceleration for $BB\to LS$-transition, $\Gamma_{BB}^{LS}$, increases with increasing $F$.
This dependence on $F$ is expected since increasing $F$ increases the weight of the granular bed which, in turn, requires
a higher shaking intensity ($\Gamma$) to get transition.

\begin{figure}[!ht]
\centering
\includegraphics[width=3.3in,height=3.0in]{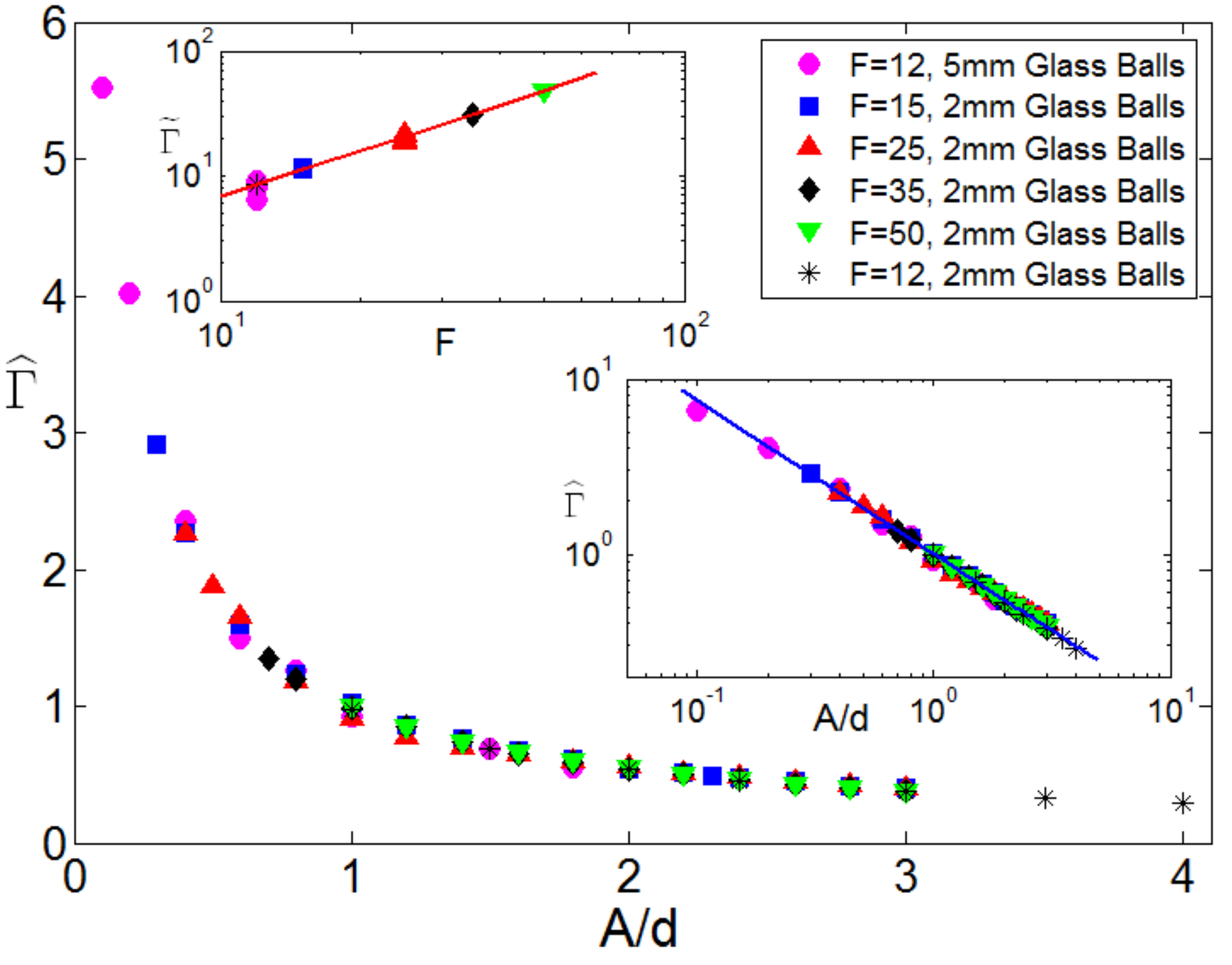}
\caption{
(Color online)
 Master phase-diagram in the ($\widehat{\Gamma}, A/d$)-plane, where $\widehat{\Gamma}=\Gamma/\alpha F^{\beta}$. The bottom inset
displays the same phase diagram in logarithmic scale, with the blue-line representing the power-law $\widehat{\Gamma}\sim (A/d)^{-7/8}$.
The top inset shows the variation of $\widetilde{\Gamma}$, Eqn. (5),  with $F$, with the red line being the best-fit curve.
The `star' symbols  refer to the data in Fig.~\ref{fig:fig9} ($L/d=40$ and $F=12$).
}
\label{fig:fig6}
\end{figure}

To determine the dependence of $\Gamma_{BB}^{LS}$ on the initial filling-height $F$,
we assume that the power-law scaling, Eqn.~(\ref{Eqn:GammaA/d1}), with $A/d$ holds for all $F$, to be verified {\it a posterioi}.
All data of Fig.~\ref{fig:fig5} are now rescaled via
\begin{equation}
 \widetilde{\Gamma} = \Gamma (A/d)^{7/8},
\label{Eqn:tildeGamma1}
\end{equation}
and its variation with $F$ is shown in the upper inset of Fig.~\ref{fig:fig6} on logarithmic scale.
It is clear that $\widetilde{\Gamma}$ has a power-law scaling with $F$:
\begin{equation}
 \label{eq:GammaTilde1}
 \widetilde{\Gamma}=\alpha F^\beta.
\end{equation}
The slope and intercept of the best fit curve (the red line in the upper inset) yields:
\begin{equation}
\alpha\approx 0.414 \quad
\mbox{and}
\quad
 \beta\approx 1.217.
\end{equation}
To demonstrate that the same power-law dependence on  the shaking amplitude [$\sim (A/d)^{-7/8}$, Eqn.~\ref{Eqn:GammaA/d1}]  holds for all $F$,
we plot the variation of the following quantity
 \begin{eqnarray}
   \label{eq:GammaHat}
   \widehat{\Gamma}=\frac{\Gamma}{\alpha F^\beta} \equiv (A/d)^{-7/8}
  \end{eqnarray}
with $A/d$ in the main panel of Fig.~\ref{fig:fig6}. It is seen that all experimental data for different $F$ and $d$ collapse very well on a single curve. 
The same data are re-plotted on logarithmic scale in the lower inset of Fig.~\ref{fig:fig6} which re-confirms the universality of the 
power-law scaling,  Eqn.~(\ref{eq:GammaHat}), of $\widehat{\Gamma}$ with $A/d$.

In summary, the critical shaking acceleration, $\Gamma_{BB}^{LS}$, for ``$BB\to LS$''-transition satisfies the following master equation:
\begin{equation}
 \label{Eqn:ScalingLaw1}
 \Gamma_{BB}^{LS} \equiv \Gamma_c=0.414 F^{1.217} \bigg(\frac{A}{d}\bigg)^{-\frac{7}{8}},
\end{equation}
representing the blue line in the lower inset of Fig.~\ref{fig:fig6}.
Rewriting Eq.~(\ref{Eqn:ScalingLaw1}) in terms of shaking strength $S=\Gamma\times(A/d)$
(which is the ratio of  the average kinetic energy injected to the system via shaking
and the potential energy of all particles~\cite{Eshuis2005,Shukla2014}),
we obtain
\begin{equation}
 \label{Eqn:ScalingLaw2}
 S_{BB}^{LS} \equiv \Gamma_{BB}^{LS}\times(A/d) =0.414 F^{1.217} \bigg(\frac{A}{d}\bigg)^{\frac{1}{8}} ,
\end{equation}
which depends on $A/d$, albeit weakly. In contrast, the previous work of Eshuis \etal~\cite{Eshuis2005} in a similar setup 
found the constancy of $S$ at $BB\to LS$-transition [their experiments correspond to lower values $A/d\sim O(0.1)$]
This weak-increase of $S$ with increasing $A/d$ might be tied to (i)  increased frictional-barrier at the front and back walls in the same limit
and/or (ii) the coupling with the ``time-dependent'' bottom boundary condition.
Ideally, the temperature boundary condition at the vibrating wall should depend on both the shaking amplitude $A/d$ and its frequency $f$ as well as on time
 -- averaging over one oscillation cycle leads to a time-independent constant temperature at the base~\cite{Eshuis2005,Eshuis2010,Shukla2014}.
Additional experiments along with theoretical analyses with time-dependent boundary condition are needed to settle the issue of the increase of 
the critical shaking intensity with increasing $A/d$, Eqn.~(\ref{Eqn:ScalingLaw2}), in a future work.

\begin{figure}[!ht]
\centering
(a)
\includegraphics[width=1.0in,height=0.7in]{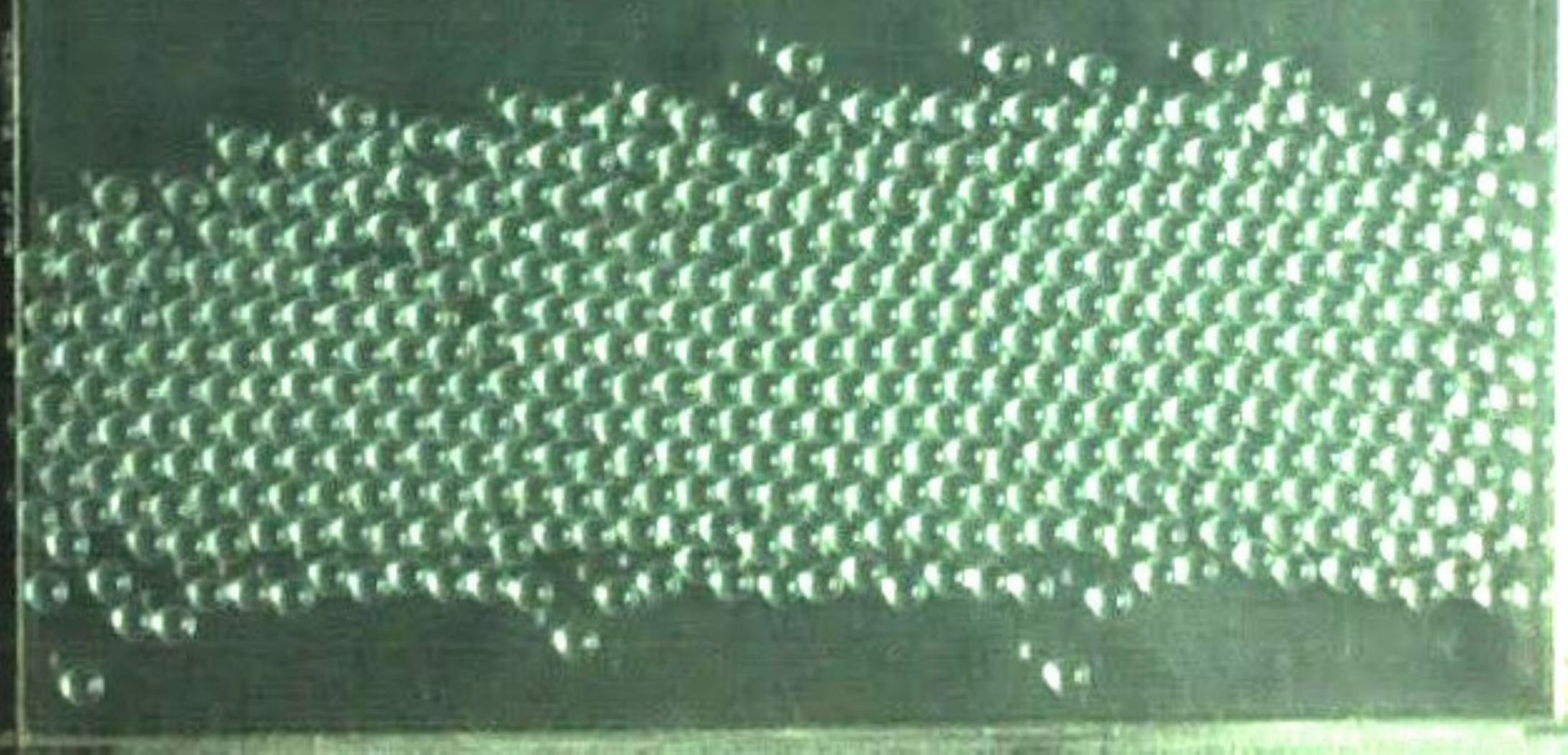}
\includegraphics[width=1.0in,height=0.7in]{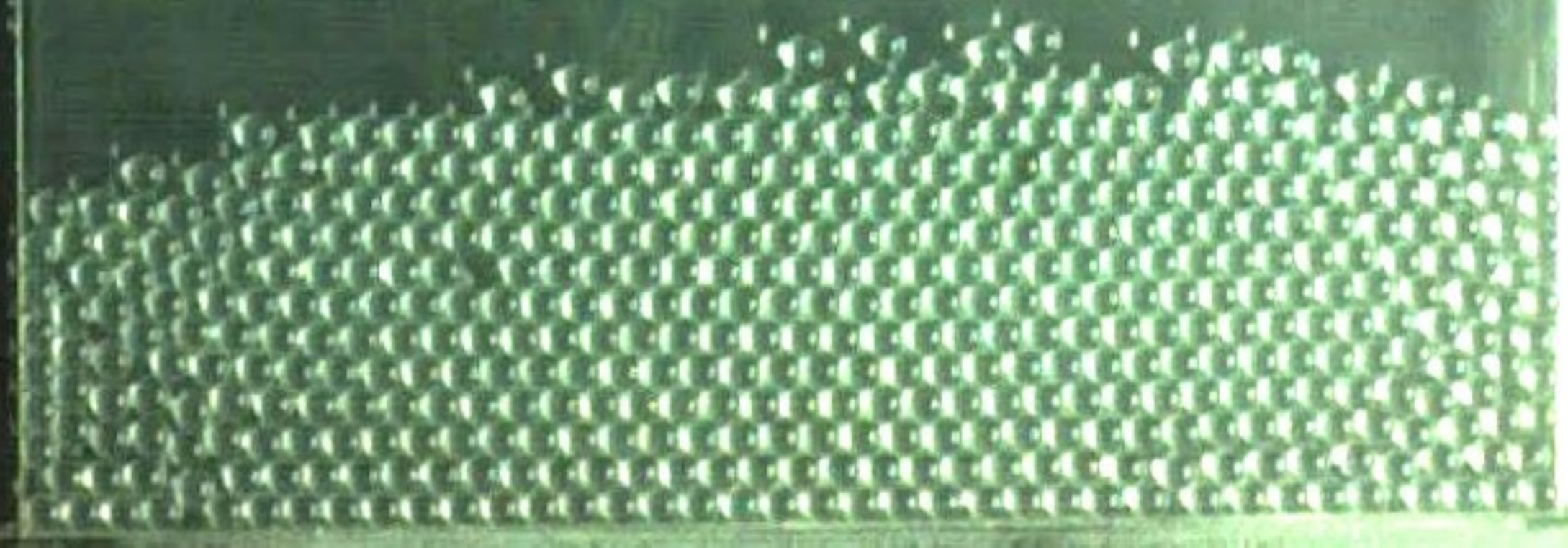}
\includegraphics[width=1.0in,height=0.7in]{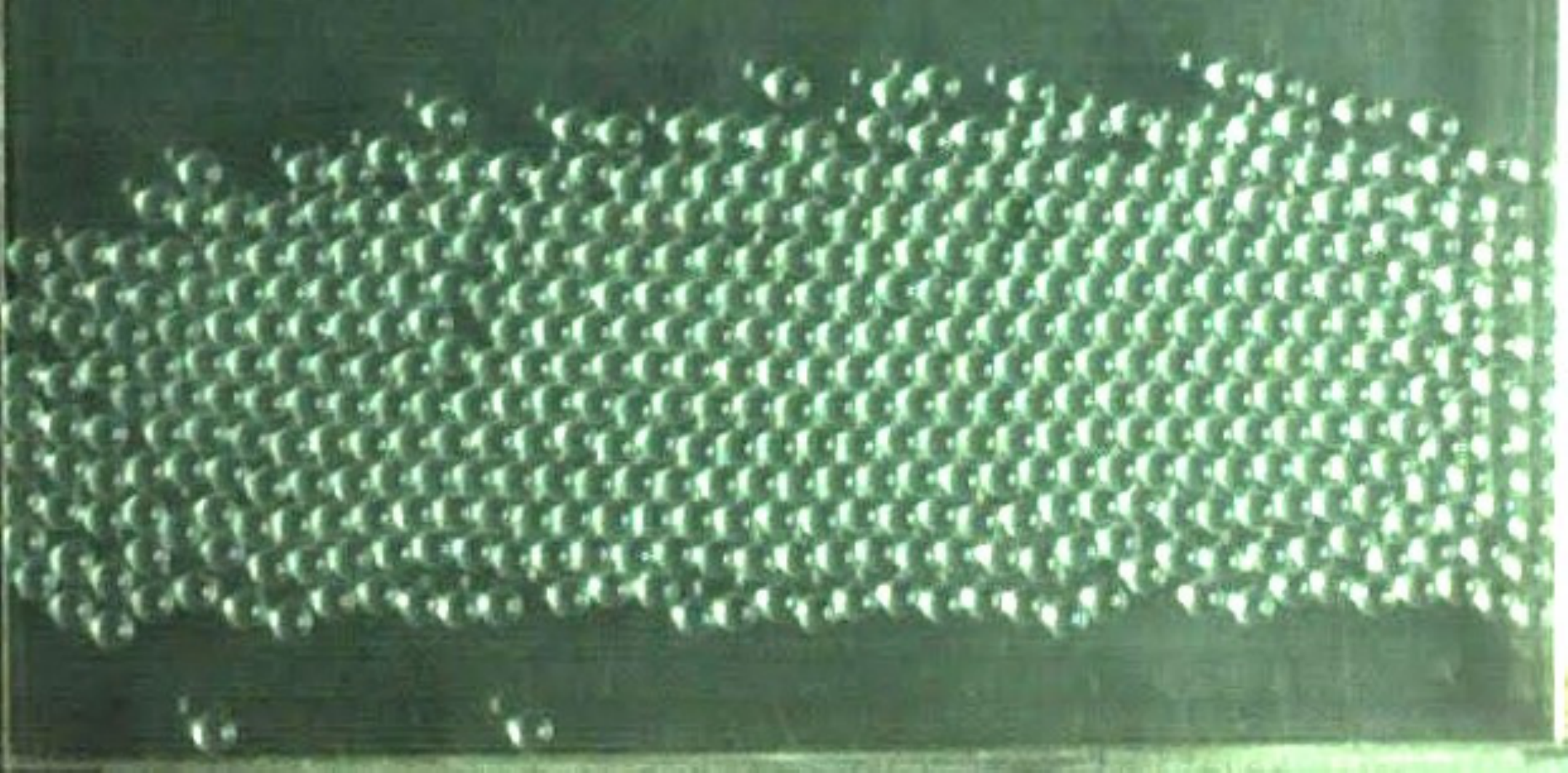}\\
(b)
\includegraphics[width=1.0in,height=0.9in]{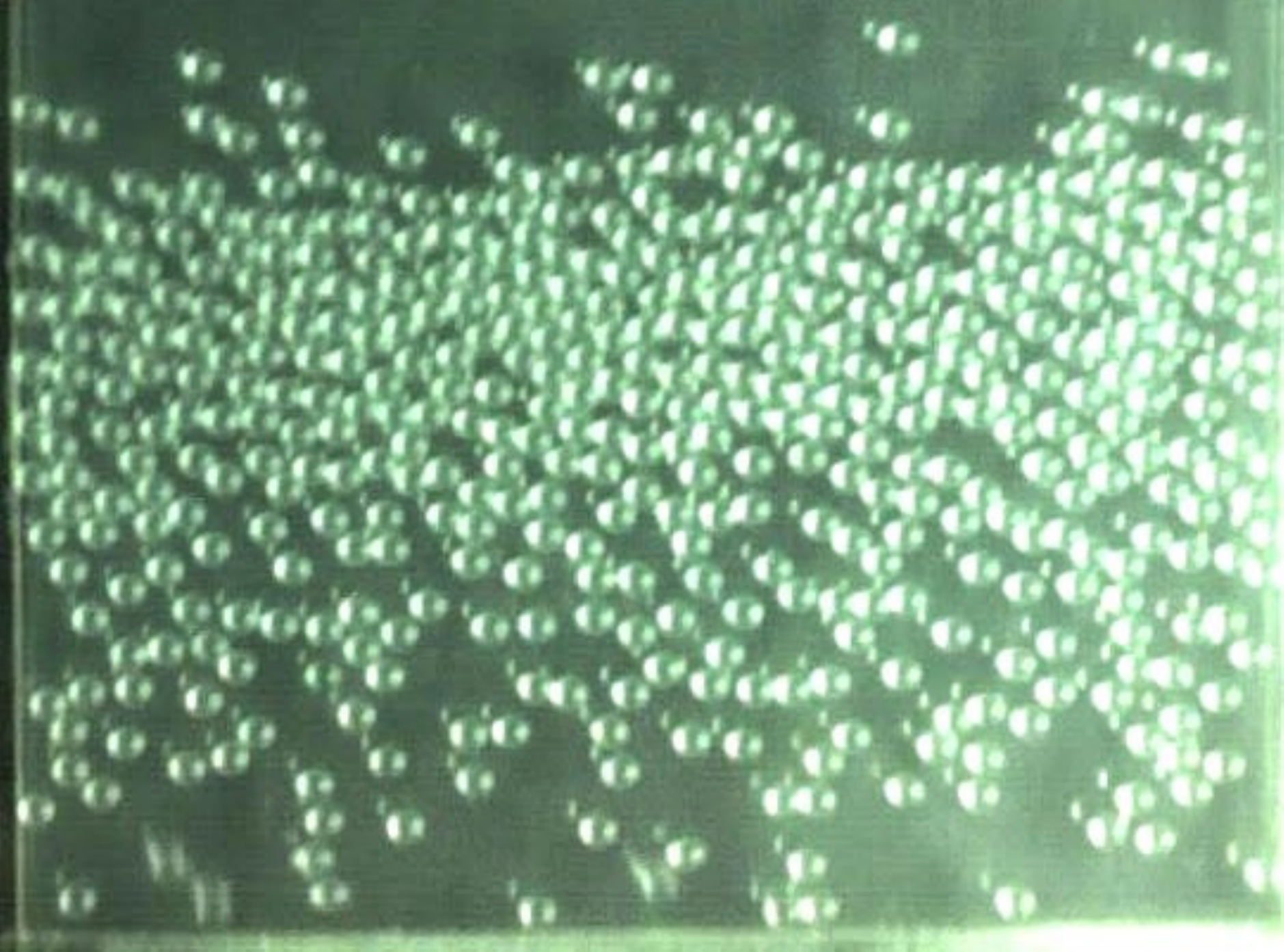}
\includegraphics[width=1.0in,height=0.9in]{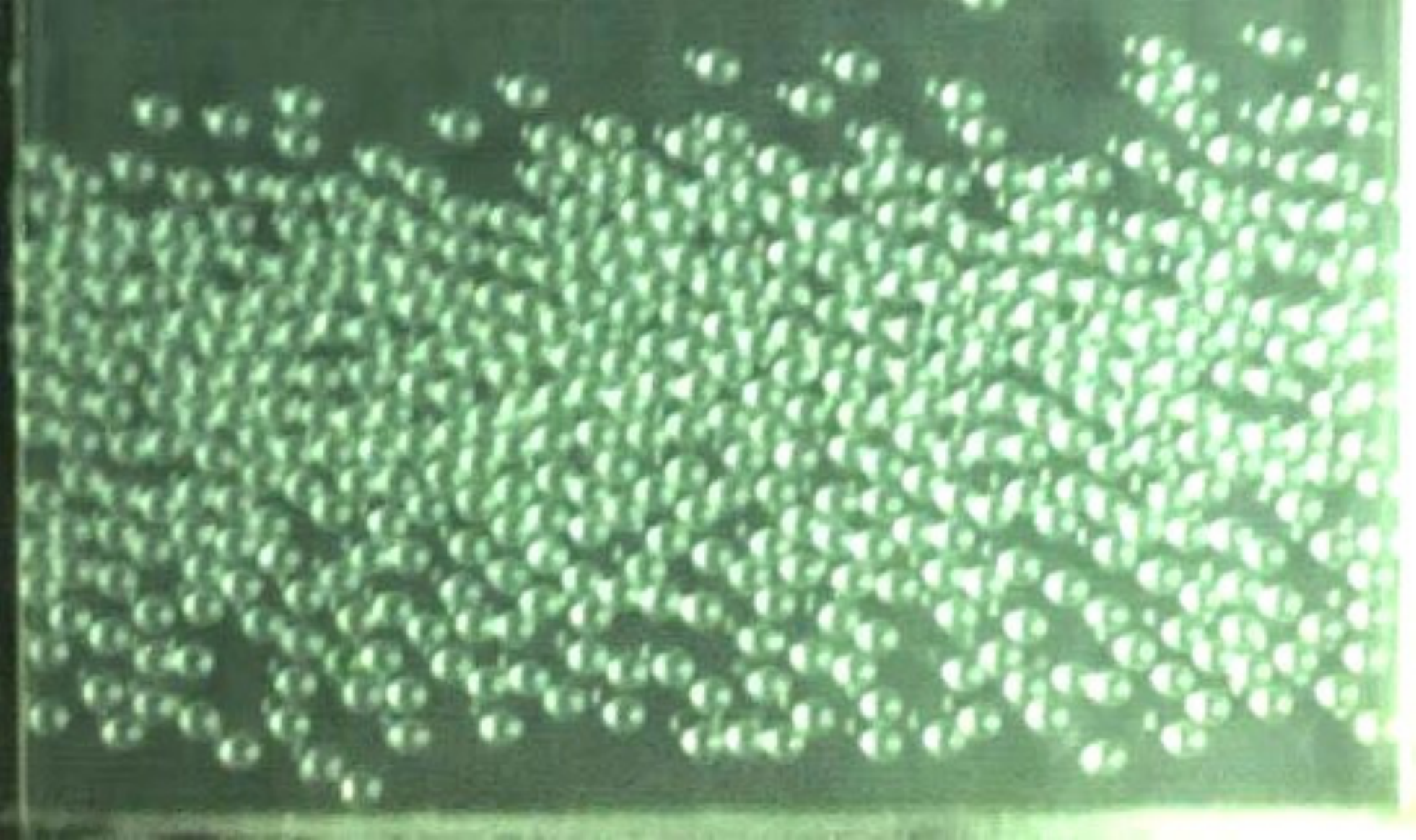}
\includegraphics[width=1.0in,height=0.9in]{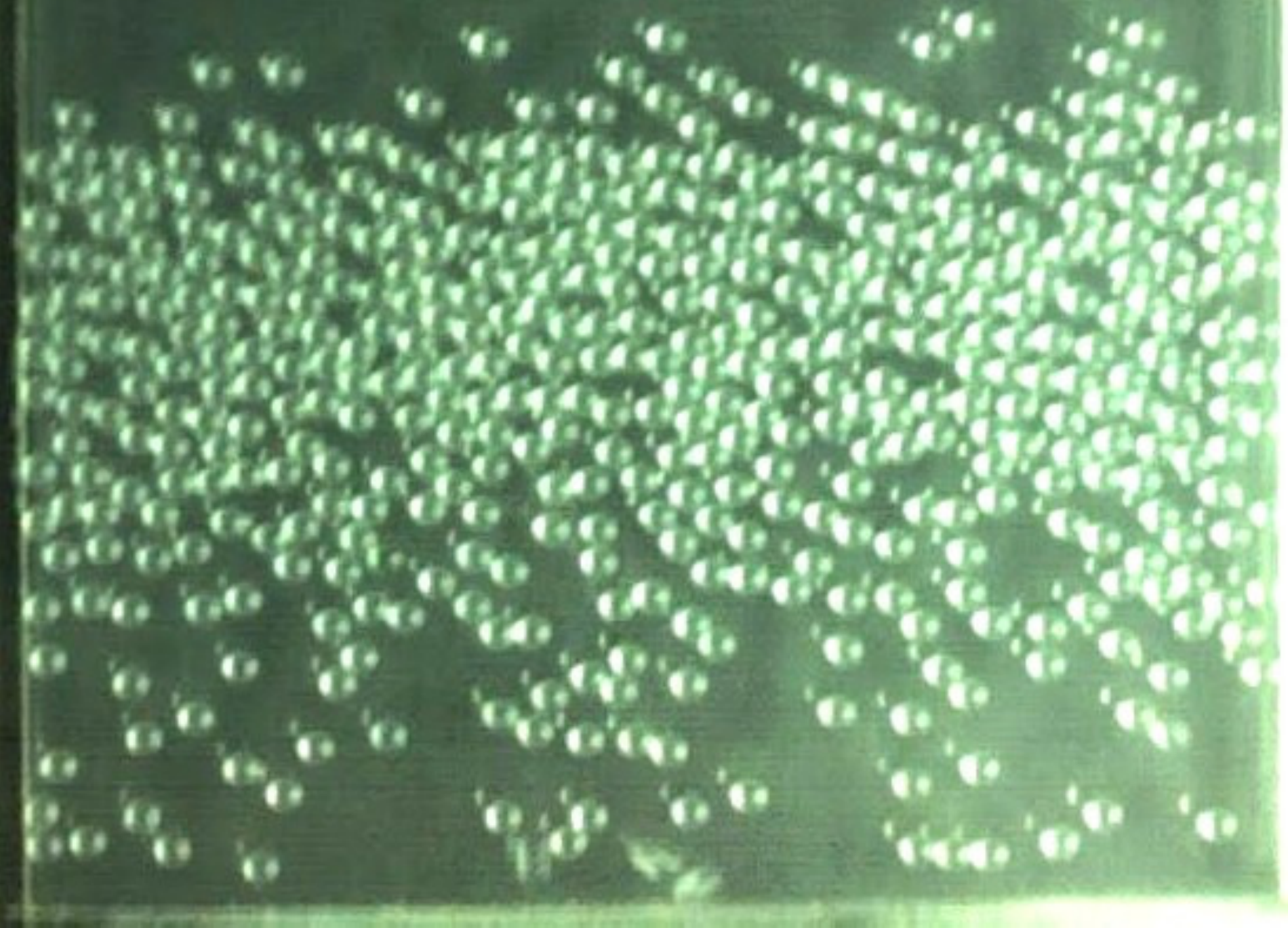}\\
(c)
\includegraphics[width=1.0in,height=1.2in]{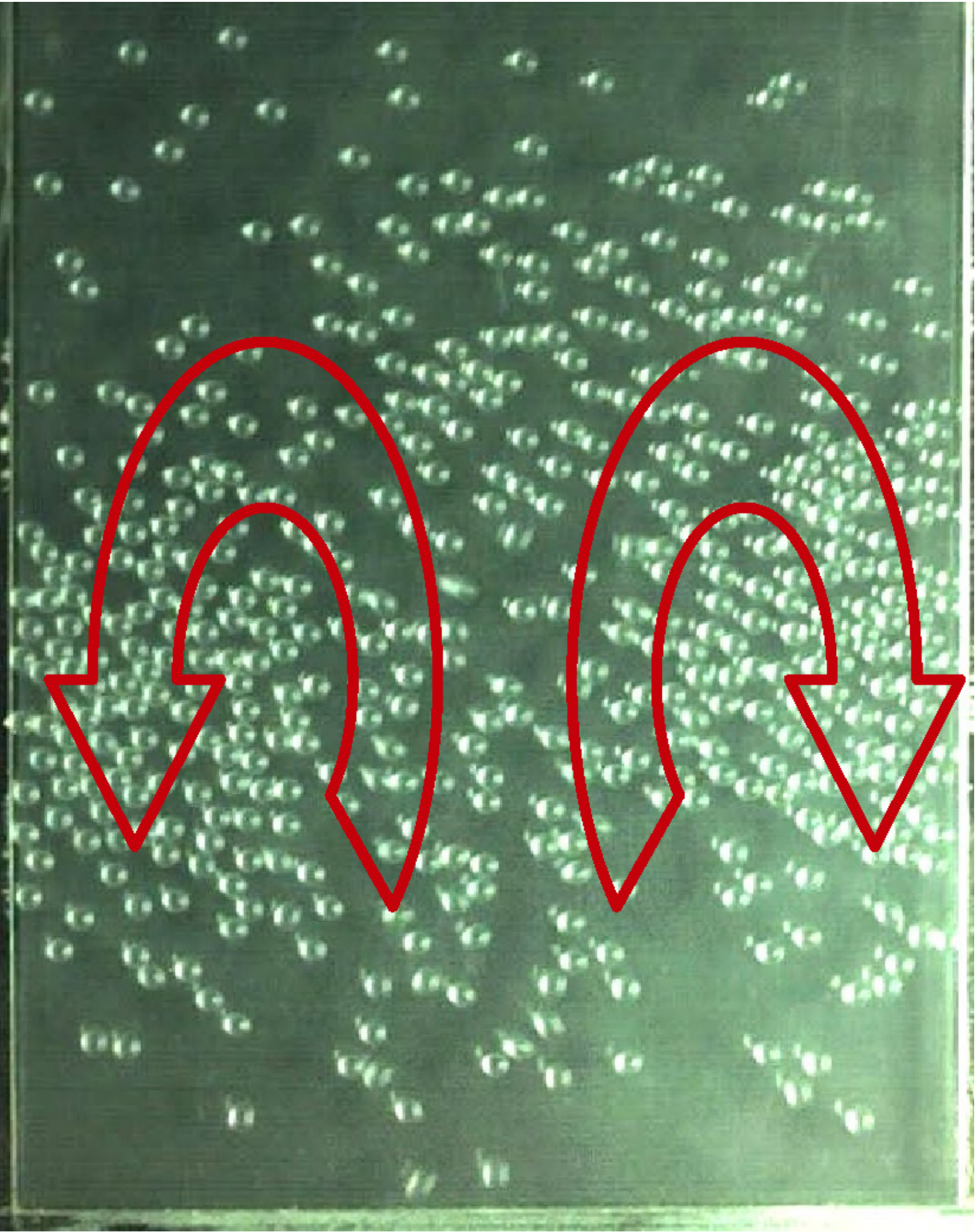}
\includegraphics[width=1.0in,height=1.2in]{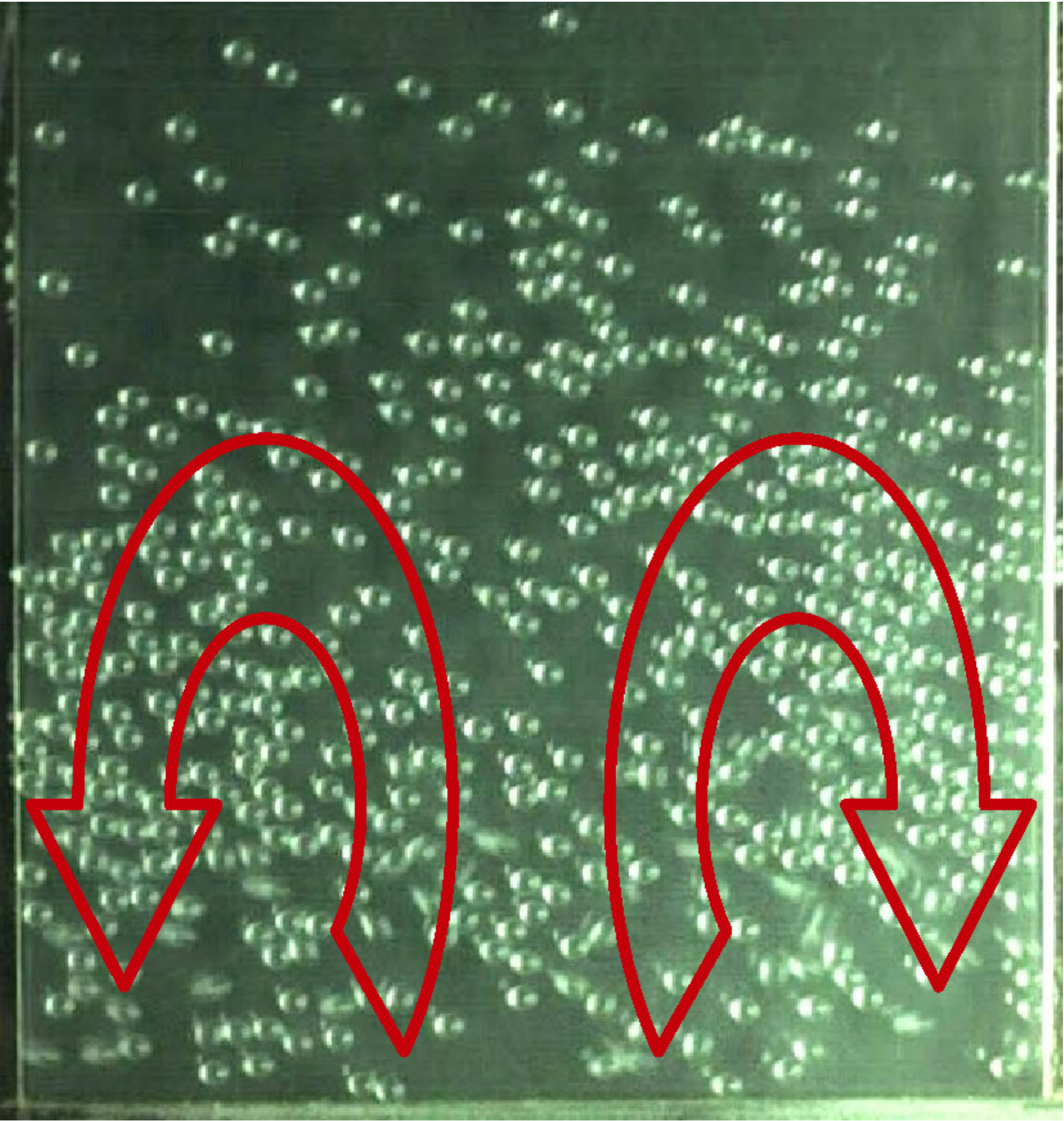}
\includegraphics[width=1.0in,height=1.2in]{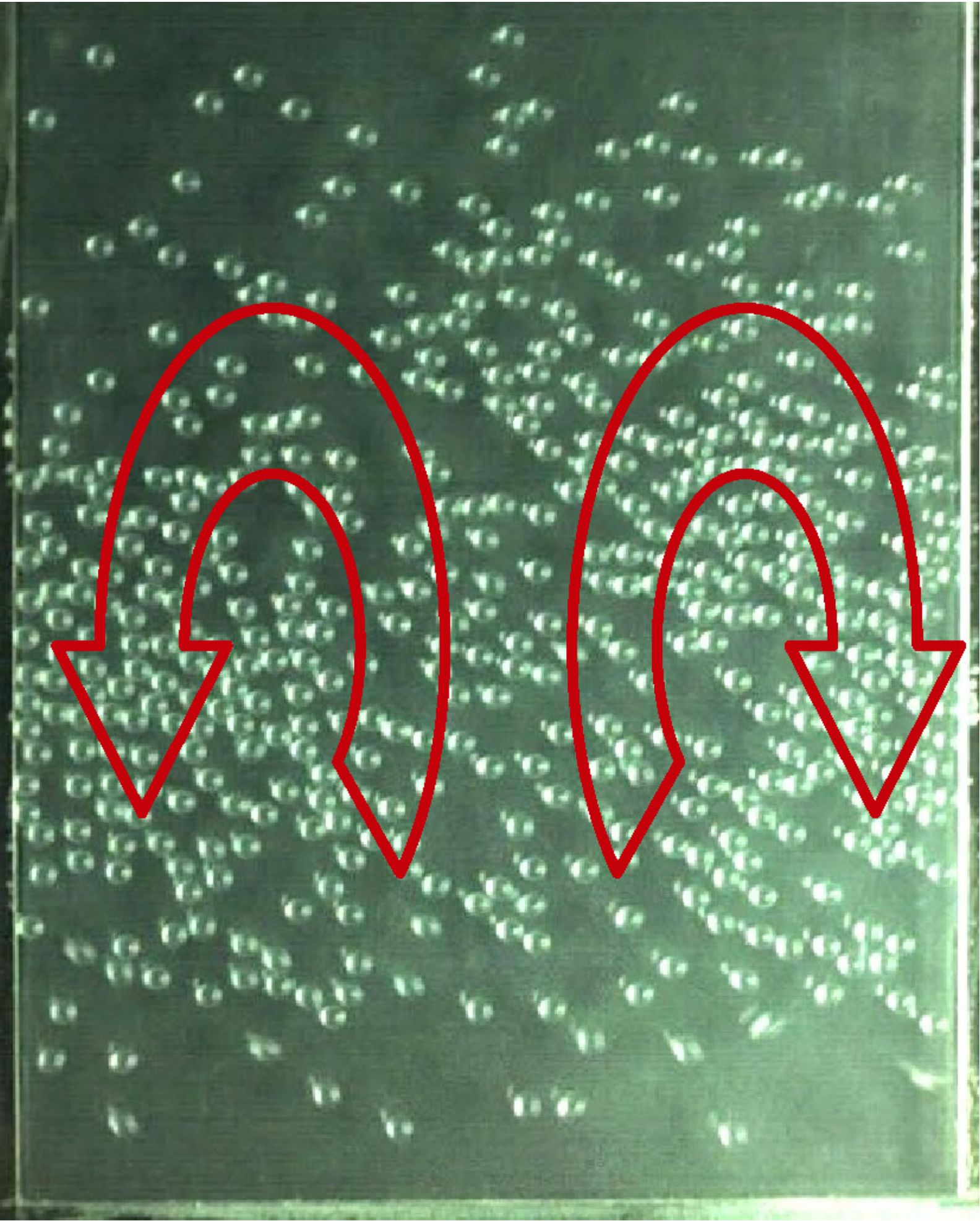}\\
(d)
\includegraphics[width=1.0in,height=1.2in]{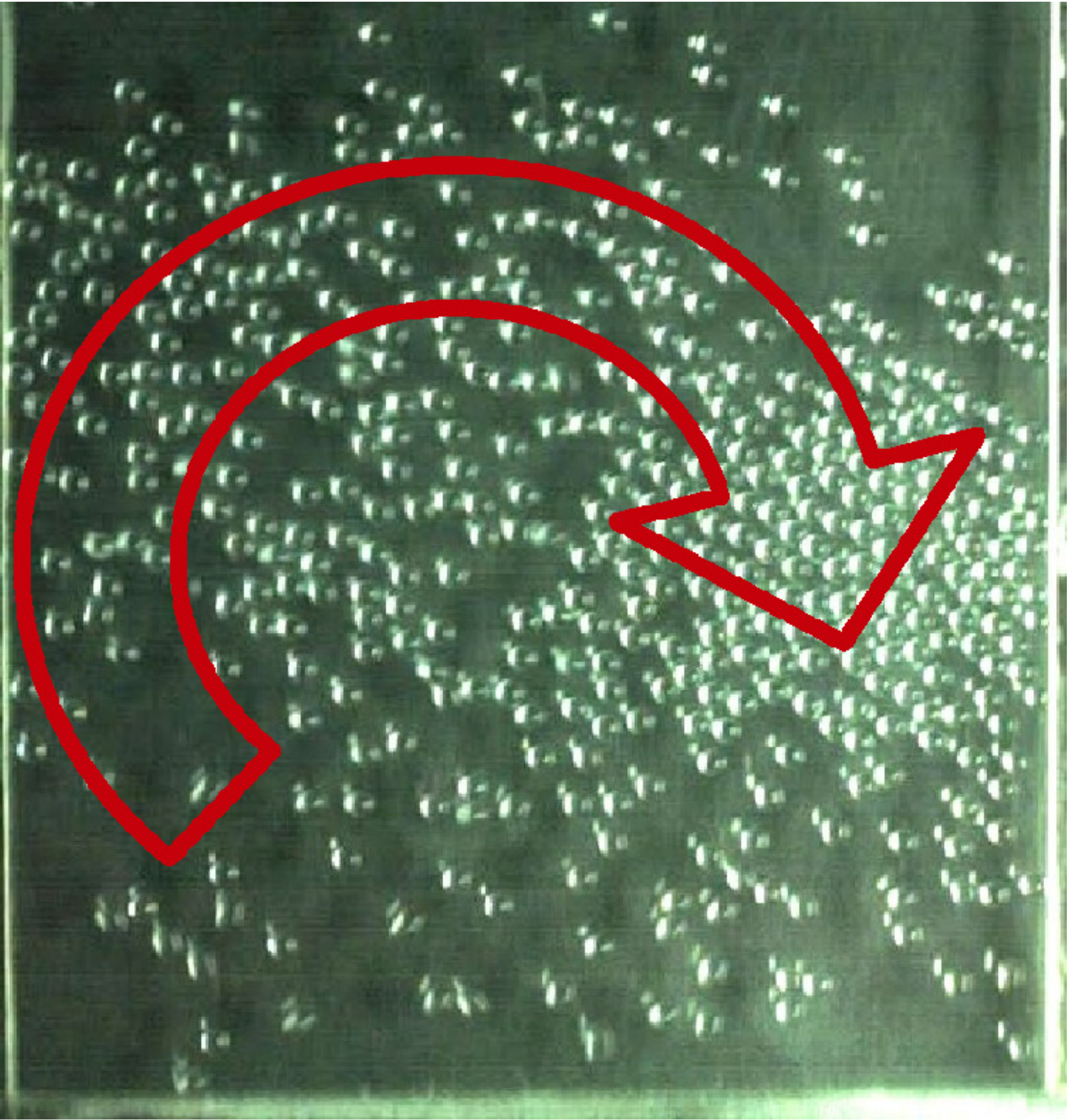}
\includegraphics[width=1.0in,height=1.2in]{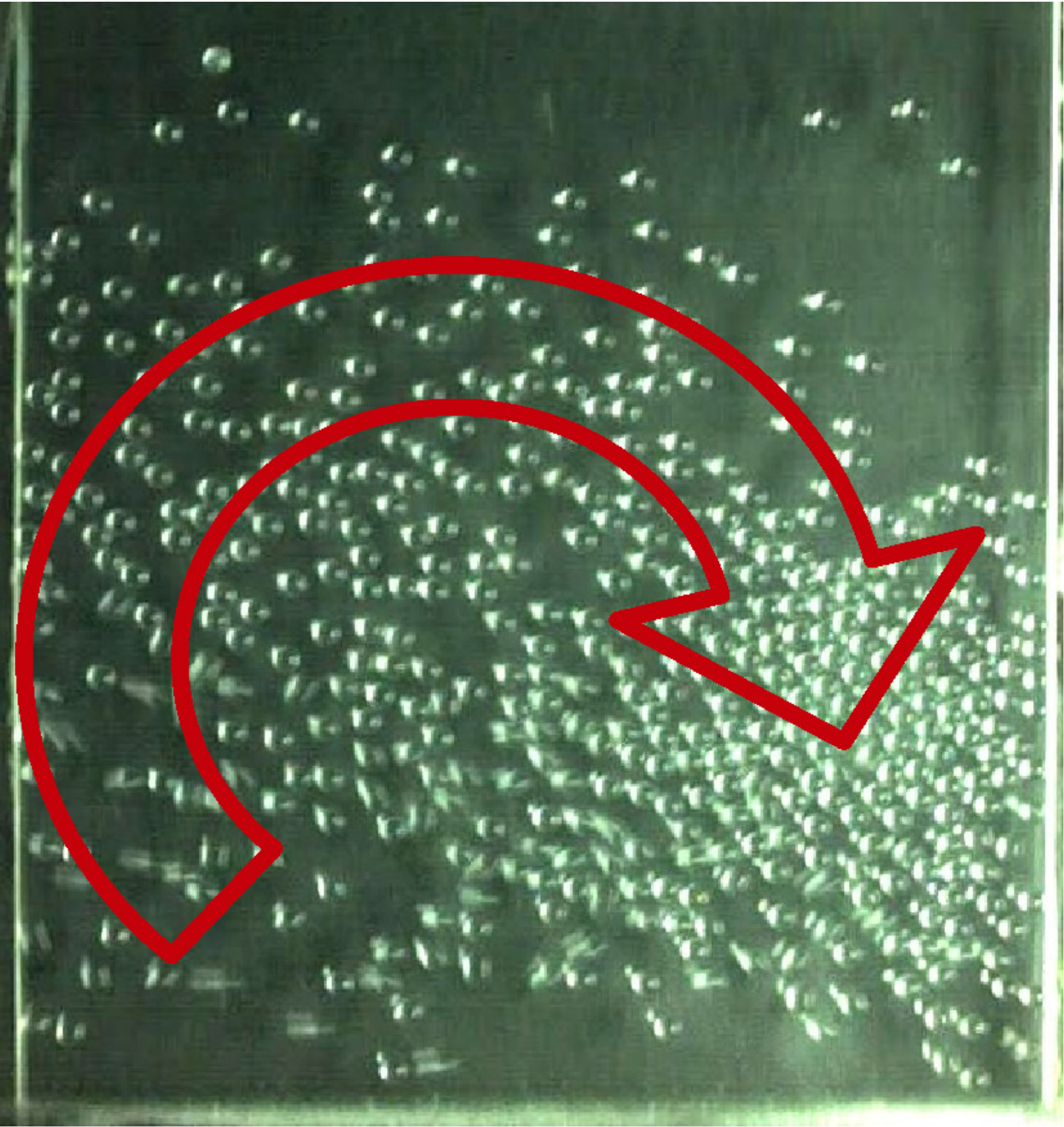}
\includegraphics[width=1.0in,height=1.2in]{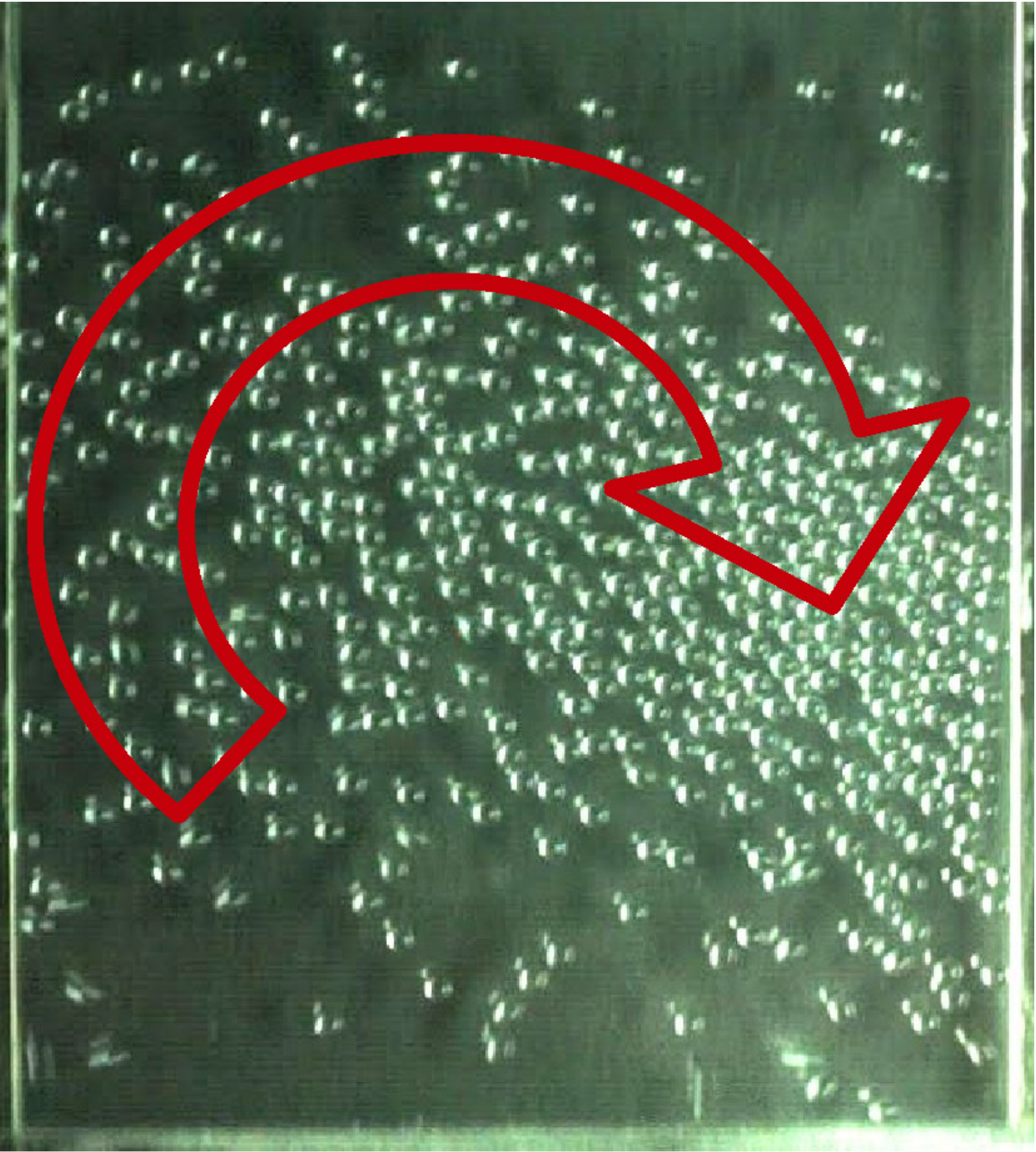}\\
(e)
\includegraphics[width=1.0in,height=1.2in]{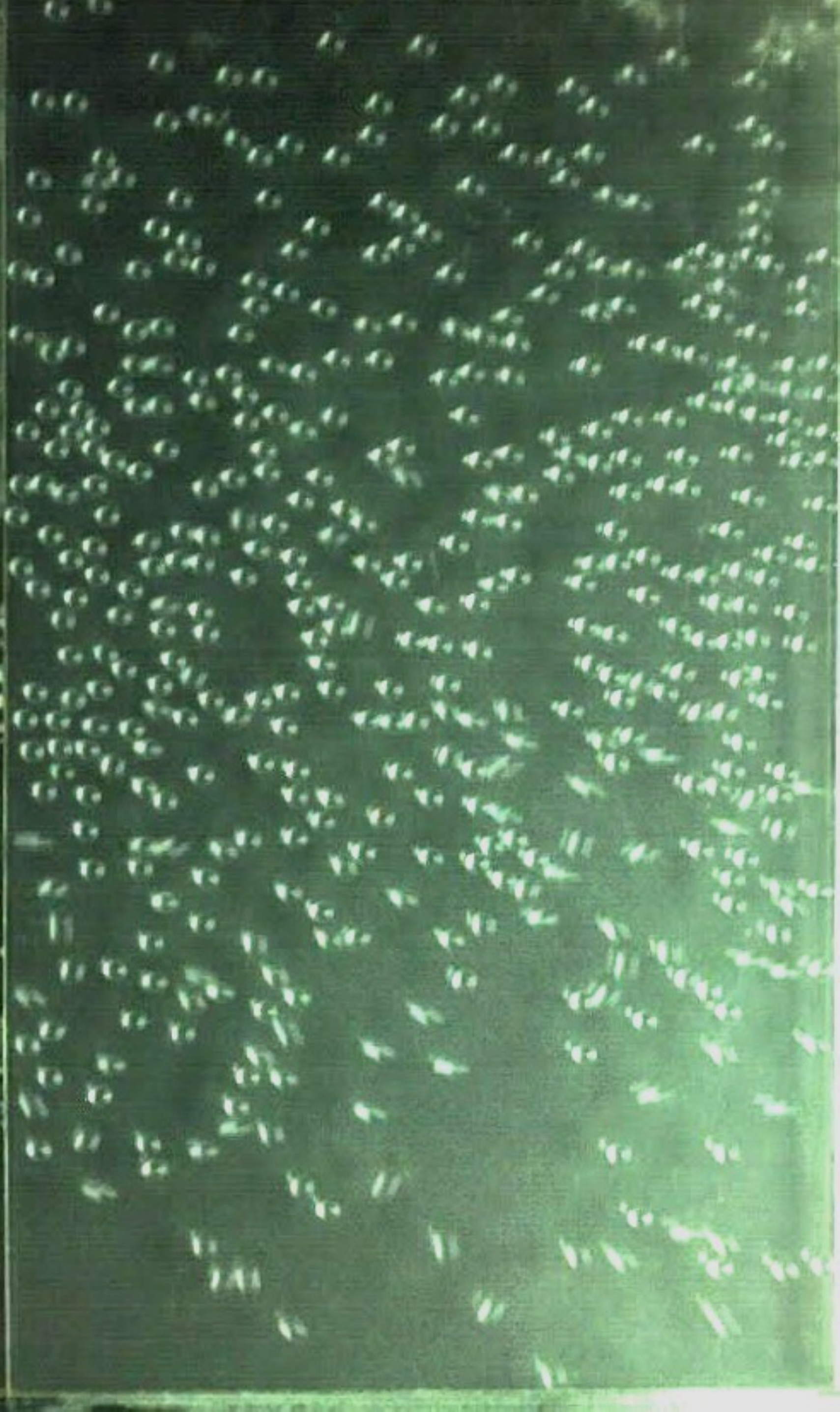}
\includegraphics[width=1.0in,height=1.2in]{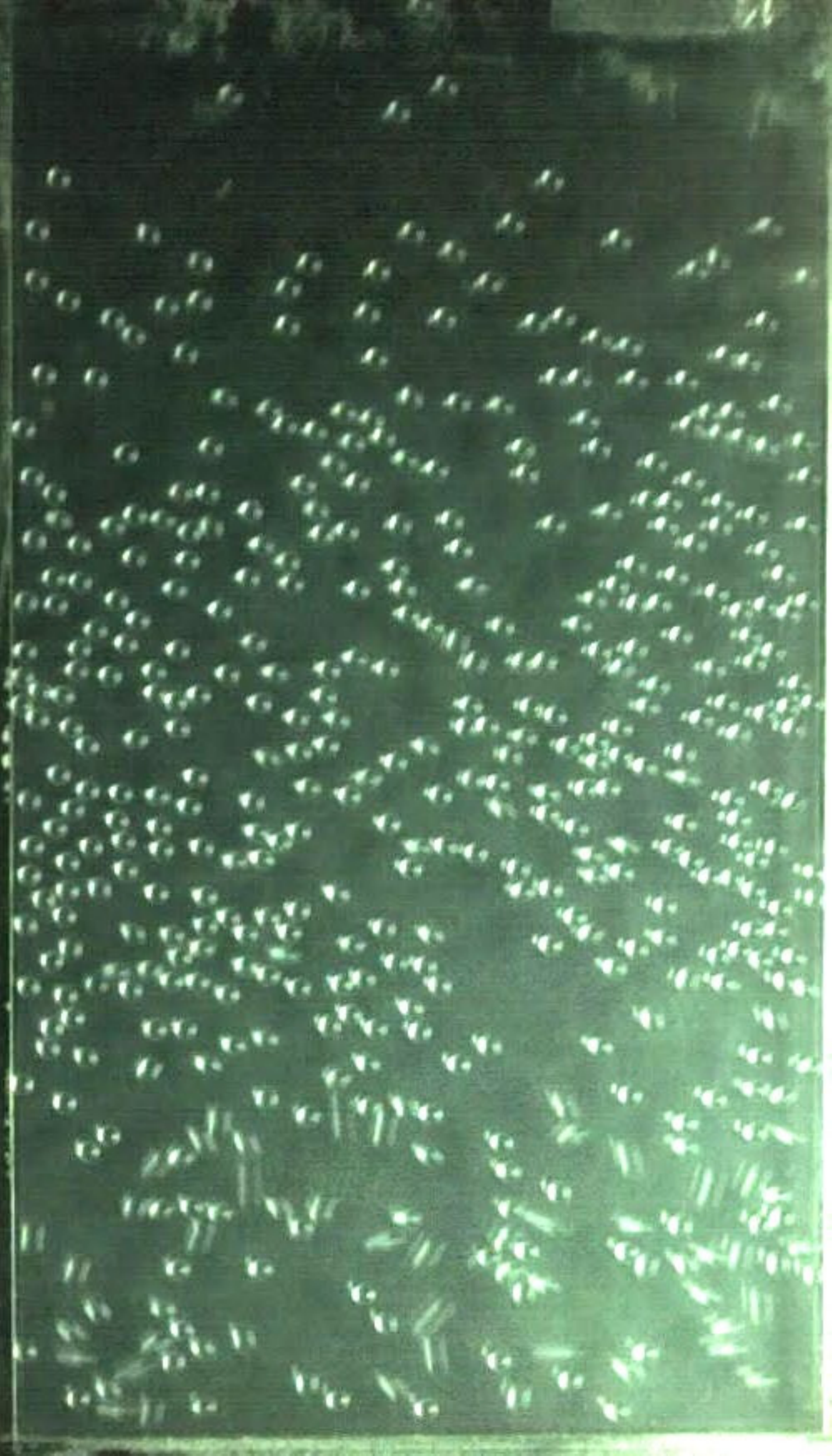}
\includegraphics[width=1.0in,height=1.2in]{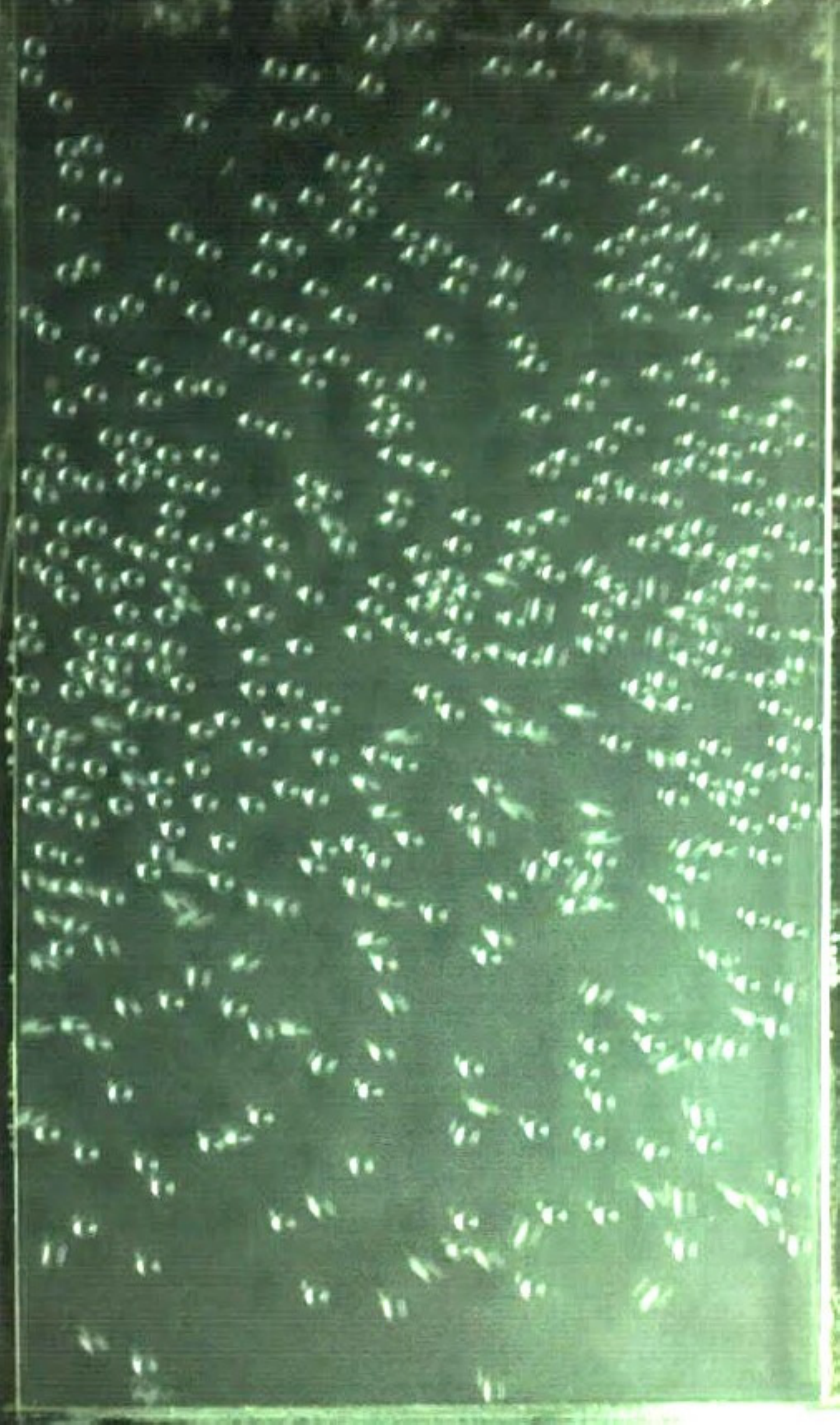}
\caption{
(Color online)
Three successive snapshots of the bed at $t=0$ (left), $t=\tau/2$ (middle) and $t=\tau$ (right):
(a) Bouncing bed (\textit{BB}) at $\Gamma=2$ ($f=7.8\;Hz$); (b) Leidenfrost state (\textit{LS}) at $\Gamma=30$ ($f=30.5\;Hz$); 
(c) \textit{Convection} with a  pair of rolls at $\Gamma=40$ ($f=35.24\;Hz$); (d) \textit{Convection} with  a single-roll at $\Gamma=45$ ($f=37.4\;Hz$); 
(e) \textit{Gas} at $\Gamma=55$ ($f=41.3\;Hz$).  The curly-arrows in panels ($c$, $d$) represent the directional  sense of convective motion.
The length of the container is $L/d=40$ and  the shaking amplitude is  $A/d=4$ for all cases; other parameters are $F=h_0/d=12$ and  $d=2.0\ mm$ diameter glass beads.
}
\label{fig:fig7}
\end{figure}

\begin{figure}[!ht]
\begin{center}
(a)
\includegraphics[width=1.00in,height=1.2in]{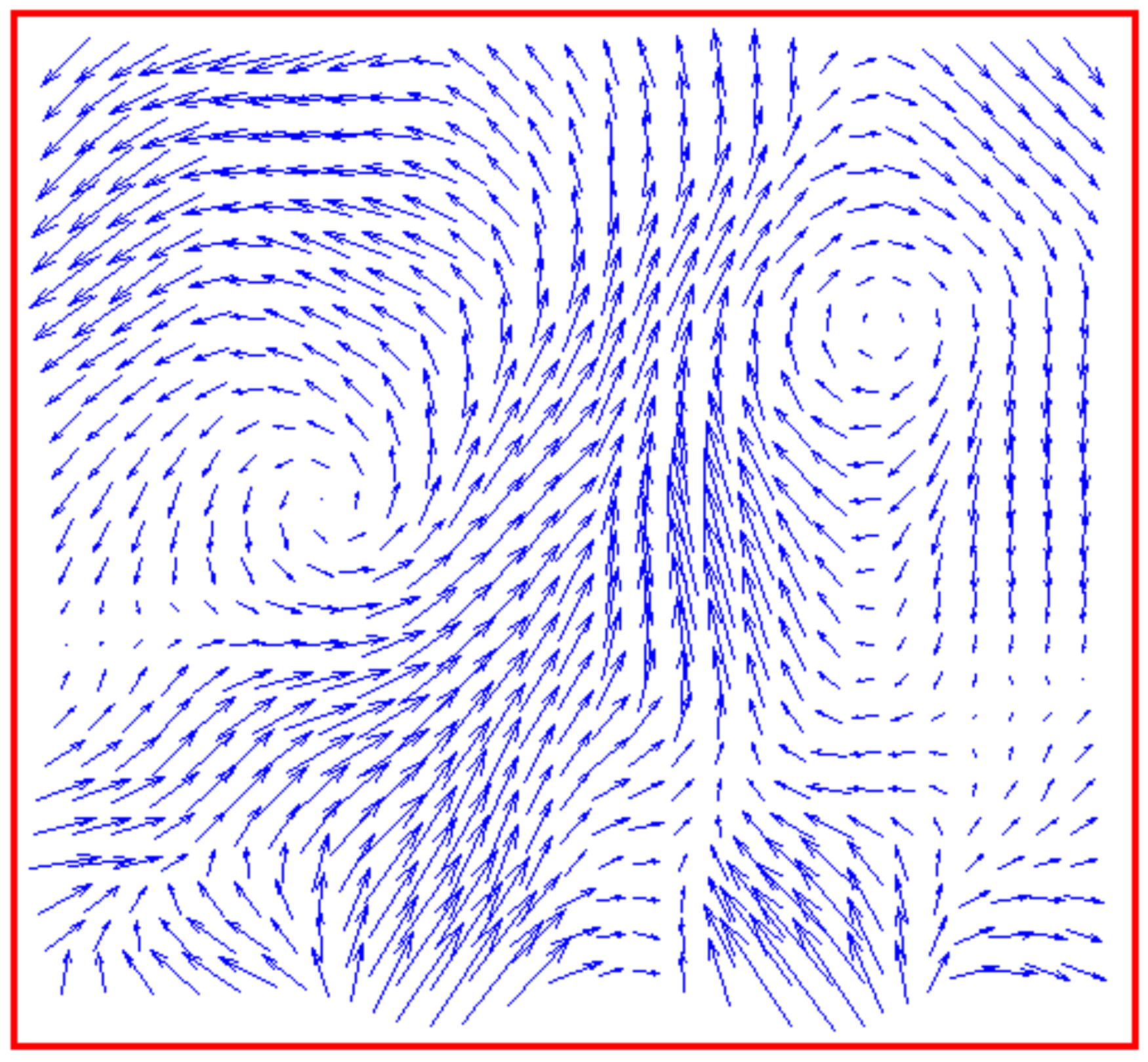}
\includegraphics[width=1.00in,height=1.2in]{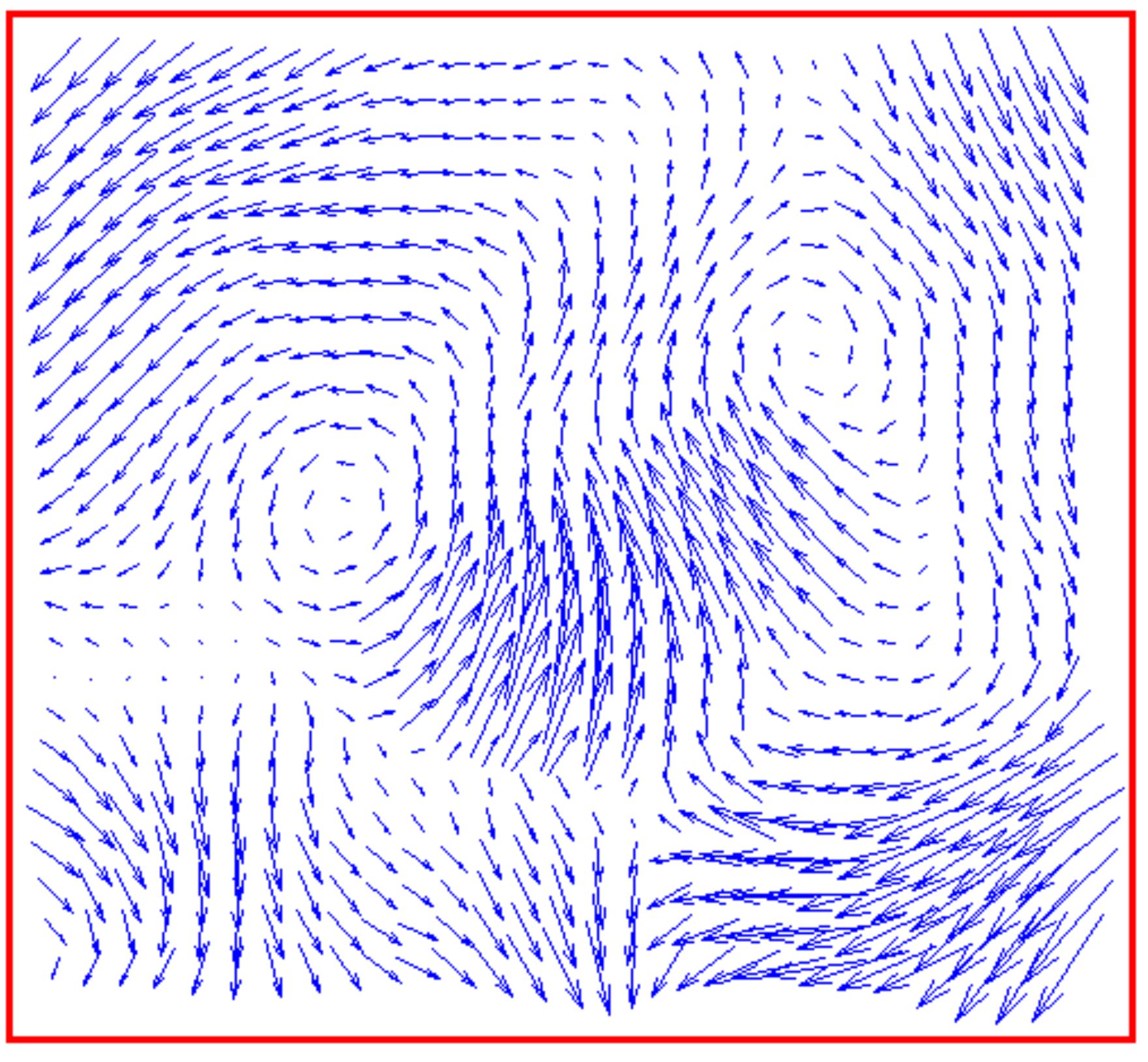}
\includegraphics[width=1.00in,height=1.2in]{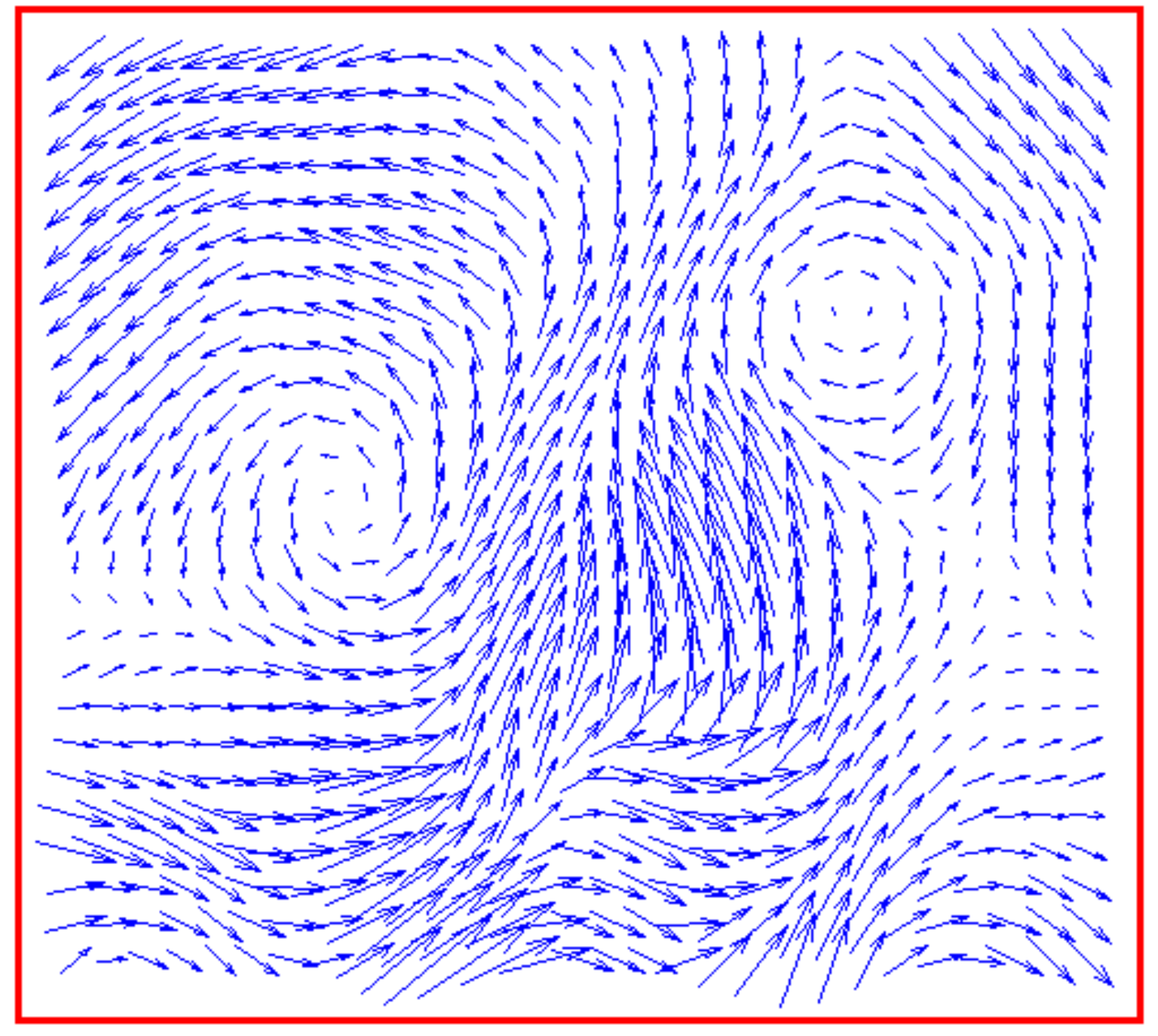}\\
(b)
\includegraphics[width=1.00in,height=1.5in]{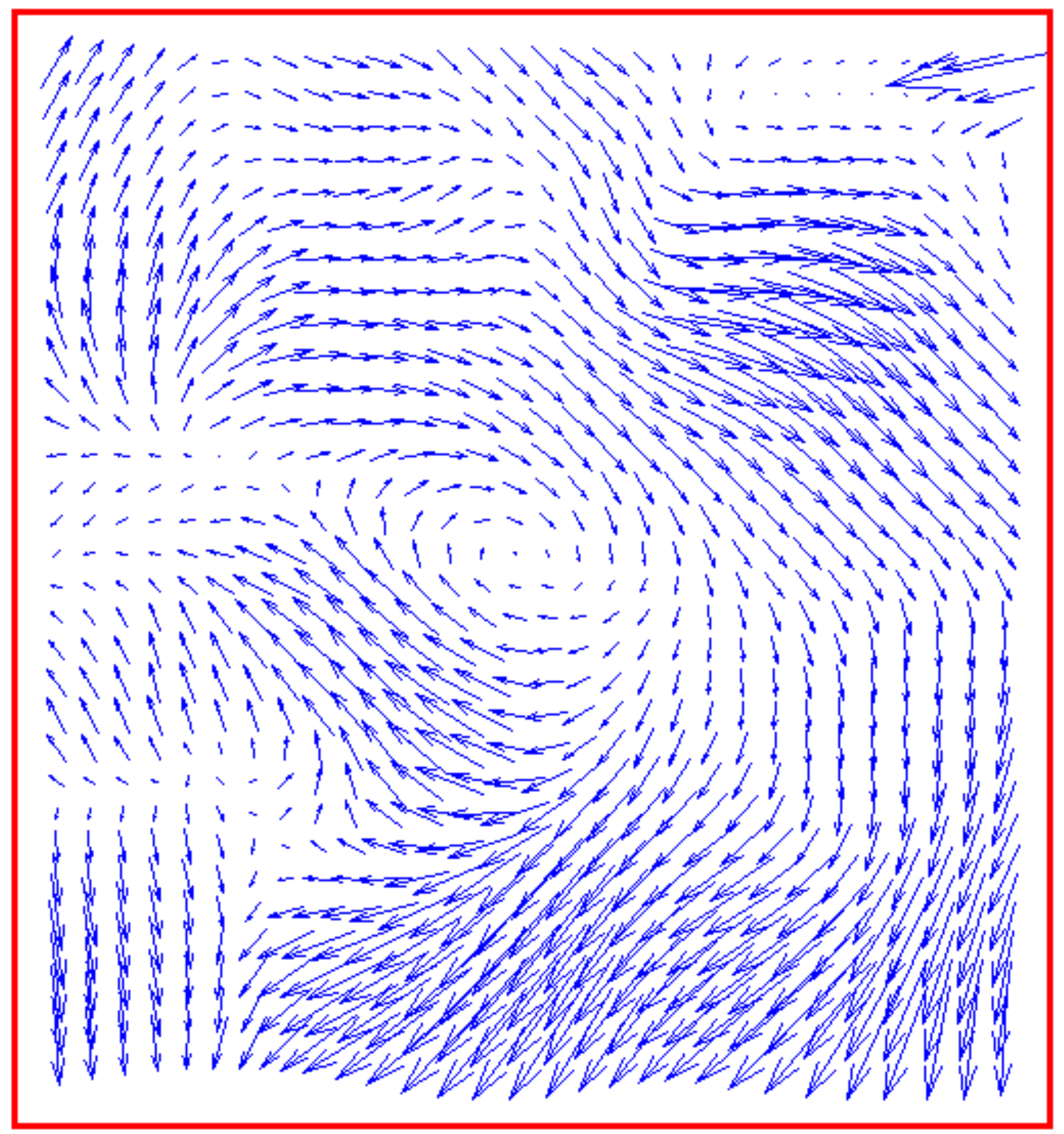}
\includegraphics[width=1.00in,height=1.5in]{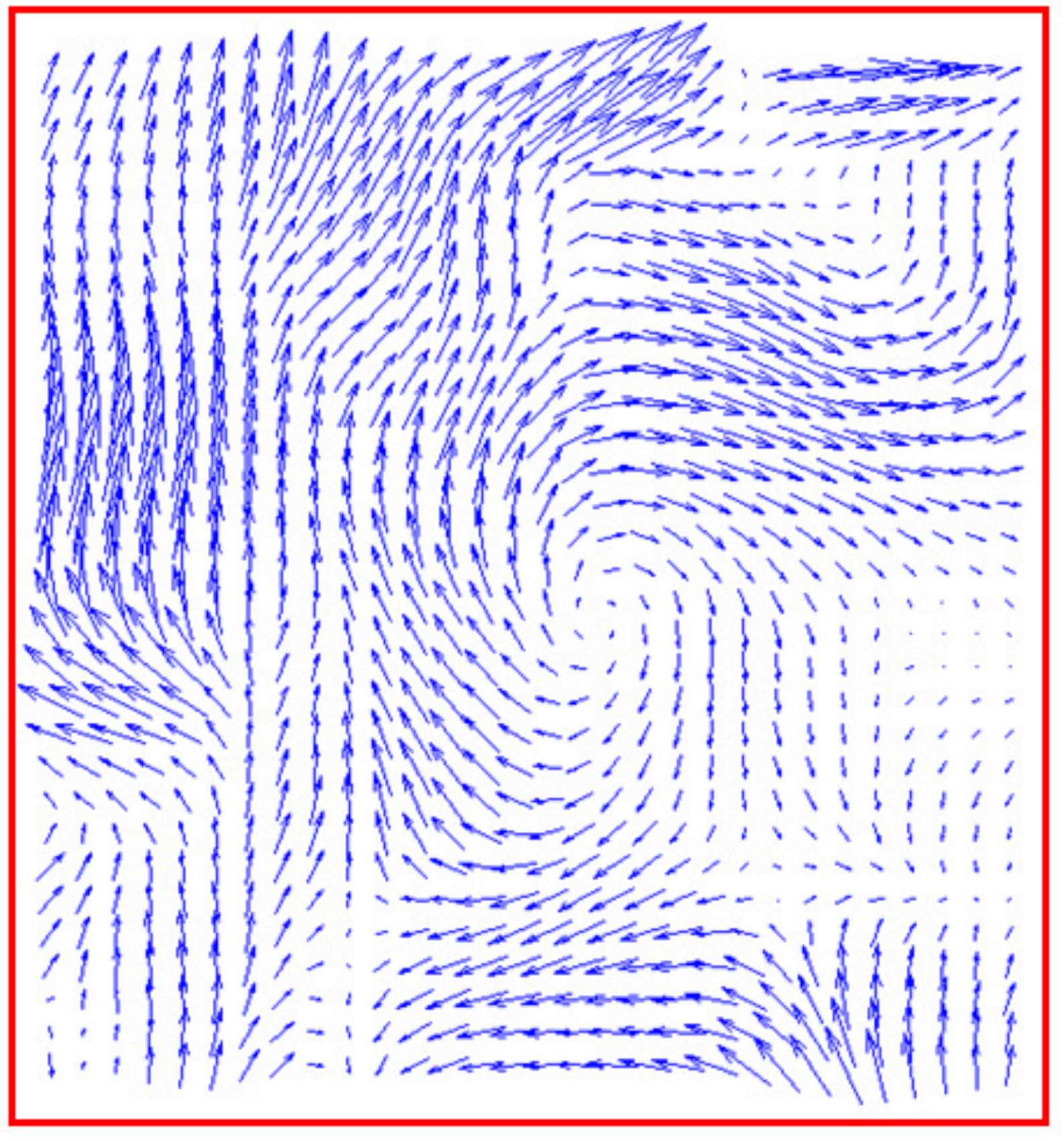}
\includegraphics[width=1.00in,height=1.5in]{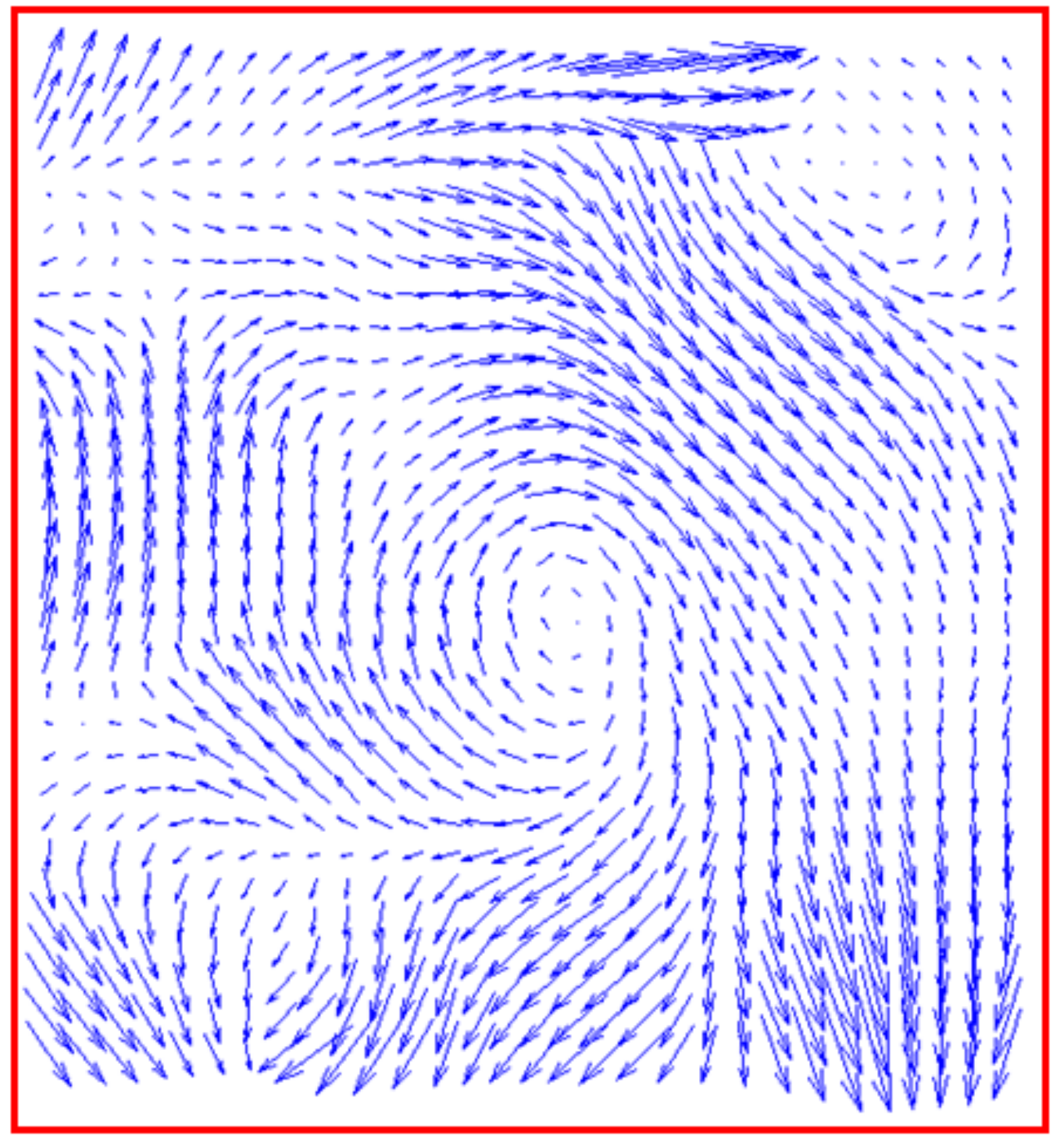}
\end{center}
\caption{
(Color online)
Coarse-grained PIV velocity fields for convection states:
(a) \textit{Convection} with a pair of rolls at $\Gamma=40$ and (b) \textit{Convection} with a single roll at $\Gamma=45$. 
The left, middle and right panels correspond to $t=0\tau$, $t=\tau/2$, and $t=\tau$, respectively.
Parameter values for panel $a$ and $b$  are same as in Fig.~\ref{fig:fig7}(c) and ~\ref{fig:fig7}(d), respectively. 
}
\label{fig:fig8}
\end{figure}

\subsection{Genesis of convection from Leidenfrost state: Multi-roll transition}

Now  we probe the effect of the length of the container on phase transition and related patterns ($SB$, $BB$ and $LS$) observed in the previous section.
The primary motivation of using a larger-box is to ascertain
whether the  ``mono-layer''  vibrofluidized bed admits convective motion whose origin may then be tied to the instability of the Leidenfrost state.
Similar transition has been reported previously~\cite{Eshuis2007,AAlam2012,AAlam2013}, however, in 
a ``quasi-2D'' box (with a depth of few particle diameters, $W/d\approx 5$) as well as in a 3D-box.
It was speculated in Ref.~\cite{Eshuis2007} that the collective motion of particles along the depth of the container could be a prerequisite 
for the onset of convection motion at strong shaking.

We used the same glass beads (of diameter $d=2.0\ mm$) in a box of length $L/d=40$ 
which is twice as that used ($L/d=20$) in Sec.~III.A, but other dimensions of the container ($W/d=1.1$ and $H/d=100$) remain the same as before.
For the sake of demonstration, we present one set of results for a filling height of $F=12$ layers at a shaking amplitude of $A/d=4$. 
The snapshots of patterns with increasing shaking $\Gamma$ are shown in Fig.~\ref{fig:fig7}(a-e).
The top-row (panel $a$) of Fig.~\ref{fig:fig7} depicts  three successive snapshots of the bouncing-bed (\textit{BB}) state at $\Gamma=2$, 
whereas the second-row (panel $b$) displays the same temporal-sequence of the Leidenfrost-state (\textit{LS}) at $\Gamma=30$.
At a higher shaking intensity of $\Gamma=40$, the system shows a pair of convection rolls (Fig.~\ref{fig:fig7}$c$).
Interestingly, the ``2-roll'' convection  coarsens into a ``single-roll'' at even higher shaking-intensity,
an example of which is shown in Fig.~\ref{fig:fig7}($d$) at $\Gamma=45$.
There is a dense cluster of particles at one-side of the container and a relatively dilute
region at the other side; the hotter particles go up from one-side, and rain-down from the other-side
forming a dense cluster. To our knowledge, such ``1-roll'' convection pattern has not
been reported in previous experiments on vibrofluidized beds at strong shaking.

The ``1-roll'' convection persists even at $\Gamma=50$ (Supplementary Movie-3), but the cluster on the right-side
becomes relatively dilute (compared to the case at $\Gamma=45$). Further increasing $\Gamma$  leads to extreme agitation of beads, which in turn
destroys  the convective motion and vapourizes the system into a gaseous state (a \textit{Granular Gas}) --
the snapshots of such a pattern  are  shown in Fig.~\ref{fig:fig7}(e) at $\Gamma=55$.

The coarse-grained  velocity fields of the snapshots of Fig.~\ref{fig:fig7}(c) and ~\ref{fig:fig7}(d) are displayed  in Fig.~\ref{fig:fig8}(a) and ~\ref{fig:fig8}(b), respectively. 
The  velocity of the collective motion of particles has been determined by analyzing the acquired images using a commercial PIV (Particle Image Velocimetry) software:
``Dynamic Studio Software, Version 3.3'' of Dantec Dynamics A/S, Denmark \citep{Dantec}.  For this purpose, the adaptive correlation technique \citep{Dantec,SR2000} 
was used in which the size of the interrogation window was  varied adaptively from $64\times64$ to $16\times16$ pixels, with $50\%$ overlap.
It is clear from Fig.~\ref{fig:fig8}(a) that the PIV-velocity field exhibits a pair of convection rolls.
Note that this represents an instantaneous velocity field, calculated over two frames separated by $1\,ms$,
however, due to the small number of particles in the system, the accuracy of the calculated velocity field
is limited. Here we are interested only about the gross features of  hydrodynamic velocity field, i.e. whether it contains a circulating motion or not.
Further increasing the shaking intensity to $\Gamma=45$, the ``2-roll'' convection degenerates  
into a single roll as it is evident from the PIV-velocity filed in Fig.~\ref{fig:fig8}(b).
It is clear that the circulation of this roll is in the clockwise sense; we have confirmed by repeating experiments that the ``1-roll'' convection 
can  also have a counter-clockwise circulation for which the dense-cluster is formed on the left-side of the container.

\begin{figure}[!ht]
\centering
\includegraphics[width=3.3in,height=3.3in]{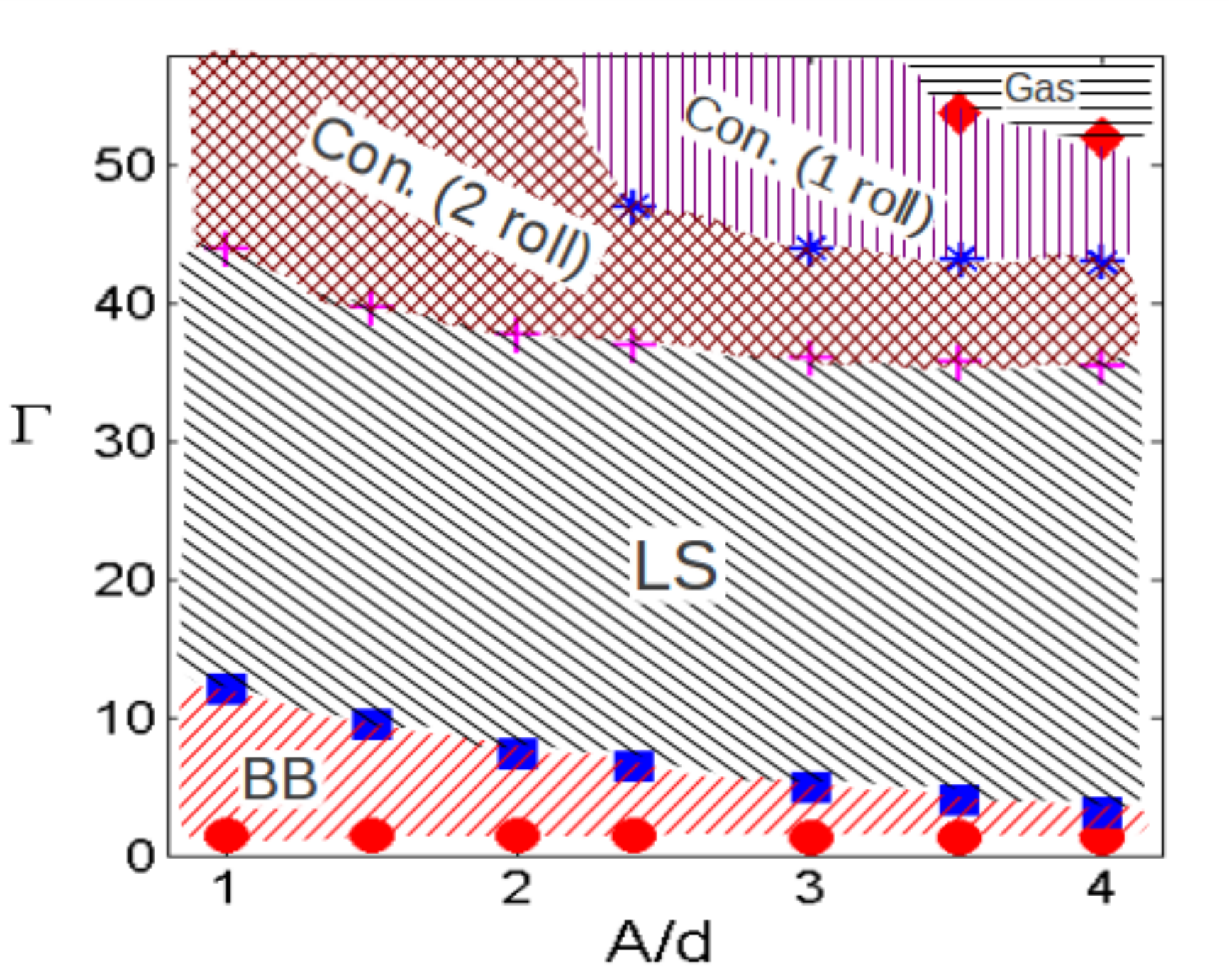}
\caption{
(Color online)
Phase diagram in ($\Gamma,A/d$)-plane for $F=12$ layers of $2\ mm$ diameter glass beads confined in $L/d=40$ cell.
Regions of bouncing bed \textit{(BB)}, granular Leidenfrost state \textit{(LS)}, \textit{``2-roll'' Convection}, \textit{``1-roll'' Convection} (1-roll) and \textit{Gas}  are marked.
The symbols represent approximate locations of transition while upsweeping at a specified shaking amplitude $A/d$ with a linear frequency-ramping of $0.01$Hz/s.
}
\label{fig:fig9}
\end{figure}

For a filling-height of $F=h_0/d=12$, we have conducted a series of experiments
covering a range of $\Gamma$ and $A/d$; various patterns observed and their transitions 
have been assimilated to construct a  phase diagram in the $(\Gamma,A/d)$-plane as shown in Fig.~\ref{fig:fig9}.
It is seen that the critical shaking intensity corresponding to the onset of \textit{LS} ($\Gamma_{BB}^{LS}$) 
follows the same decaying-trend  with increasing shaking amplitude $A/d$  as reported for the previous set of experiments (refer  to `star' symbols in Fig.~\ref{fig:fig6}).
Further increasing $\Gamma$, first we find a transition of the \textit{LS} to the \textit{convection} motion with a pair of rolls spanning the length of the container,
and susequently to a ``single-roll'' convection pattern. It mat be noted that  
the ``1-roll'' convection patterns are observed at $A/d>2$ and $\Gamma>42$ and the gaseous state are observed for $A/d\geq 3.5$ and $\Gamma>50$.
We reckon that the absence of the  ``$LS\to Convection$'' transition in the phase
diagram in Fig.~\ref{fig:fig2} can be tied to the fact that a pair of rolls (such as in Fig.~\ref{fig:fig8}$a$)  cannot be fitted into a container of length $L/d=20$.
In summary, a 2D monolayer vertically vibrated granular system (of sufficient length $L$) admits Rayleigh-Benard-type convection rolls at strong shaking:
the convection sets in from an instability of the LS (beyond a critical shaking intensity), leading to a pair of rolls, which degenerates into 
a novel ``single-roll'' pattern with increasing $\Gamma$ and subsequently to a ``granular'' gas.
The route to the gaseous state from a ``2-roll'' convection to ``1-roll'' convection is a new finding of our experiments -- a similar
transition has been reported in a recent simulation work~\cite{RLT2013}, see the discussion below.

It is interesting to compare our findings on ``$LS\to Convection$'' transition in a monolayer system with
previous experiments of Eshuis {\it et al.}~\cite{Eshuis2007} in a quasi-2D box of width $W/d=5.5$ and length $L/d=100$.
Returning to  Fig.~\ref{fig:fig9}  we find that the critical shaking intensity for the onset of ``2-roll'' convection, $\Gamma_{LS}^{Con}$ (denoted by the plus-symbols),
decreases with increasing shaking amplitude,
\begin{equation}
        \Gamma_{LS}^{Con}= 42.7 (A/d)^{-0.157}
\end{equation}
 and therefore the related shaking strength,
 \begin{equation}
        S_{LS}^{Con} \equiv \Gamma_{LS}^{Con}(A/d) = 42.7 (A/d)^{0.843},
        \label{eqn:LS2Con}
\end{equation}
increases strongly with increasing shaking amplitude.  A careful analysis of the  experimental data~\cite{Eshuis2007} 
(compare their data for $A/d=2$ and $4$ in their Fig.~11 for a range of $F\in(4,12)$)  reveals that $S_{LS}^{Con} $   increases weakly with increasing $A/d$.
The soft-particle Molecular dynamics simulations (see Fig.~2 in Ref.~\cite{Eshuis2010} and Fig.~9 in Ref.~\cite{Eshuis2013}) of the same quasi-2D system seem to support the 
dependence of $S_{LS}^{Con} $   on $A/d$; on the other hand, the recent event-driven simulations of Rivas {\it et al.}\cite{RLT2013} (in 
a quasi-2D box as in experiments of  Eshuis {\it et al.}~\cite{Eshuis2007}) indicate  that $S_{LS}^{Con}$ is almost independent of the shaking amplitude if  $L/d>20$ (see their Fig.~3).
The differences between our finding, Eqn.~(\ref{eqn:LS2Con}), and the previous quasi-2D experiments ~\cite{Eshuis2007,Eshuis2013} 
remain unresolved at present; this calls for additional experiments and simulations.

Let us now discuss about our finding of the ``$2Rolls\to 1Roll$'' transition  with increasing $\Gamma$:
since the length of the container is held fixed ($L/d=40$) in Fig.~\ref{fig:fig9}, the above-found coarsening of convection-rolls
is fundamentally different from the appearance of different number of  rolls with increasing $L/d$~\cite{Eshuis2007}.
Previous quasi-2D experiments (Fig.~12 in \cite{Eshuis2007} and  Fig.~4 in \cite{Eshuis2013})  suggest that the convection rolls
appear in pairs as the container-length $L/d$ is increased when the shaking strength is larger that some minimum value.
For given $L/d$ and $F$, they found that  the number of rolls decreases stepwise with increasing  $S$: ``{\it the steps involve two rolls at a time, 
since the pattern always contains an even number of rolls due to the downward motion imposed by the sidewalls}''~\cite{Eshuis2007};
moreover, their phase diagram (Fig.~14) does not indicate the presence of a gaseous state at larger values of $S$ beyond the convection regime.
Interestingly, however, the  quasi-2D simulations of Rivas {\it et al.}~\cite{RLT2013} found that the above type of coarsening-transition with increasing $S$
can  occur via a decrease in  the number of rolls in step-one (see their Fig.~2 for $A/d=4$), depending on the container-length $L/d$,
as follows: (i) from ``$4R\to 3R\to 2R\to Gas$''  for $80<L/d<100$, (ii) from ``$3R\to 2R\to Gas$''  for $50<L/d<80$, (iii) from ``$2R\to Gas$''  for $20<L/d<50$,
(iv) from ``$2R\to 1R\to Gas$''  for $15<L/d<20$, (v) from ``$1R\to Gas$''  for $10<L/d<15$ and (vi) $``LS\to Gas$'' for $L/d<10$.
Our 2D experiments with $L/d=40$ corresponds to  case (iii) and hence a ``$2Rolls\to Gas$'' transition is expected which is different from 
``$2R\to 1R\to Gas$'' transition that we found at $A/d=4$ (see Fig.~\ref{fig:fig9}). On the other hand, our experiments with $L/d=20$ did
not show any transition to convection (Figs.~\ref{fig:fig2} and \ref{fig:fig5}) even at $\Gamma=55$ and $A/d=4$ (i.e. $S=220$).

It may be noted that all related simulations~\cite{Eshuis2010,Eshuis2013,RLT2013} have been done in a quasi-2D box,  closely following the experimental
setup  of Eshuis {\it et al.}~\cite{Eshuis2007}, and, moreover, the driving in Ref.~\cite{RLT2013} is {\it bi-parabolic} (rather than via the sine function as in Eqn.~1).
The recent simulations~\cite{RLT2013} predicted a transition-route of ``$4R\to 3R\to 2R\to Gas$'' [the case-(i) above, with increasing $\Gamma$] 
for a container of length $L/d=100$ which may be contrasted with previous experimental finding~\cite{Eshuis2007} of ``$6R\to 4R\to 2R$'' convection,
indicating an anomaly in the simulated-predictions on the number of rolls  in the same setup having similar parameter values.
In addition, our recent experiments~\cite{Shukla2014} in a quasi-2D box ($W/d=5.5$) with $L/d=80$ found that the primary bifurcation from the 
Leidenfrost state is a 4-roll ($4R$) convection pattern in contrast to the $3R$-pattern observed in simulations~\cite{RLT2013}.

Overall, the above comparative discussion suggests that,
to make a one-to-one comparison with our experimental findings on the ``multi-roll'' transition scenario (via ``$2R\to 1R\to Gas$'' with increasing $\Gamma$ at $L/d=80$),
 the future simulations should be  carried out  in a mono-layer box under harmonic shaking with parameter values as in present experiments.
We also recommend additional experiments by increasing the length of the container $L/d$ to see whether the
``multi-roll'' transition scenario as depicted in Fig.~2 of  Ref.~\cite{RLT2013} survives in a two-dimensional vibrofluidized bed.

 \begin{figure*}[!ht]
\centering
(a)
\includegraphics[width=2.1in,height=2.8in]{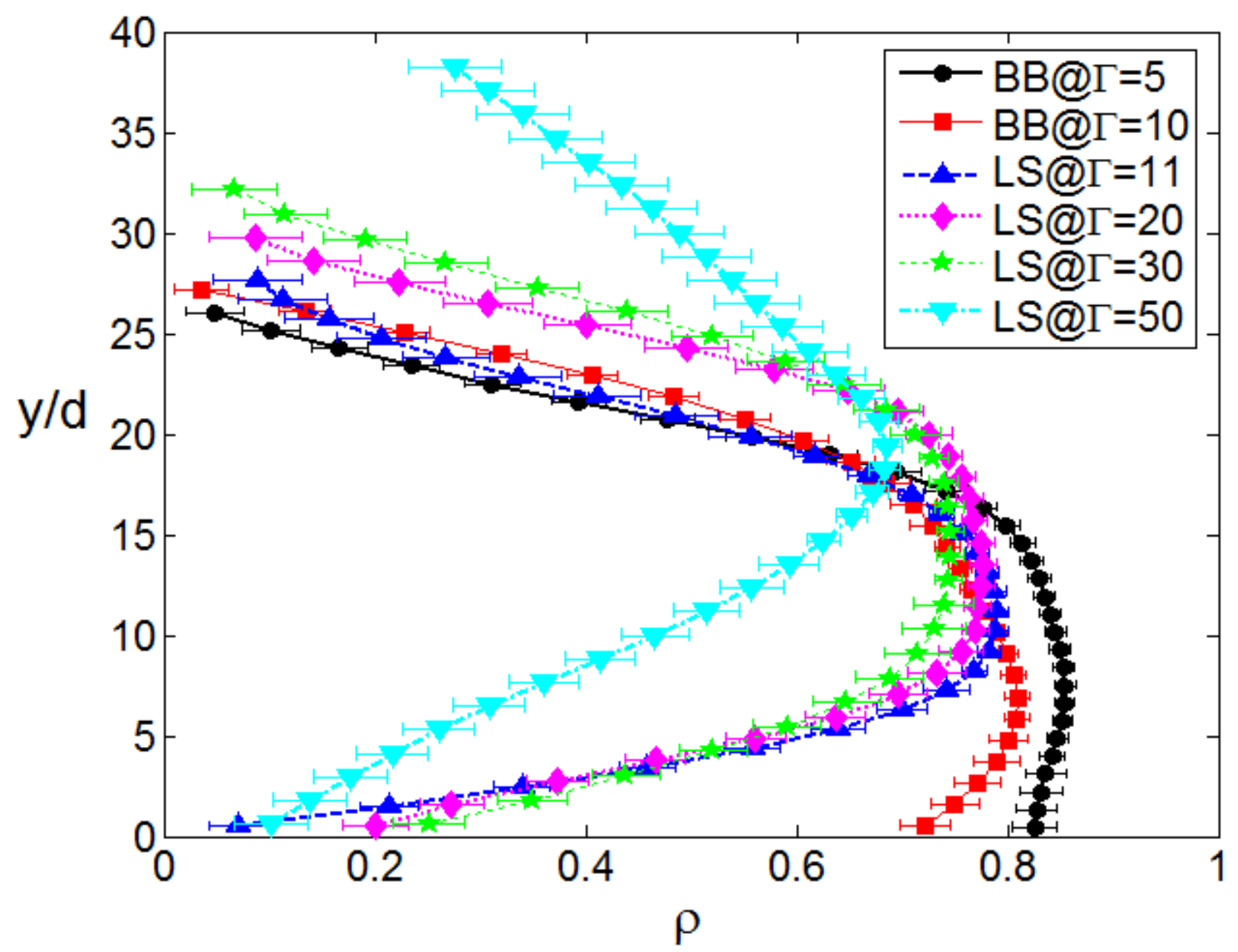}
(b)
\includegraphics[width=2.1in,height=2.8in]{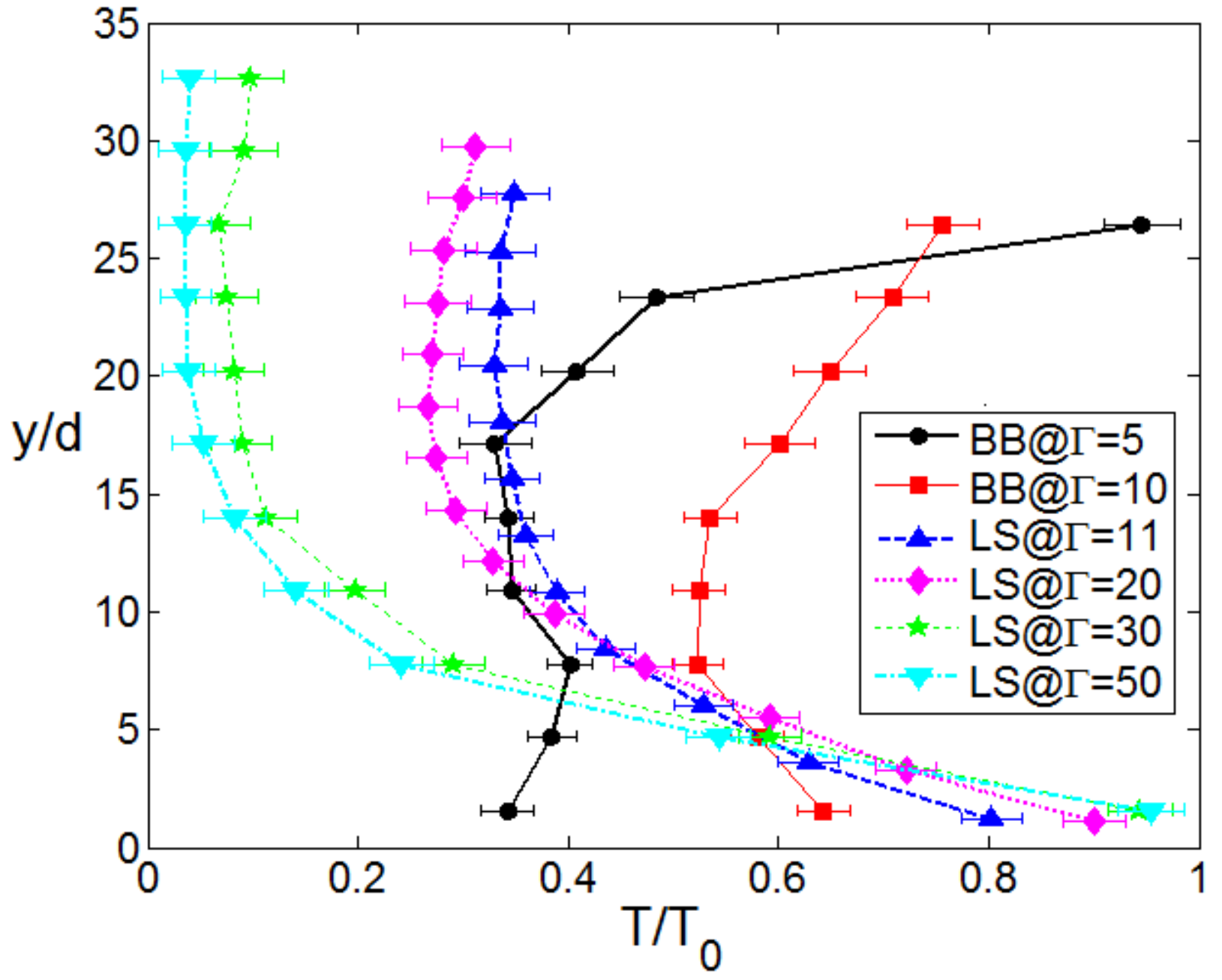}
(c)
\includegraphics[width=2.1in,height=2.8in]{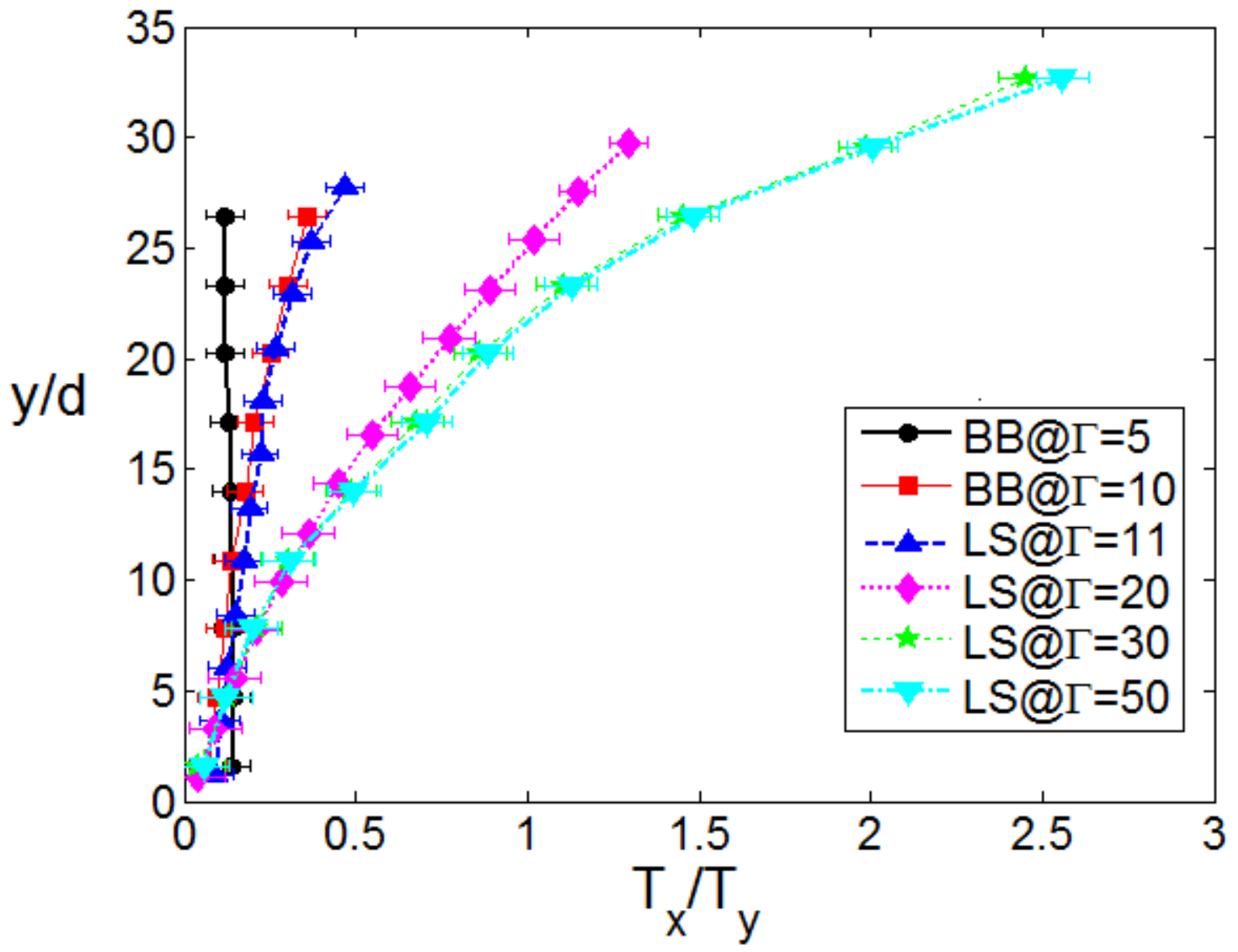}
\caption{
(Color online)
Variations of (a) density and (b) granular temperature with height for the various states in $F=25$ layers of $2.0\ mm$ diameter
glass-beads at constant $A/d=2.4$ with increasing shaking intensity $\Gamma$. Here, $T_0$ is the input shaking-energy (per particle) at the base.
(c) Profiles of the temperature ratio $T_x/T_y$ such that $T=(T_x+T_y)/2$ -- note that the data for $\Gamma=30$ and $50$ almost overlap with each other.
}
\label{fig:fig10}
\end{figure*}

\subsection{Density, temperature and temperature-anisotropy}

To assess the structure of the shaken granular-bed with varying $\Gamma$ and $A/d$,
we measured the `coarse-grained' density and temperature profiles from image analysis using the Image-J software as described in Sec.~II.B.
The density profiles along the vertical direction have been calculated from the digitized (binary) version
of the experimental snapshots by determining an ``effective'' normalized pixel density (i.e.,~by counting pixels that constitute the beads) over a box of height 5 pixels and
width equal to the image-width in pixels. The density profile is subsequently smoothed by fitting the data via a polynomial.

The ``granular'' temperature is defined as the mean-square of the velocity fluctuations around the mean flow velocity:
\begin{equation}
 \label{eq:temperature}
   T = \frac{1}{2}m\left\langle(v-u)^2\right\rangle, 
\end{equation}
where $v$ is the instantaneous particle velocity and $u=\langle v \rangle$ 
is the hydrodynamic/flow velocity (which vanishes in the present case of harmonically-shaken bed)
and the angular bracket denotes a suitable averaging over many snapshots of the system.
Once the individual particle position is extracted for a batch of snapshots using the particle-tracking
routines, the individual particle velocity can be determined from 
two successive frames, which is fed into Eqn.~(\ref{eq:temperature}) to obtain temperature field.
More specifically, the granular temperature at a specified height, $T(y)$,
is calculated by dividing the system into a series of horizontal bins/layers of height of $2$ particle diameters
and width as that of snapshot (assuming horizontal homogeneity). 
In each bin (say, at $y=y_i$), the averaging is carried out   (i) over all particles inside the bin in each snapshot as well as (ii) over
a batch of $400$ snapshots or more that span over many oscillation cycles.

Figure~\ref{fig:fig10}(a) displays the density profiles with increasing shaking intensity $\Gamma$ at a shaking amplitude of $A/d=2.4$ (see Fig.~\ref{fig:fig2}).
The density profile in the BB-state observed at $\Gamma=5$  is indicated by the red curve in Fig~\ref{fig:fig10}(a) -- the corresponding 
snapshots of the system are shown in Fig.~\ref{fig:fig3}.  The density shows a slight increase from the base of the container to a height corresponding
to the `maximum' density, beyond which it remains nearly constant over a certain height,
representing a region wherein the particles are packed hexagonally (see the snapshots in Fig.~\ref{fig:fig3}). 
Beyond a certain height the density shows a rapid fall indicating the existence of a dilute gas-like layer of fast moving
particles on the top of bed. It is noteworthy that the density profile of the {BB}-state
at a higher shaking intensity $\Gamma=10$ (just below the transition to \textit{LS}, see Fig.~\ref{fig:fig2}),
shows a  weak ``density inversion''.  Furthermore, the `maximum' density
which the bed has achieved is  slightly smaller than the `maximum' density attained at  $\Gamma=5$ --
this is expected since the bed at higher $\Gamma$ is likely be loosely packed due to the relatively higher degree of fluidization. 
The density profile of \textit{BB} at $\Gamma=10$ (see the light blue curve in Fig.~\ref{fig:fig10}$a$)
follows almost the same trend of the density decay as that of \textit{BB} at $\Gamma=5$ beyond the maximal density height. It is also noticeable that the density
profile at higher $\Gamma$ spans  a larger height due to the higher input shaking-energy
(due to which the particles on the top of the bed bounces off to a greater height).

Increasing the shaking intensity from $\Gamma=10$ to $\Gamma=11$, 
an extreme \textit{density inversion} (see the magenta curve in Fig.~\ref{fig:fig10}a) is found to occur.
This signals that the system has transited to the Leidenfrost state (\textit{LS}) in which a dense cluster floats over a dilute gaseous layer
and the maximum density within \textit{floating} cluster  is much lower than that in the bouncing bed (at $\Gamma=10$).
Further increasing $\Gamma$ leads to an overall expansion of the granular bed in which the extent of the floating cluster
gradually reduces and  the dilute gaseous region beneath grows in size.
Comparing the black ($\Gamma=30$) and blue ($\Gamma=50$) curves in Fig.~\ref{fig:fig10}(a),
we find that the density reduction above the floating cluster becomes more gradual with increasing $\Gamma$, implying that
there exists a `saltating' layer of particles above the dense-cluster where the particles move ballistically -- this is dubbed \textit{ballistic layer}.

To summarize Fig.~\ref{fig:fig10}(a), we found that  the granular Leidenfrost state is characterized by three distinct regions:
(i) a dense \textit{floating cluster}, (ii) a dilute gaseous \textit{collisional layer} adjacent to the base of vibrating container 
and (iii) a \textit{ballistic layer} consisting of fluidized particles on the top of the floating cluster.
Note that the \textit{ballistic} layer is also present in the \textit{BB}-state above the crystal-packed layer (see the snapshots in Fig.~\ref{fig:fig3}).

 \begin{figure*}[!ht]
\centering
(a)
\includegraphics[width=2.1in,height=2.8in]{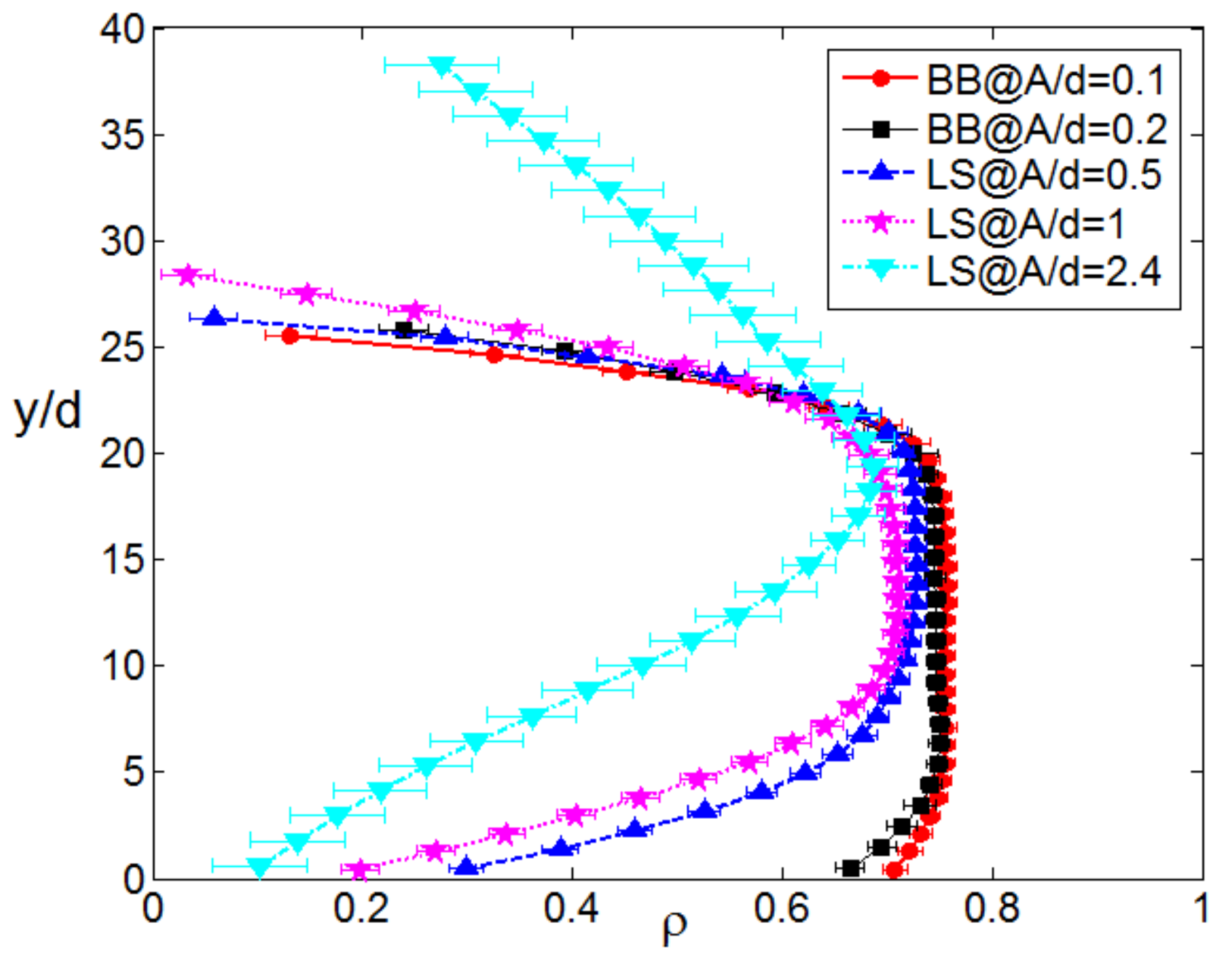}
(b)
\includegraphics[width=2.1in,height=2.8in]{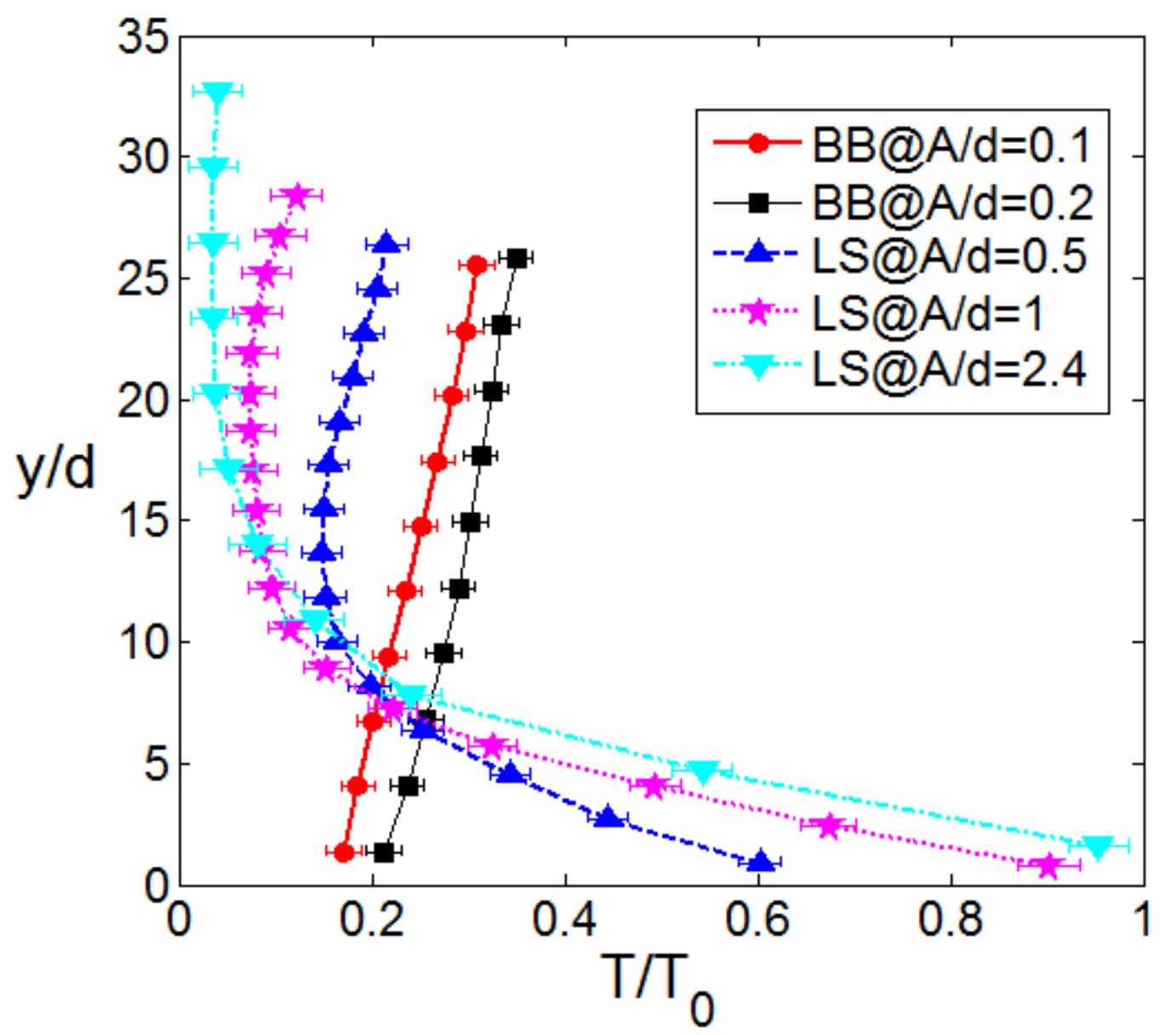}
(c)
\includegraphics[width=2.1in,height=2.8in]{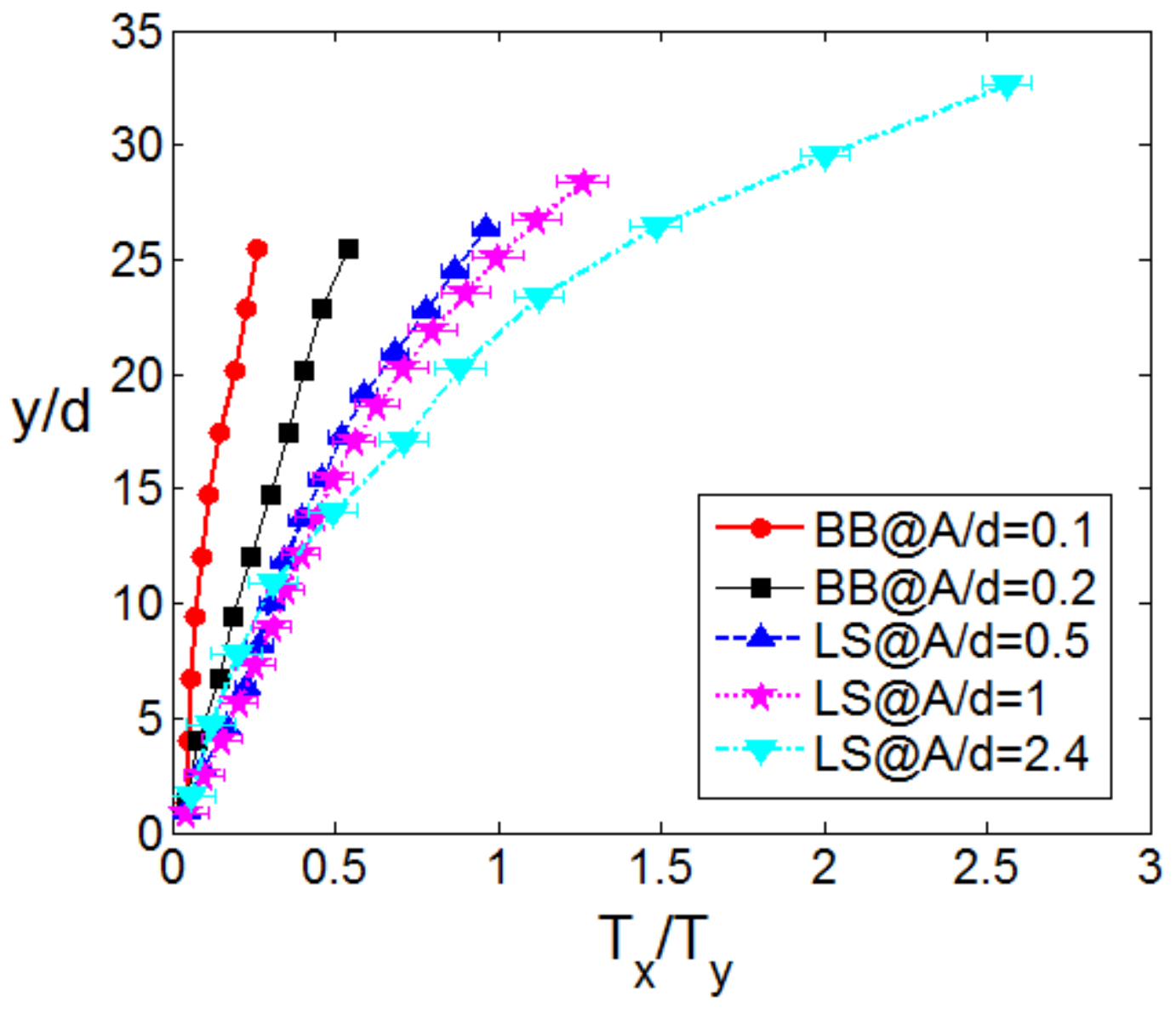}
\caption{
(Color online)
Same as Fig.~\ref{fig:fig10} but with increasing shaking amplitude $A/d$ at a constant shaking intensity $\Gamma=50$.
}
\label{fig:fig11}
\end{figure*}

Corresponding to the density profiles in Fig.~\ref{fig:fig10}(a), the profiles of granular temperature $T(y)$ and the temperature ratio ($T_x/T_y$,
where $T_i$ is the temperature along $i$-th direction such that $T=(T_x+T_y)/2$) are shown in Fig.~\ref{fig:fig10}(b) and Fig.~\ref{fig:fig10}(c), respectively. 
The temperature has been normalized by the average input energy $T_0$ at the base. In the case of \textit{BB}-states at $\Gamma=5$ and $10$, the temperature remains nearly
constant up-to the height of the ``crystalline'' packed bed, but increases at higher elevations due to the higher kinetic energy possessed by the fluidized particles at the top.
In the case of \textit{LS}, however, the temperature monotonously decreases from  the vibrating base up-to the top of the floating cluster. 
This confirms that a dense cluster of particles floats over fast-moving/hotter particles in the granular Leidenfrost state. 
Comparing the temperature profiles between BB and LS in Fig.~\ref{fig:fig10}(b), we find that the granular temperature in \textit{BB}-states is lower near the vibrating base 
in contrast to the \textit{LS} for which maximum-temperature occurs  at the base -- this is a distinguishing criterion between \textit{BB} and \textit{LS}. 
The origin of the comparatively higher temperature near the base in the \textit{LS} can be tied to the hotter dilute region which is absent in the BB-state.

Figure~\ref{fig:fig10}(c) indicates that the temperature ratio in the \textit{BB} at $\Gamma=5$ is small 
and remains almost  constant ($T_x/T_y\approx0.15$); however, at $\Gamma=10$, this ratio increases with elevation from the base.
The latter observation also holds in the case of \textit{LS} at $\Gamma=11$ and higher $\Gamma$. Overall, the temperature in the vertical direction is larger
than that in the horizontal direction ($T_y > T_x$, which is expected since the energy is imparted to particles via shaking along the vertical direction) except 
at very high values of $\Gamma$ where $T_x> T_y$ near the top of the bed that constitutes the ballistic-layer.
The latter finding is intriguing since it implies that the vertical component of temperature can be lower than its horizontal component in the ballistic-layer of the vibro-fluidized bed.
It is conceivable  that at large enough $\Gamma$  the particles in the ballistic layer can loose their momentum in the vertical direction much easily
due to collisions with the  floating-cluster; on the other hand, their collisions along the
horizontal direction are less likely due to the dilute nature of the ballistic region, resulting in higher values of $T_x$ 
and consequently $T_x/T_y>1$ in the ballistic region.

The above-discussed  characteristic features of the density, granular temperature and temperature-ratio profiles hold even if we 
traverse the phase diagram in Fig.~\ref{fig:fig2} at a constant shaking intensity $\Gamma$ while increasing the shaking amplitude $A/d$,
see Figs.~\ref{fig:fig11}(a,b,c) at $\Gamma=50$ for various values of $A/d$ spanning both BB and LS.
Increasing the shaking amplitude ($A/d$) beyond a critical value causes a \textit{density inversion} (see Fig.~\ref{fig:fig11}(a)), marking the onset of
\textit{LS} in the system. The maximum  density occurs  at a certain height away from the base once it crosses the dilute \textit{collisional-layer} and
the density remains nearly constant at the maximal value unto a certain thickness spanning \textit{floating cluster}, and
subsequently drops rapidly across the rarefied \textit{ballistic layer} at the top of the bed. Other features of the density profile with increasing $A/d$
are similar to those found for increasing $\Gamma$ as in Fig.~\ref{fig:fig10}(a).
Figure~\ref{fig:fig11}(b) indicates that the temperature in the \textit{BB}-states increases monotonically
from the base to the top which is different from the non-monotonic temperature profiles found in the BB-states
in Fig.~\ref{fig:fig10}($b$) for the case of increasing shaking frequency $f$ at constant shaking amplitude.  
 On the contrary, in the \textit{LS}, the temperature shows a decaying behaviour away from the base, attaining a minimum value 
at the end of the collisional-layer, and then increases, albeit mildly, at higher elevations.
The degree of increase of $T(y)$ at higher elevations, however, decreases with increasing $A/d$ (compare
the $T$-profiles at $A/d=0.5$ and $1$) and the temperature remains almost constant near the top of the bed at $A/d=2.4$.
The latter observations are also evident in Fig.~\ref{fig:fig10}(b) -- see the $T$-profiles at $\Gamma=20$, $30$ and $50$.
The temperature-ratio ($T_x/T_y$) profiles in Fig.~\ref{fig:fig11}(c)  show similar characteristic features as those presented in Fig.~\ref{fig:fig10}(c).

It may be noted that the density profiles were first measured in a 2D vibrofluidized bed by Warr {\it et al.}~\cite{Warr1995},
and some of their  density profiles did show signatures of a density inversion (see their Figs. 2a and 2b).
The overall shapes of our density and granular temperature profiles (Figs.~\ref{fig:fig10}$a$ and ~\ref{fig:fig10}$b$) with increasing $\Gamma$
are found to be  similar to those in the 2D-simulation of Yang and Hsiau~\cite{YH2000} (see their Figs.~5 and 6) -- 
in particular, the increase of granular temperature at higher elevations in the BB-state  looks strikingly similar to that in Fig.~6 of Ref.~~\cite{YH2000}.
The related NMR-experiments by Huan {\it et al.}\cite{Huan2004} in a 3D-vibrofluidized bed identified
similar variations of density and temperature with height. The latter experiments also measured the height-profiles
of horizontal ($T_x$) and vertical ($T_y$) temperatures (see their Figs.~12 and 13), but they always found  $T_x<T_y$ for all case studies 
with $T_x/T_y\approx \mbox{constant}$ with elevation.

\begin{figure}[!ht]
\begin{center}
(a)
\includegraphics[width=3.2in,height=2.9in]{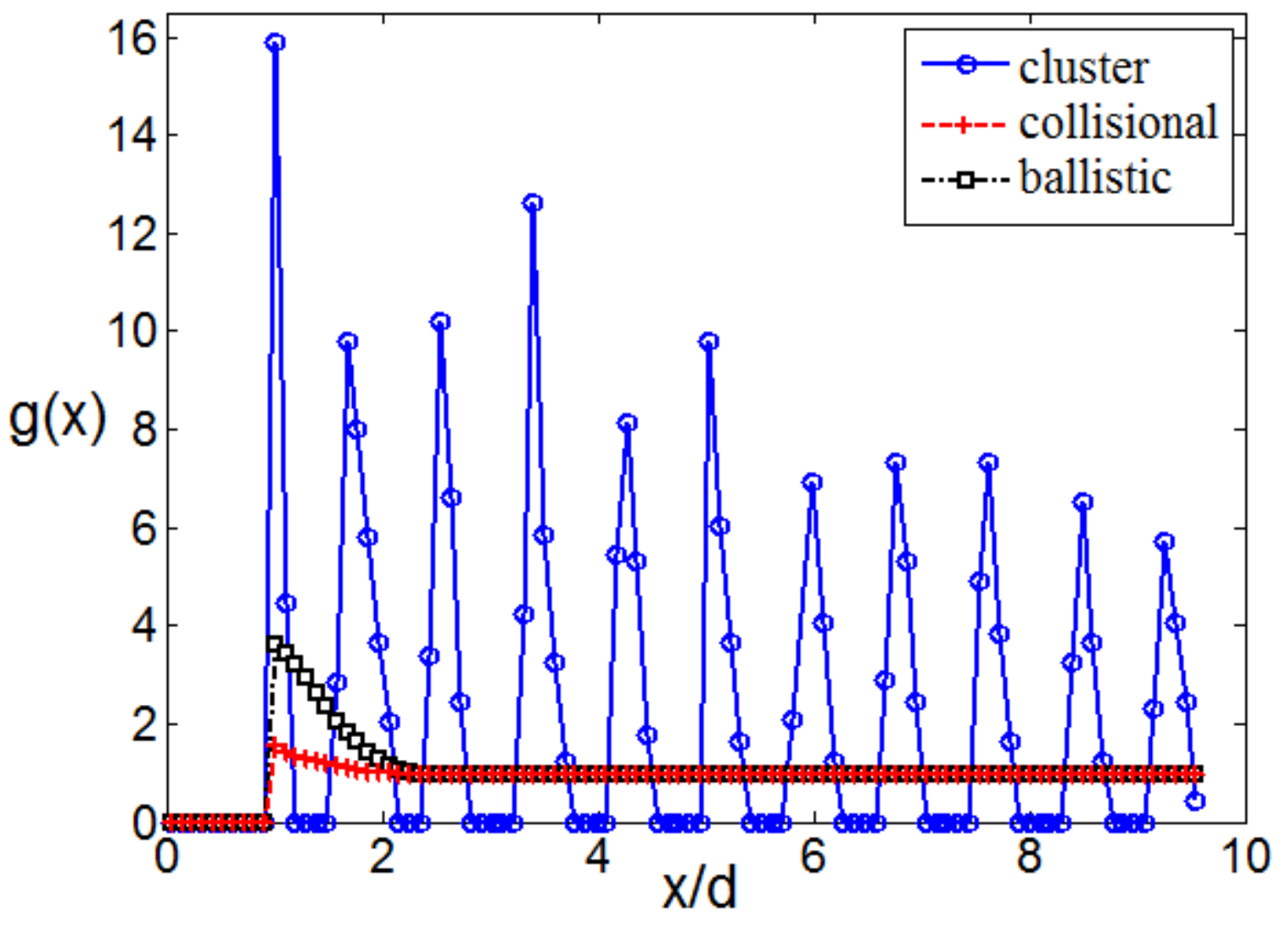}\\
\includegraphics[width=3.2in,height=2.9in]{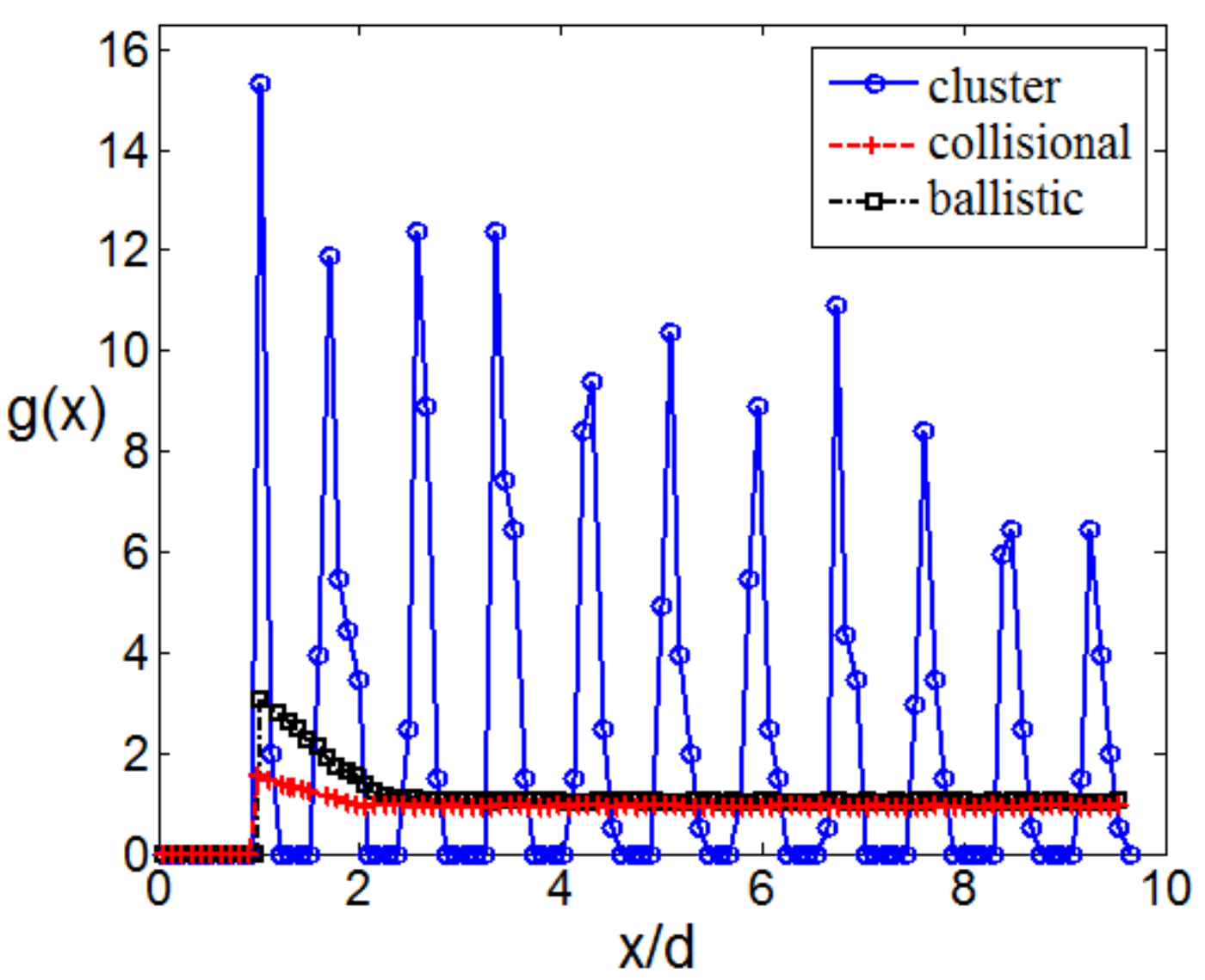}
(b)
\end{center}
\caption{
(Color online)
One-dimensional pair-correlation function, $g(x|y_i)$, in three regions of the Leidenfrost-state (LS) 
at (a) $\Gamma=30$ and (b) $\Gamma=50$,  with other parameters being  $A/d=2.4$, $F=h_0/d=25$ and  $d=2.0\;mm$.
Note that $g(x|y_i)$ is evaluated in a horizontal-stripe [of thickness $\delta y=1.5d$] 
whose vertical location, $y=y_i$, is approximately at the center of each region.
In each panel,  the open blue circles refer to the floating-cluster, the lower curve marked by red plus-symbols refers to the
collisional layer and the middle curve marked by black squares refers to the ballistic layer.
}
\label{fig:fig12}
\end{figure}

\section{Microstructure and Dynamics in Granular Leidenfrost State}

\subsection{Microstructure and spatial-ordering of particles}

An important quantity employed for studying the micro-structural characteristics and  the spatial-order of a particulate system is the
so-called pair correlation function  which describes how, on average, the
particles are radially packed around each other. Mathematically, the pair-correlation  function is given by \citep{AT1989}
\begin{equation}
 g({\boldmath r}) = \frac{1}{N\rho({\boldmath r})}\sum_{i=1}^{N}\sum_{j\neq i}^{N}\langle\delta
                ({\boldmath r}+ {\boldmath r}_j - {\boldmath r}_i)\rangle,
\end{equation}
where $\rho({\boldmath r})g({\boldmath r})$ is the conditional probability of finding a particle at a distance
${\boldmath r}$ away from the reference particle such that
\begin{equation}
  \int\rho({\boldmath r}) g({\boldmath r}) d{\boldmath r}=N-1.
\end{equation}
Since the hydrodynamic fields are inhomogeneous along the vertical direction, first we calculated the one-dimensional pair-correlation function $g(x)=g(x|y_i)$ 
in a horizontal-stripe which is  located at $y=y_i$. In the granular Leidenfrost state, the  particles belonging to  three regions
[(i) collisional layer, (ii) floating cluster and (iii) ballistic layer] are considered separately;  we then evaluate  $g(x)=g(x|y_i)$
at the center of each region ($y=y_i$) over a horizontal-stripe of thickness $\delta y=1.5d$.

Figures~\ref{fig:fig12}(a) and ~\ref{fig:fig12}(b) display the pair-correlation function $g(x|y_i)$ in three regions of the \textit{LS} 
at $\Gamma=30$ and $50$, respectively; other parameters are as in Fig.~\ref{fig:fig10}.
In both cases, the $g(x)$ in the \textit{floating-cluster} contain an array of peaks: the first peak is located at the contact point, the second peak at $r/d\approx\sqrt{3}$
and other subsequent peaks occurring at regular spacings, signifying a nearly hexagonal-packed  structure.
The $g(x)$ in the \textit{collisional} and \textit{ballistic} regions of the \textit{LS}  indicates that both are  gaseous  in nature:
there is a peak at $r=d$ and then it decays rapidly with distance until it asymptotes to unity at $r/d\approx 2$, suggesting that particles are uncorrelated at large distances. 
Comparing the $g(x)$ between the \textit{ballistic} and \textit{collisional} regions
of the LS in Fig.~\ref{fig:fig12}, we find that the first peak of $g(x)$ in the \textit{ballistic} layer is larger  than that in the collisional layer.
This suggests that the center of the ballistic region (i.e.~$y=y_i$ at which $g(x|y_i)$ is calculated) is comparatively denser than that of the collisional region.

With increasing shaking intensity $\Gamma$, the \textit{collisional}  and \textit{ballistic} layers of \textit{LS} expand and
become relatively dilute  due to the higher degree of fluidization, consequently there is a decrease  in the contact value of $g(x)$ 
as seen in the respective red and black curves in Figs.~\ref{fig:fig12}(a) and ~\ref{fig:fig12}(b). In a similar manner, the packing within the  \textit{floating-cluster} also 
becomes  more and more loose with increasing $\Gamma$, causing a minor drop in the contact value of $g(x)$ --
compare the blue curves in  Figs.~\ref{fig:fig12}(a) and ~\ref{fig:fig12}(b).

\begin{figure*}[!ht]
\begin{center}
(a)
\includegraphics[width=1.8in,height=1.75in]{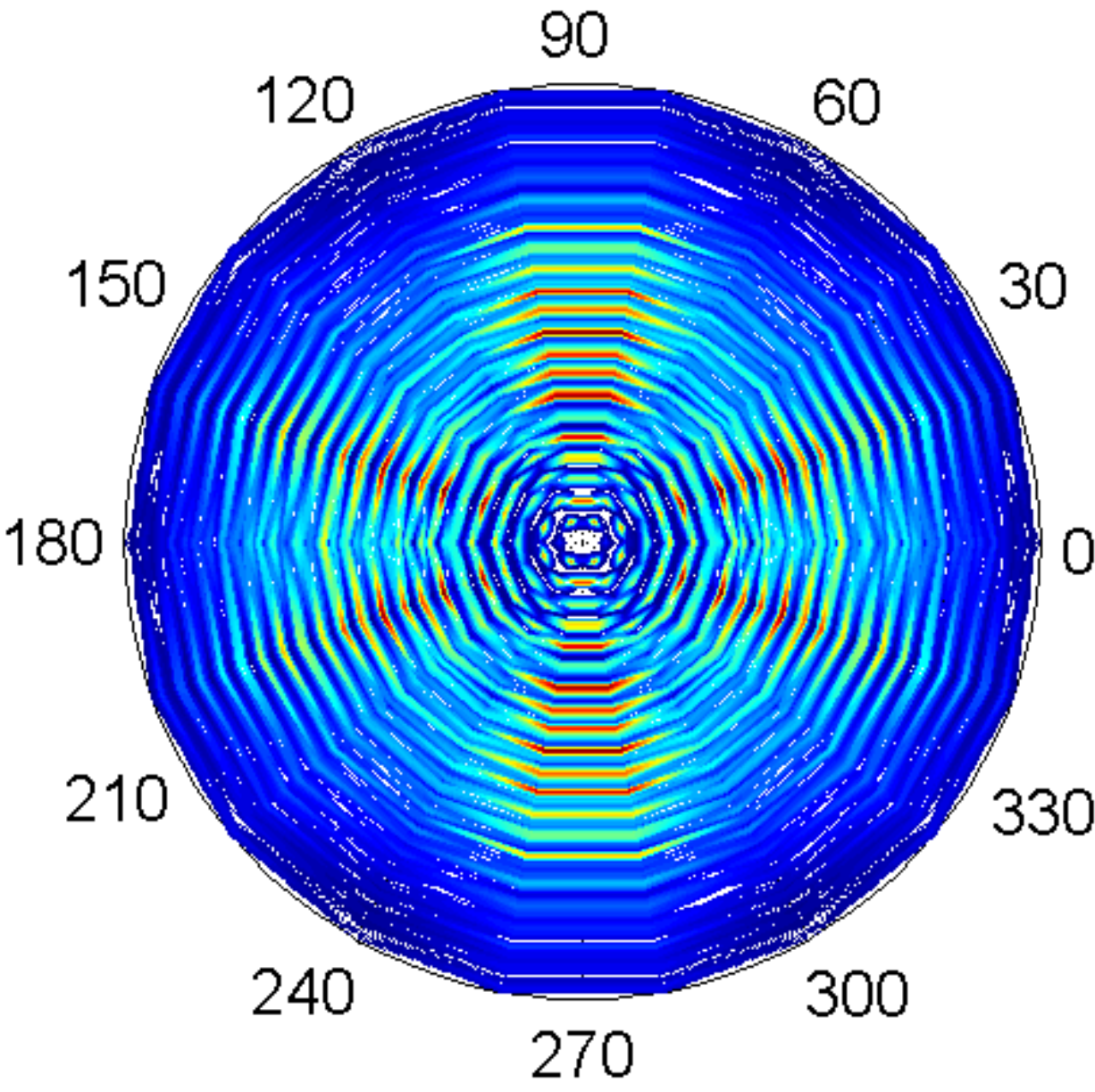}
(b)
\includegraphics[width=1.8in,height=1.75in]{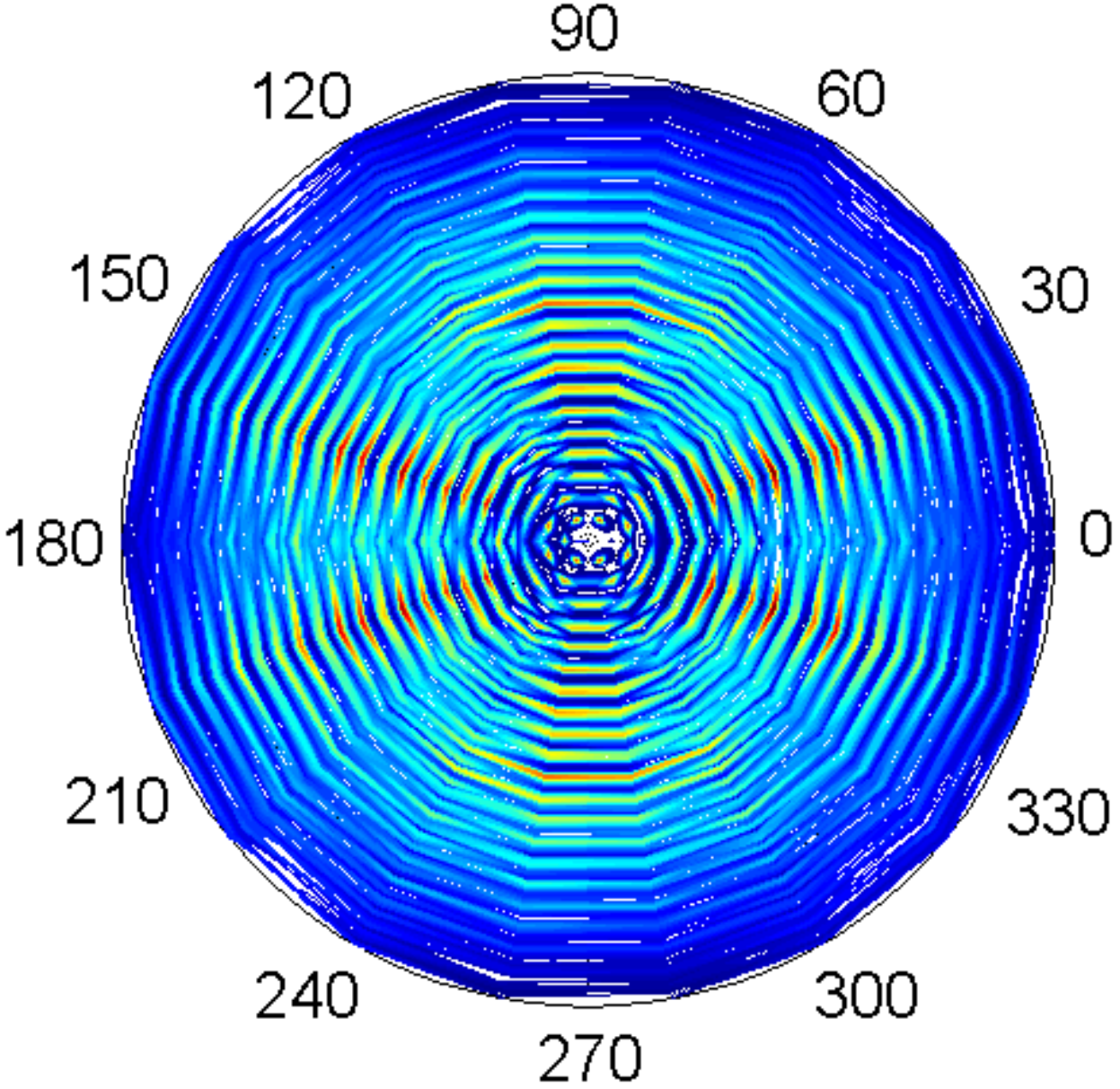}
(c)
\includegraphics[width=2.0in,height=1.75in]{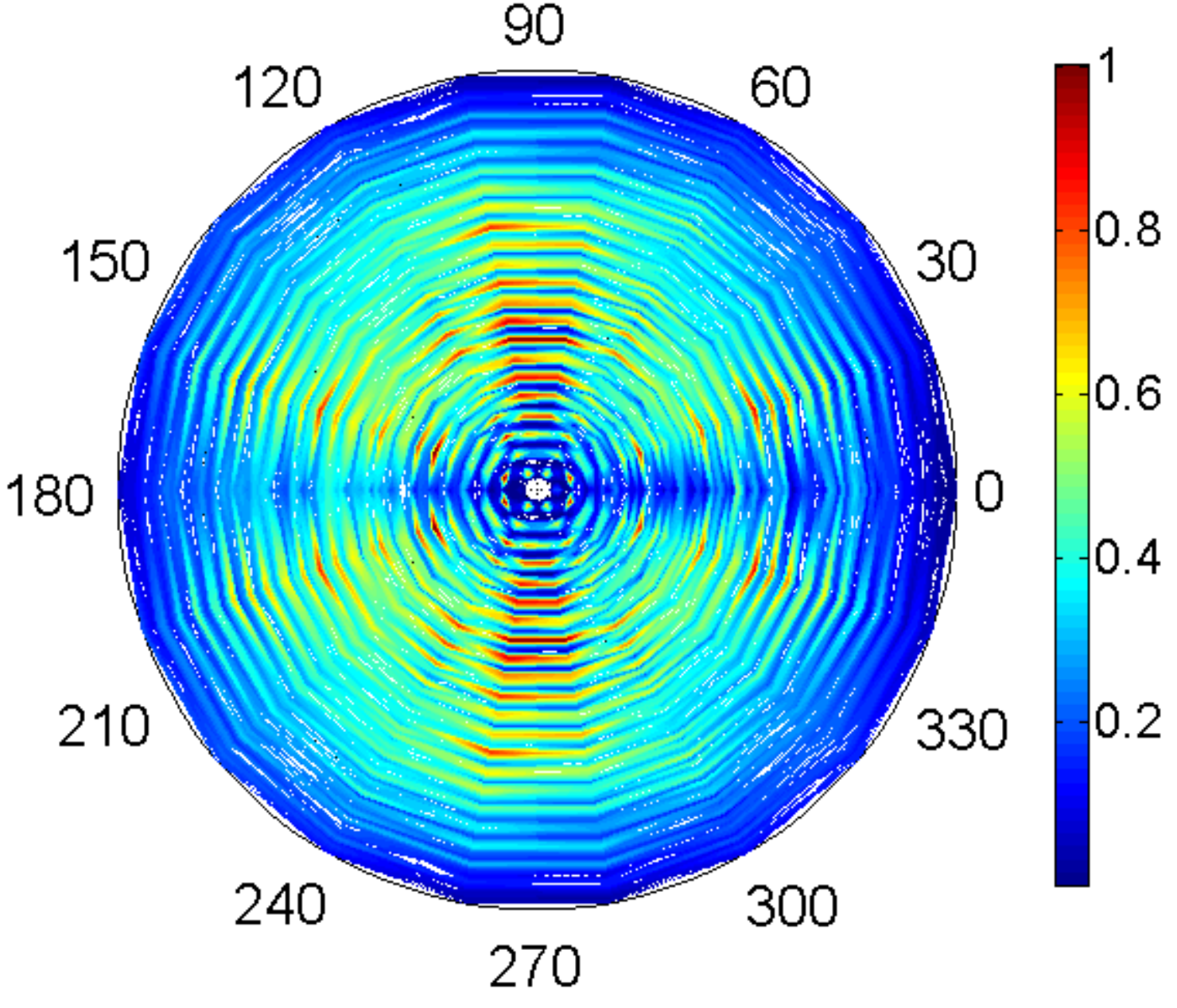}
\end{center}
\caption{
(Color online)
The radial-angular distribution function $g(r,\theta)$  in the ``Bouncing Bed''-state at a shaking-acceleration $\Gamma=5$ 
with increasing shaking amplitudes: (a) $A/d=0.5$, (b) $A/d=1$ and (c) $A/d=2.4$.  Other parameters are as in Fig.~\ref{fig:fig12}.
}
\label{fig:fig13}
\end{figure*}

\begin{figure*}[!ht]
\begin{center}
(a)
\includegraphics[width=2.0in,height=1.7in]{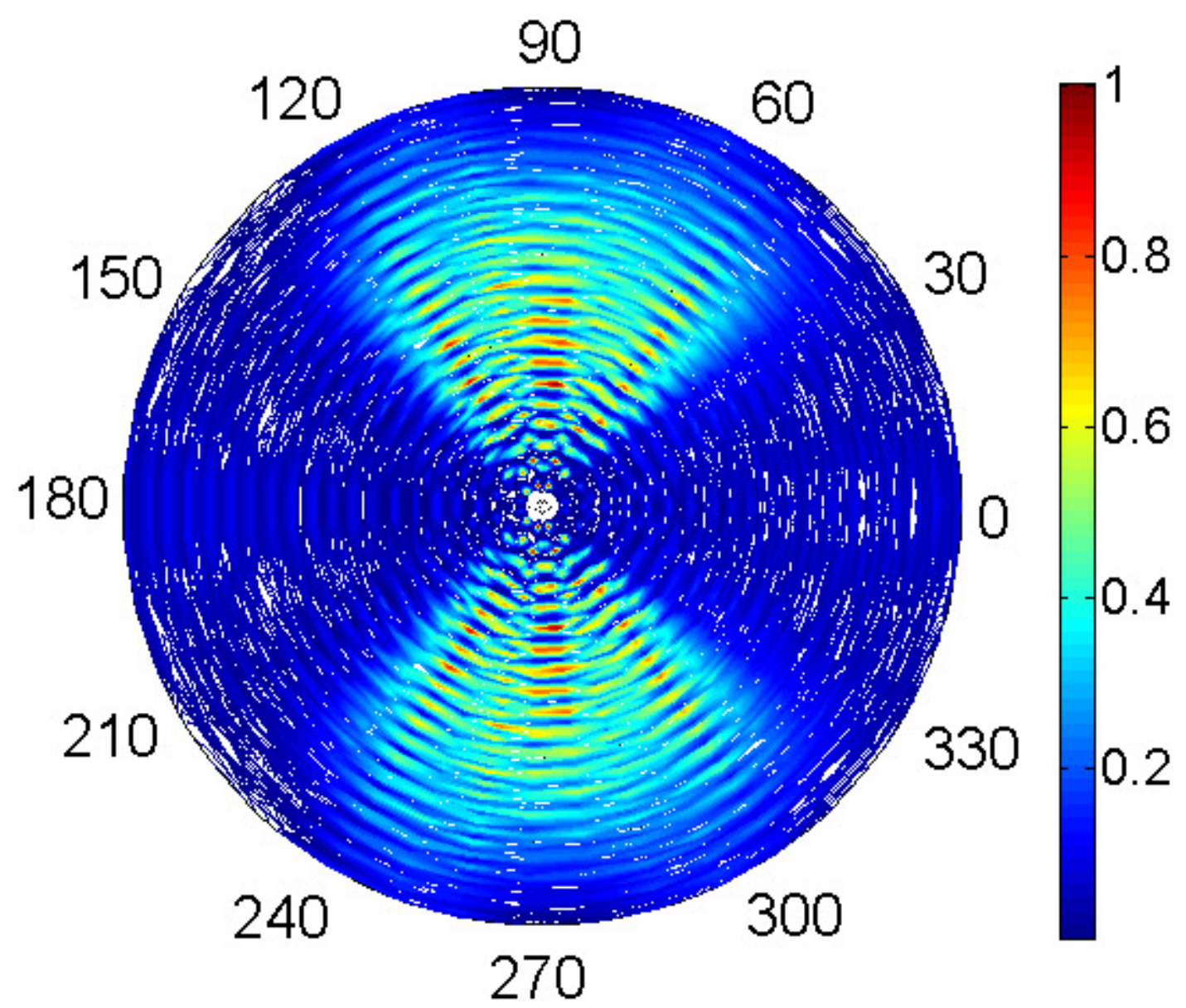}
(b)
\includegraphics[width=2.0in,height=1.7in]{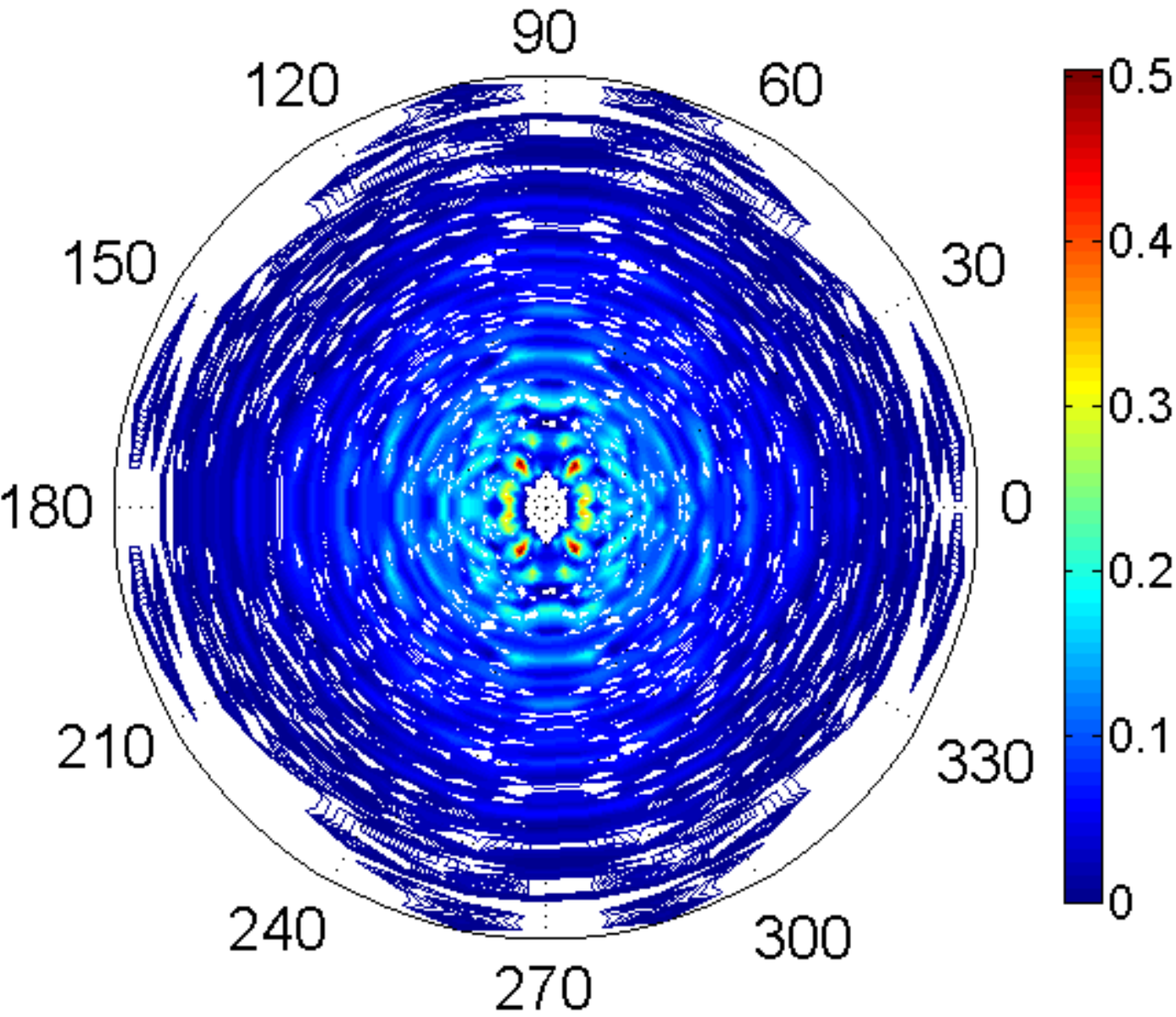}
(c)
\includegraphics[width=2.0in,height=1.7in]{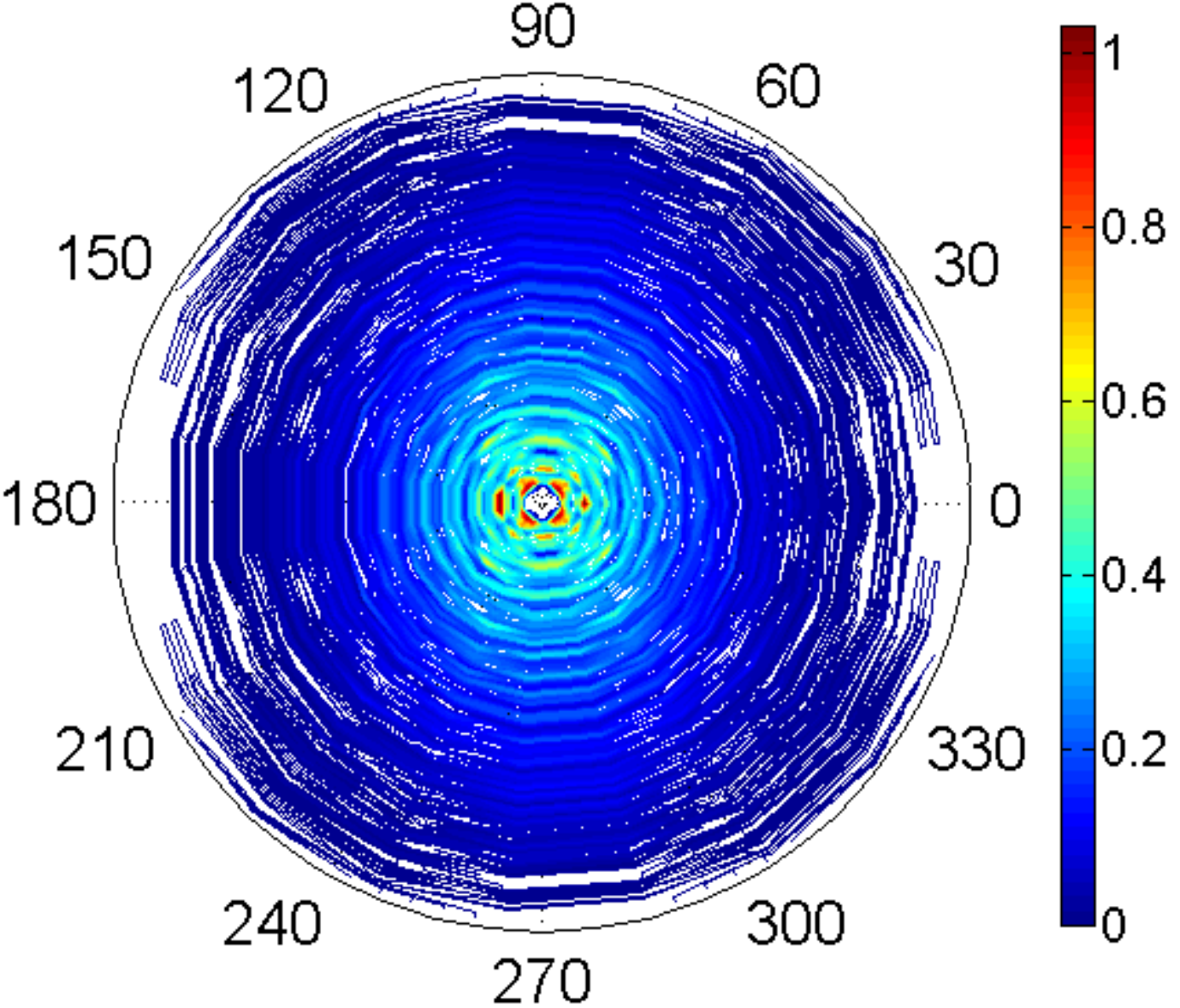}\\
(d)
\includegraphics[width=3.0in,height=2.6in]{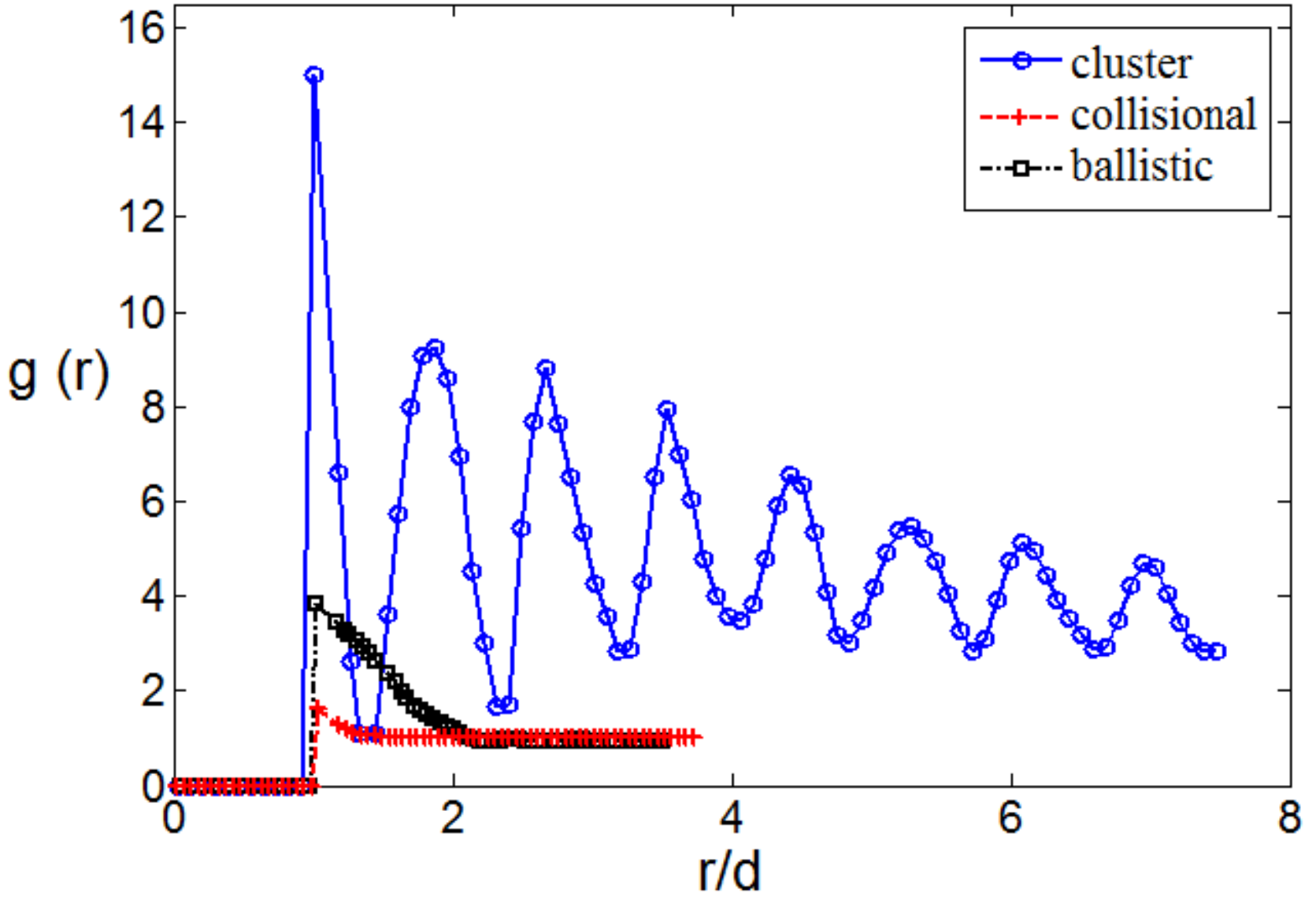}
\end{center}
\caption{
(Color online)
The radial-angular correlation function \boldmath{$g(r,\theta)$} in various regions of \textit{LS}: (a) floating-cluster, (b) collisional layer and (c) ballistic layer; 
other parameters are same as in Fig.~\ref{fig:fig12}.
(d) The radial component of the $g(r,\theta)$ in the floating-cluster (blue circles), collisional layer (the lower curve marked by red plus) and ballistic layer 
(the middle curve marked by black squares). 
}
\label{fig:fig14}
\end{figure*}

For a better understanding of the orientational-ordering of particles in the ($x,y$)-plane,
we probed the two-dimensional pair-correlation function $g(r,\theta)$ by mapping $g({\boldmath r})$ onto polar coordinates ($r,\theta$).
The radial-angular correlation distribution $g(r,\theta)$ is defined as~\cite{AT1989,Bogdan2006}
\begin{equation}
 g(r,\theta)=\frac{1}{N\rho(r,\theta)}\sum_{i=1}^{N}\sum_{j\neq i}^{N}\langle\delta
 (r-r_{ij})\delta(\theta-\theta_{ij})\rangle,
\end{equation}
which gives correlations for the pair of particles $i$ and $j$,  the distance between whose centers of mass is $r_{ij}$ and the angle between the plane
containing particles $i$, $j$ and the horizontal plane is  $\theta_{ij}$;  $\rho(r,\theta)g(r,\theta)$ is the conditional probability
of finding a particle at a distance $r$ from a reference particle and in a plane containing reference particle
which makes an angle $\theta$ with respect to the horizontal plane.   The contour plot of the $g(r,\theta)$ in the ($r,\theta$)-plane is likely to  reveal the
``contact-network'' of particles in the granular-bed, indicating the most probable spatial-configuration of particles around a reference particle.

To calculate $g(r,\theta)$, in addition to binning the particles in the radial direction, we also binned them in the angular direction --
the number of bins considered in radial and angular directions are $100$ and $20$, respectively.
Let us first probe the $g(r,\theta)$ in the bouncing-bed state. The contour-plots of $g(r,\theta)$ for different shaking amplitude $A/d$
are shown in Figs.~\ref{fig:fig13}(a,b,c) -- the shaking acceleration is set to $\Gamma=5$ such that the system is in the BB-state (see Fig.~\ref{fig:fig2}). 
The six-fold symmetry of the contact-network, resembling the hexagonally-packed crystalline
structure of the \textit{Bouncing Bed}, is clearly evident in Fig.~\ref{fig:fig13}(a).
With increasing shaking amplitude ($A/d$), the lattice points of the hexagonal-packing structure get perturbed, thereby breaking 
the `exact' directional symmetry as it is evident in Figs.~\ref{fig:fig13}(b) and ~\ref{fig:fig13}(c). 
This is because, as $A/d$ increases, the particles get more loosely packed making them more mobile to move around each other
and  thus destroying the angular anisotropy of their positions.

The $g(r,\theta)$ in three  regions of the \textit{LS}, namely, \textit{collisional layer}, \textit{floating cluster} and \textit{ballistic layer}, are  presented in Fig.~\ref{fig:fig14}.
The reference particle considered in these polar-plots is located at the center of the circle.
The $g(r,\theta)$ in Fig.~\ref{fig:fig14}(a) indicates that the \textit{floating-cluster} is highly anisotropic in nature, 
showing directional dependence with increased probability of ``head-on'' collisions (i.e. $\theta=\pi/2, 3\pi/2$).
On the contrary, the \textit{collisional} and \textit{ballistic} layers in  Figs.~\ref{fig:fig14}(b) and ~\ref{fig:fig14}(c), respectively, show
angular-isotropy which implies that the particles are more  likely to be found at any angular orientation on average -- this  is expected since these regions are in a gaseous phase.
A closer look at Fig.~\ref{fig:fig14}(a) reveals that the the six-fold symmetry of the contact network still survives in the LS, but the `hexagonal'' lattice
structure seem to have been significantly modified, with more collisions  likely to occur at $\theta=2\pi/3$ and $4\pi/3$ in additions to ``head-on'' collisions.
This indicates that the packing of particles in the LS is much looser than the ``ideal'' hexagonal-packing. The latter observation can be further rationalized
if we analyze the  radial component of the $g(r,\theta)$ [of Figs.~\ref{fig:fig14}(a,b,c)] which is displayed in Fig.~\ref{fig:fig14}(d). 
Note that $g(r)\equiv\langle g(r,\theta)\rangle_{\theta}$, and hence it provides information on the radial-configuration of particles around a test particle over
a circular region of diameter equal to the height of the floating-cluster. The $g(r)$ in the floating-cluster region (the blue curve in Fig.~\ref{fig:fig14}d)
resembles more a liquid-like structure, with its second peak being located at $r/d\approx 1.9$.
Therefore the floating-cluster region is in a liquid-state, which hovers over a gas-like collisional layer underneath,
and this makes the connection with the original Leidenfrost-state~\cite{Leidenfrost,Eshuis2005} more appropriate.

\subsection{Height oscillations  in granular Leidenfrost state}

\begin{figure}[!h]
\begin{center}
\includegraphics[width=3.2in,height=2.8in]{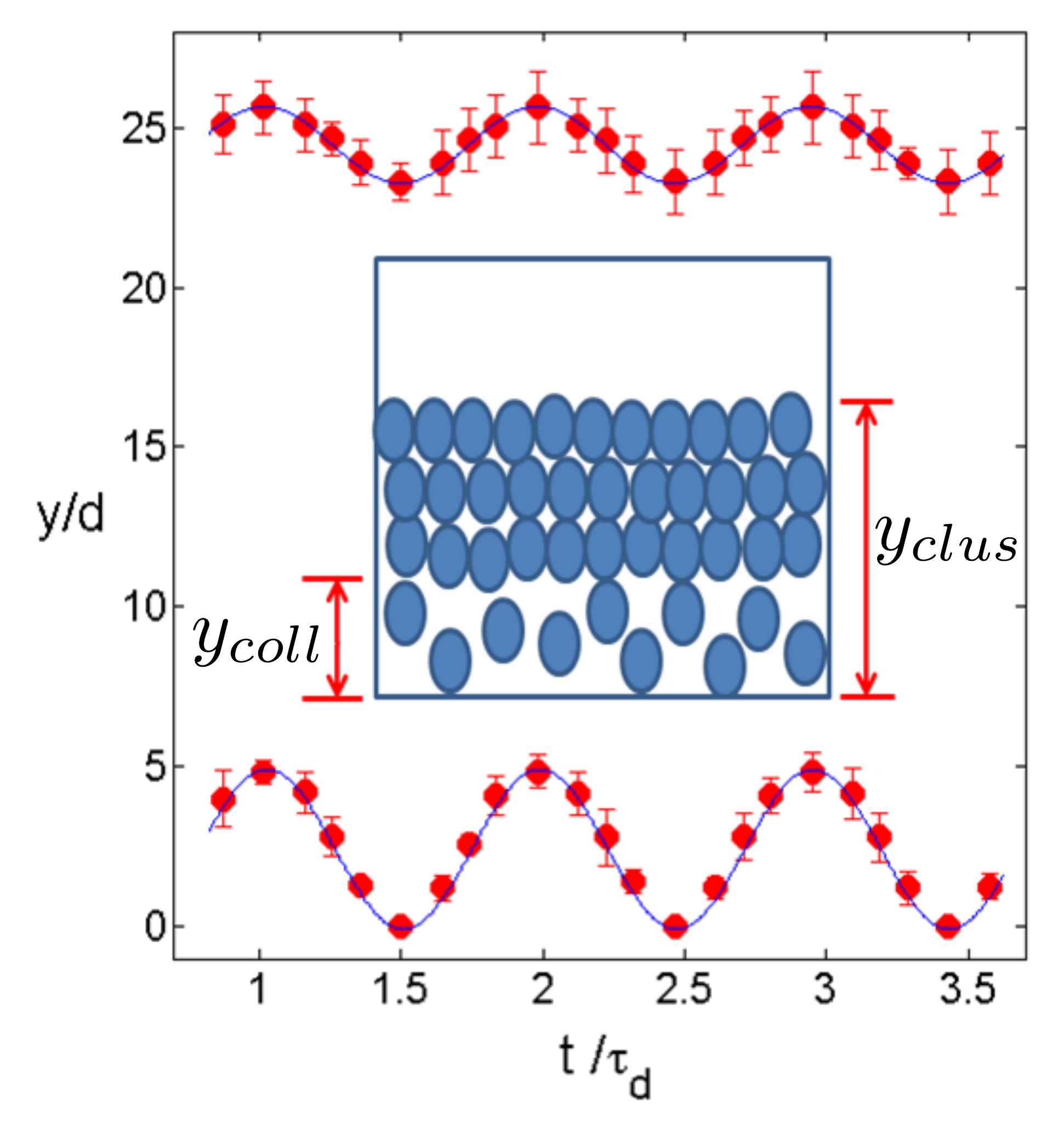}
\end{center}
\caption{
(Color online)
Temporal variations of  $y_{coll}(t)$ and  $y_{clus(t)}$ at $\Gamma=30$ and $A/d=1.6$, with other parameters being $F=h_0/d=25$ and $d=2 mm$ diameter glass beads.
The red circles represent experimental data (with error bar) and the best fitted  curve is denoted by the blue line.
Inset depicts an sketch of the Leidenfrost state (\textit{LS}):  $y_{coll}(t)$ is the instantaneous height of the collisional-layer
and $y_{clus}(t)$ is the  vertical location of the top of the floating-cluster.
}
\label{fig:fig15}
\end{figure}

A closer look at the snapshots (of the Leidenfrost state) in Fig.~\ref{fig:fig4} reveals  that the top-surface of the granular bed as well as the interface
separating the dense floating-cluster and the dilute collisional-layer do vary with time within an oscillation cycle of the external driving.
It is interesting to find out if there is a definite  frequency associated with such height oscillations in the granular Leidenfrost state.
To this end, we have tracked the temporal evolution of two characteristic heights:
(i) the location of the  top surface of the floating-cluster region $y_{clus}(t)$, and  (ii) the height of the  collisional layer $y_{coll}(t)$ 
(i.e.~the vertical location of the interface between the floating-cluster  and the  collisional-layer beneath), with both being measured from the base of the container,
see the sketch in the inset of Fig.~\ref{fig:fig15}.  The high-speed images of the bed 
were analyzed to measure $y_{coll}(t)$ and $y_{clus}(t)$ at various time instants over a few shaking cycles.
The unsteadiness of the LS, if any, is likely to be implicated in the temporal variations of $y_{coll}(t)$ and $y_{clus}(t)$ as we demonstrate  below.

Let us  consider the case of \textit{LS} observed in  experiments with $F=25$ layers of $d=2$mm diameter glass beads
at a shaking intensity of $\Gamma=30$ with $A/d=1.6$ ($f=48.26\;Hz$). 
The time evolutions  of $y_{coll}(t)$ and $y_{clus}(t)$ are shown in Fig.~\ref{fig:fig15} --  the time has been scaled by time-period of driving ($\tau_d=1/f\approx 20.72 ms$).
The experimental data (denoted by red circles) for both $y_{coll}$ and $y_{clus}$ are best fitted by sinusoids (the blue curves in Fig.~\ref{fig:fig15}) of the following form
\begin{eqnarray}
 y_{coll}(t)/d=2.5 + 2.4\sin(0.305t+1.159),\\
 y_{clus}(t)/d=24.46 + 1.2\sin(0.307t+1.231),
\end{eqnarray}
where $t$ is the time measured in $ms$.  Note that the oscillation-amplitude of the top surface ($\langle y_{clus}(t)\rangle$)
is smaller  than that of the interface ($\langle y_{coll}(t)\rangle$).  From the above equations, the angular frequencies $\omega_{coll}$ and $\omega_{clus}$
are $305\;rad/s$ and $307\;rad/s$, respectively, which closely agree with
the driving angular-frequency ($\omega_{d}=2\pi f=303$). Thus, the interface and top surface of the \textit{LS} oscillates harmonically
and is synchronized with the frequency of the external vibration.

\begin{figure}[!h]
\begin{center}
(a)
\includegraphics[width=3.2in,height=2.8in]{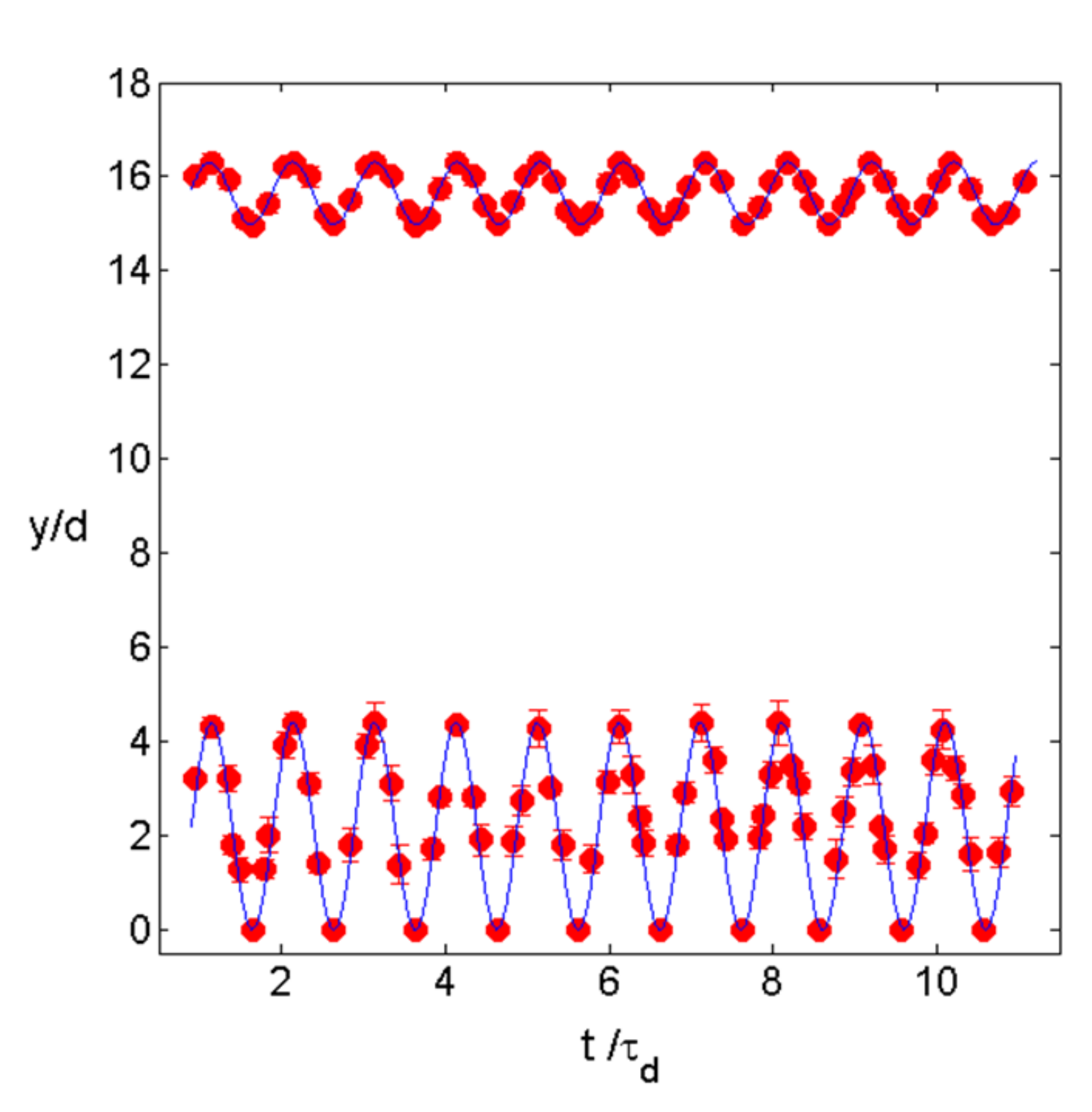}\\ 
\includegraphics[width=3.2in,height=2.8in]{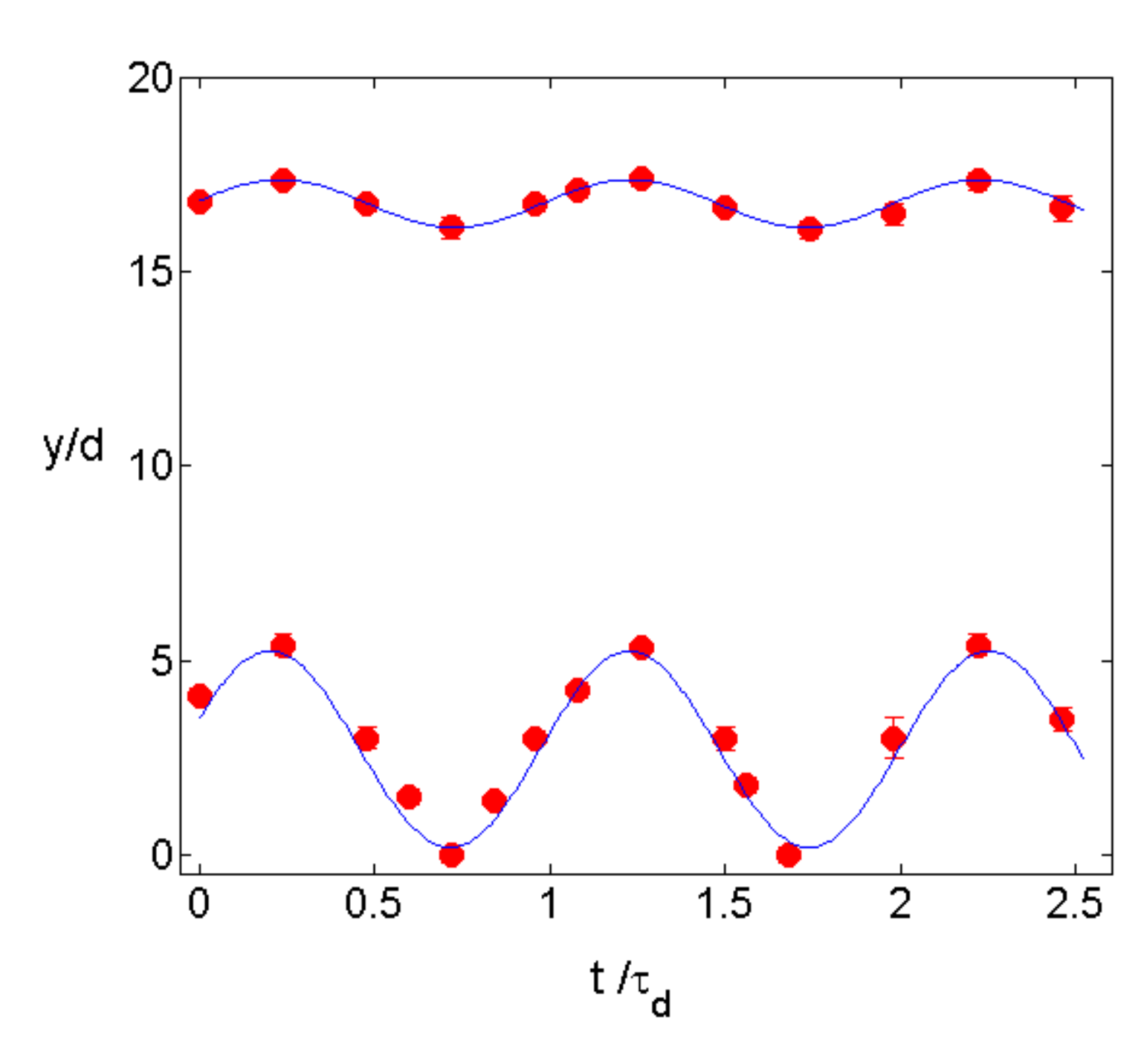}
(b)
\end{center}
\caption{ 
(Color online)
Same as Fig.~\ref{fig:fig15}, but for $F=12$ layers of {$d=5\;mm$} diameter glass-beads with  $A/d=0.6$:
(a) $\Gamma=30$ and (b) $\Gamma=43.4$.
}
\label{fig:fig16}
\end{figure}

To check the robustness of above finding, we analyzed the images of the \textit{LS} observed in  a second set of experiments with $5\ mm$ diameter beads.
The temporal variations of $y_{coll}(t)$ and $y_{clus}(t)$ at $\Gamma=30$ and $43.4$ are shown in Figs.~\ref{fig:fig16}(a) and ~\ref{fig:fig16}(b), respectively;
other parameters are $A/d=0.6$ and $F=h_0/d=12$ (refer to the phase-diagram in Fig.~\ref{fig:fig5}).
While the experimental data for $\Gamma=30$ are best fitted via the sinusoids of the following form
\begin{eqnarray}
 y_{coll}(t)/d=2.2+1.84\sin(0.315t+0.62)\\
 y_{clus}(t)/d=15.7+0.66\sin(0.31t+0.864),
\end{eqnarray}
the data for $\Gamma=43.4$ are best fitted by 
\begin{eqnarray}
 y_{coll}(t)/d=2.7+2.53\sin(0.37t-0.322)\\
y_{clus}(t)/d=16.8+0.616\sin(0.377t+0.118),
\end{eqnarray}
where the time is measured in ms.
At $\Gamma=30$, we found that $\omega_{coll}\approx 314.7\;rad/s$ and $\omega_{clus}\approx 310\;rad/s$
that are very close to the driving frequency $\omega_{d}=313$rad/s.
At $\Gamma=43.4$, $\omega_{coll}=370\; rad/s$ and $\omega_{clus}=377\;rad/s$ and
$\omega_{d}=376\;rad/s\approx \omega_{coll} \approx \omega_{clus}$.
Another noteworthy feature in Fig.~\ref{fig:fig16}(a) is the absence of low-frequency modulation 
(even over 12 oscillation cycles) in both  $y_{coll}(t)$ and $y_{clus}(t)$.

Collectively, the above analysis indicates  that both the interface between the collisional-layer and the floating-cluster and the top surface of the floating-cluster
oscillates harmonically and are synchronized with the frequency of the external vibration.
Therefore, the granular Leidenfrost state is a {\it period-1} wave (i.e.~an {\it f-wave}) as is the case for the bouncing-bed.

Although the `synchronous'  height-oscillations of the LS were never quantified previously,
there are simulation works in the  same direction. The  most recent simulation~\cite{RLT2013} identified a low-frequency (semi-periodic) oscillation
in the density-inverted state (in a quasi-3D setup of a narrow column of length and width $L/d=5-W/d$); they also found 
delta-like peaks at the driving frequency and its harmonics in the power-spectra of the temporal-variation of the center-of-mass of the system (see their Fig.~\ref{fig:fig4}$b$)--
while the former may be connected to  the $f$-wave nature of the LS as found in our work, the implications of the peaks at the harmonics of $f$ remain unclear.
On the other hand, the  simulation work of Bougie \etal~\cite{BPP2012} probed the time-dependence and density inversion
in a vertically vibrated 3D-box filled with shallow layers ($F=h_0/d=4.3$) of granular materials.  While they found that the density-inverted state
is indeed an $f$-wave in the low-$\Gamma$ regime ($\Gamma=5.68$), the density profiles become time-invariant in the high-$\Gamma$ regime ($\Gamma=56.5$). 
The latter observation is in stark contrast to our finding in Fig.~\ref{fig:fig16} that increasing $\Gamma$ 
from $30$ to $43.4$ does not seem to have an effect on the  $f$-wave nature of the LS.
(It may be noted that  the low- and high-$\Gamma$ regimes in simulations of Ref.~\cite{BPP2012}  correspond to the same
 shaking strength ($S=\Gamma(A/d)=constant$) but having small- and large-$A/d$, respectively.)
To isolate the possible effects of the two-dimensionality of our experiments on height-oscillations, 
additional experiments (not shown) were conducted in a quasi-2D box (of width of about  5 particle diameter, $W/d=5.5$, and length $L/d=100$)
with a shallow granular layer ($F=h_0/d=6$);  increasing the shaking intensity from $30$ to $40$, however,  we found that the  Leidenfrost state remains a ``period-1'' wave,
implying that the LS is synchronized with the shaking frequency even in a quasi-2D vibro-fluidized bed.
Simulations in a 2D-box and/or additional  experiments may help to resolve the disagreement between the above simulations~\cite{BPP2012,RLT2013} and our
experiments about the synchronous time-dependence of the Leidenfrost state.

\section{Summary and conclusions}

We carried out detailed experiments on the pattern-formation dynamics in a vertically shaken two-dimensional ``mono-layer'' granular system.
A collection of spherical glass beads were held in a Hele-Shaw-type container of certain length ($L$) and height ($H$),
having a width ($W$) such that it can accommodate only one layer of beads across its width ($W/d\approx 1.1$);
the containers with two different lengths, $L/d=20$ and $40$, were investigated.
The particle-filled container was vertically vibrated harmonically, $y=A\sin(2\pi f t)$, via an Electromagnetic shaker, where $A$ is the amplitude of shaking and $f$ is its frequency.
The experimental results were presented for a wide range of (i) shaking intensities $\Gamma=A\omega^2/g\in(0,55)$, (ii) amplitude ratio $A/d\in (0.1, 4)$
and (iii) particle loading or filling-height $F=h_0/d$ (where $h_0$ is the number of particle layers at rest);
In addition to showing the raw images and movies of various patterns, the quantitative
measurements have been made on (i) the density and granular temperatures and (ii) the pair-correlation functions   using particle tracking algorithms.

For shaking accelerations $\Gamma\leq  1$, the granular bed moves with the container base without detaching from it
and this is the regime of  {\it solid-bed} which gave birth to the well-known {\it bouncing bed} state
(in which the bed detaches from the base and starts bouncing like a single particle) at $\Gamma>1$.
At $\Gamma\sim O(10)$, the BB-state transitioned into the so-called Leidenfrost-state~\cite{Leidenfrost,Eshuis2005}
in which a dense cluster of particles floats over a dilute gaseous layer~\cite{LR1995,MPB2003,Eshuis2005}. 
The critical shaking acceleration for  the transition from \textit{BB} to \textit{LS} was found to have
a power-law dependence, $\Gamma_{BB}^{LS}\sim F^{1.217} (A/d)^{-7/8}$, on the particle loading depth ($F=h_0/d$) and the shaking amplitude ($A/d$).
Therefore, the critical shaking strength [$S=\Gamma(A/d)$ which is a measure of the input kinetic energy via shaking] increases weakly
with increasing shaking amplitude [$S_{BB}^{LS}\propto (A/d)^{1/8}$] for a specified particle loading $F$;
this result  is in contrast to the findings of Eshuis \etal~\cite{Eshuis2005} who showed $S=const$
at ``$BB\to LS$''-transition.  We speculate that the frictional-barrier at the front and back walls increases
with increasing shaking amplitude which might be responsible for the  weak-increase of $S_{BB}^{LS}$ with $A/d$.

Carrying out experiments in a container of larger length $L/d=40$, we uncovered the complete sequence of bifurcations with increasing $\Gamma$:
SB (solid bed) to BB (bouncing bed) to LS (Leidenfrost state) to ``2-roll Convection'' to ``1-roll Convection'' and finally to a granular gas.
While the first two transitions ($SB\to BB\to LS$) were also reported  previously [by Eshuis {\etal}~\cite{Eshuis2005}], however, the ``$LS\to Convection$'' transition
and the ``$Convection\to Gas$'' transition are new findings in the context of  a mono-layer vibrofluidized system at strong shaking.
In particular, for a given length of the Hele-Shaw container, 
the coarsening of a pair of convection rolls leading to the genesis of a ``single-roll'' structure (i.e. the {\it multi-roll transition})
and its subsequent transition to a granular-gas  were never reported in previous experiments
(although a recent simulation study~\cite{RLT2013} did report on similar multi-roll transition in a quasi-2D box).
The shaking strength for the onset of ``$LS\to Convection$'' transition, $S_{LS}^{Con}\sim (A/d)^{0.84}$, is found to increase strongly with
increasing shaking amplitude. The latter finding is in contrast to the very weak-dependence of the same found in the quasi-2D experiments~\cite{Eshuis2007},
and the related simulations~\cite{RLT2013}  also indicate that  $S_{LS}^{Con}$ is almost independent of $A/d$.

The density and temperature profiles, obtained via particle tracking algorithms,
revealed clear signatures of the transition from the bouncing-bed state to the density-inverted Leidenfrost state.
The Leidenfrost-state is characterised by three distinct regions: (i) a dilute ``collisional-layer''
of particles near the vibrating base, (ii) a dense ``floating-cluster'' above the collisional layer
and (iii) a ``ballistic-layer'' on the top of the floating-cluster (where the particles move around ballistically).
While the ballistic-layer was also found to exist in the bouncing-bed state (at large shaking strength),
the crucial distinction of the LS from the BB-state is the existence of the dilute collisional-layer
that acts as  a cushion over which a dense cluster floats~\cite{Eshuis2005}.
Another distinguishing criterion between LS and BB is that the granular temperature
in the BB-state is lower near the vibrating base in contrast to the LS for which the maximum temperature occurs near the base. 
The temperature in both BB and LS was found to be  ``anisotropic'' in the sense that
the vertical component ($T_y$) of temperature is, in general, larger than its
horizontal component ($T_x$). Interestingly, in the LS, the temperature ratio $T_x/T_y$
increased with increasing elevation, and can even exceed unity (i.e.~$T_x>T_y$) in its ballistic layer at very strong shaking.

The microstructure within the bouncing-bed and the Leidenfrost state  were probed 
by evaluating the two-dimensional pair-correlation function $g(r,\theta)$ which provided information on the particle-configuration 
(i.e.~both the spatial and orientational ordering of particles).
The six-fold symmetry of the $g(r,\theta)$ in the BB-state was tied to the hexagonally-packed crystalline structure of the bed, and the degree of anisotropy
of this crystalline state was found to decrease with increasing shaking amplitude ($A/d$) at a fixed  shaking intensity $\Gamma$ and vice versa.
The $g(r,\theta)$ in the ``collisional-layer'' and the ``ballistic-layer'' of the LS displayed angular-isotropy, and are therefore in a gas-like state.
The analysis of the radial distribution function, $g(r)\equiv \langle g(r,\theta)\rangle_{\theta}$, in the ``floating-cluster'' region of the LS
revealed a clear liquid-like structure (the blue curve in Fig.~\ref{fig:fig14}d).
Therefore the floating-cluster is indeed in a liquid-state, which hovers over a gas-like collisional layer underneath,
and this makes the connection with the original Leidenfrost-state~\cite{Leidenfrost,Eshuis2005} more succinct.

We uncovered a novel unsteady  behaviour associated with the Leidenfrost state, wherein the height of the 
collisional-layer (i.e.~the interface that separates the floating-cluster from the dilute 
collisional-layer underneath) as well as  the height of the floating-cluster oscillated sinusoidally with time. The oscillation frequencies closely matched the frequency of the shaker. 
Therefore, the granular Leidenfrost state is not a stationary state, rather it is a period-1 or $f$-wave, i.e., the temporal-order of the LS is the same  as that of the Bouncing-bed state.
This finding has important implications on the theoretical analysis~\cite{Eshuis2010,Shukla2014}  that has  been carried out in the recent past.
On the other hand, the  recent simulations~\cite{RLT2013} (in a narrow quasi-3D column)
identified  a low-frequency (semi-periodic) oscillation in the density-inverted state which we did not observe in our experiments,  presumably due to 
very low values of the underlying frequency and/or  due to the lateral-confinement  of the vibrated column -- 
these  issues require future investigations.

In this work  we have restricted our experiments to a relatively narrow aspect-ratio ($L/h_0 < 7$) container, and
did not observe sub-harmonic patterns like $f/2$ undulatory waves, $f/4$ spikes, etc.
Such patterns are known to appear~\cite{DFL1989,Sano2005,Eshuis2007,AAlam2013} 
in experiments with large aspect-ratio ($L/h_0>15$) container for a range of $\Gamma$ lying between the BB-state and the LS,
and hence they may be expected in the present 2D-setup too if we further increase the length of the container.
These experimental issues can be taken up in a future work.
 A recent simulation work~\cite{RTLM2015} has probed the role of noise on the ``$LS\to Convection$'' transition,
and modelled it via a quintic-order stochastic amplitude equation. It may be possible to derive such a quintic-order 
equation from the underlying hydrodynamic equations~\cite{AS2013,SA2009}.
Lastly, we recall that  the theoretical works on granular convection based on linear and nonlinear stability analyses~\cite{Eshuis2010,Shukla2014}
do not show any dependence of the critical  shaking strength $S_{LS}^{Con}$ (for the onset of convection) on the shaking amplitude $A/d$ (in contrast to present findings, Eqn.~12). 
This may be due to the over-simplified boundary conditions imposed at the vibrating wall (constant temperature)  in all theoretical analyses.
Therefore, the present work also opens up important theoretical issues that need to be addressed in the future.

\begin{acknowledgments}
This work has been generously funded by the Department of Atomic Energy,  Government of India 
via ``DAE-SRC Outstanding Research Investigator Award'' to MA (Project No. 2010/21/06-BRNS).
The experiments  were carried out  by Mr. Ansari under the supervision of MA; 
the final manuscript with interpretations and its revision were prepared by MA.
We sincerely thank Mr. Krishnoji Rao for fabricating the experimental setup and the related accessories, and Mr. Raghavendra 
and Mr. Venkateshulu (Jost Engg. Ltd.) and Mr. Ramakrishna (Data Physics Pvt. Ltd.) for tutoring the Student on the shaker-system.
\end{acknowledgments}



\begin{thebibliography}{99}
\bibitem{Chladni1787}
E.~F.~F. Chladni,
{\it Entdeckungen \"{u}ber die Theorie des Klanges}.
Kessinger (1787).
\bibitem{Faraday1831}
M. Faraday,
On a peculiar class of acoutic figures and certain forms assumed by a group of particles upon vibrating
elastic surfaces.
Phil.~Tran.~.~Soc.~Lond. {\bf 52}, 299-318 (1831).
\bibitem{DFL1989}
S. Douady, S. Fauve, and C. Laroche,
Subharmonic instabilities and defects in a granular layer under vertical vibrations.
Europhys.~Lett  {\bf 8}, 621-626 (1989).
\bibitem{CR1991}
E. Clement and  J. Rachenbach,
Fluidization of a bidimensional powder.
Europhys.~Lett. {\bf 16}, 133 (1991).
\bibitem{GHS1992}
J.~A.~C. Gallas, H.~J.~Herrmann, and S. Sokolowski,
Convection cells in vibrating granular media.
Phys.~Rev.~Lett., {\bf 69}, 1371-1374 (1992). 
\bibitem{PB1992}
H.~K.~Pak and R.~P.~Behringer,
Surface waves in vertically vibrated granular materials.
Phys.~Rev.~Lett. {\bf 71}, 1832-1835 (1992).
\bibitem{Luding1994}
S. Luding, E. Clement, A. Blumen, J. Rachenbach, and J. Duran,
The onset of convection in molecular dynamics simulations of grains.
Phys.~Rev.~E  {\bf 50}, R1762--1765 (1994).
\bibitem{LR1995}
Y. Lan and A.~D. Rosato,
Macroscopic behaviour of vibrating beds of smooth inelastic spheres.
Phys.~Fluids  {\bf 7}, 1818-1831 (1995).
\bibitem{PH1995}
T. P\"{o}schel and H.~J.~Herrmann,
Size segregation and convection.
Europhys.~Lett. {\bf 29}, 124-128 (1995).
\bibitem{BM1995}
M.~Bourzutschky and  J.~Miller,
Granular convection in a vibrated fluid. 
Phys.~Rev.~Lett.  {\bf 74}, 2216-2219 (1995).
\bibitem{HYH1995}
H.~Hayakawa, S.~Yue,  and  D.~C.~Hong,
Hydrodynamic description of granular convection.
Phys.~Rev.~Lett.  {\bf 75}, 2328-2331 (1995).
\bibitem{Warr1995}
S.~Warr, J.~M.~Huntley, and G.~T.~H. Jacques,
Fluidization of a two-dimensional granular system: experimental study and scaling behaviour.
Phys.~Rev.~E  {\bf 52}, 5583-5595 (1995).
\bibitem{WB1996}
C.~R.~Wassgren, C.~E.~Brennen, and M.~L.~Hunt,
Vertical vibration of a deep bed of granular materials in a container.
J.~Appl.~Mech., {\bf 63}, 712-719 (1996).
\bibitem{Aoki1996}
K.~M. Aoki, T. Akiyama, Y. Maki, and  T. Watanabe,
Convective roll patterns in vertically vibrated beds of granules.
Phys.~Rev.~E  {\bf 54}, 874-882 (1996).
\bibitem{Knight1996}
J.~B.~Knight, E.~E. Ehrichs, V.~Y. Kuperman, J.~K. Flint, H.~M.  Jaeger, and S.~R.   Nagel,
Experimental study of granular convection.
Phys.~Rev.~E {\bf 54}, 5726-5738 (1996).
\bibitem{UMB1996}
P.~B. Umbanhower, F. Melo, and H.~L.   Swinney,
Localized excitations in a vertically vibrated granular layer.
\emph{Nature} \textbf{382}, 793-796 (1996).
\bibitem{KWG1997}
A. Kudrolli, M. Wolpert, and J.~P. Gollub,
Cluster formation due to collisions in granular materials.
Phys.~Rev.~Lett. {\bf 78}, 1383 (1997).
\bibitem{Bizon1998}
C. Bizon, M.~D. Shattuck, J.~B. Swift, W.~D., McCormick, and H.~L.  Swinney, 
Patterns in 3D vertically oscillated granular layers: simulation and experiment.
{Phys.~Rev.~Lett.} {\bf 80}, 57-60 (1998).
\bibitem{RRC2000}
R. Ramirez, D. Risso, and P. Cordero,
Thermal convection in fluidized granular systems.
Phys.~Rev.~Lett. {\bf 85}, 1230-1233 (2000).
\bibitem{YH2000}
S.~C. Yang and S.~S.~Hsiau,
Simulation study of the convection cells in a vibrated granular bed.
Chem. Engg. Sci.  {\bf 55}, 3627 (2000).
\bibitem{PSS2000}
T. P\"{o}schel, T. Schwager, and C. Saluena,
Onset of fluidisation in vertically shaken granular material.
Phys.~Rev.~E  {\bf 62}, 1361 (2000).
\bibitem{WHP2001}
R.~D. Wildman, J.~M. Huntley, and D.~J. Parker,
Convection in highly fluidized three-dimensional granular beds.
{Phys.~Rev.~Lett.} {\bf 86}, 3304-3307 (2001).
\bibitem{BRM2001}
J.~J. Brey, M.~J. Ruiz-Montero, and F. Moreno,
Hydrodynamics of an open vibrated granular system.
Phys.~Rev.~E {\bf 63}, 061305 (2001).
\bibitem{BKS2002}
N. Burtally, P.~J. King, and M.~R.~Swift,
Spontaneous air-driven separation in vertically vibrated fine granular mixtures.
Science {\bf 295}, 1877 (2002).
\bibitem{GMIZ2002}
A. Garciamartin, D. Maza, J.~L.~Ilquimiche, and I. Zuriguel,
Convective motion in a vibrated granular layer.
Phys.~Rev.~E {\bf 65}, 031303 (2002).
\bibitem{MPB2003} 
B. Meerson, T. P\"{o}schel, and Y. Bromberg, 
Close-packed floating clusters: granular hydrodynamics beyond a freezing point?
Phys.~Rev.~Lett.  {\bf 91}, 024301 (2003).
\bibitem{OO2003}
T. Ohtsuki and T.  Ohsawa, 
Hydrodynamics for convection in vibrating beds of cohesionless granular materials.
{J.~Phys.~Soc.~Japan} {\bf 72}, 1963-1967 (2003).
\bibitem{KM2003}
E. Khain and B. Meerson,
Onset of granular convection in a horizontal layer of granular gas.
Phys.~Rev.~E. {\bf 67}, 021306 (2003).
\bibitem{Huan2004}
C. Huan, X. Yang, D. Candela, R.~W. Mair, and R.~L. Walsworth,
NMR experiments on a three-dimensional vibrofluidized granular medium.
Phys.~Rev.~E  {\bf 69}, 041302 (2004).
\bibitem{Sano2005}
O. Sano,
Dilatancy, buckling and undulations on a vertically vibrating granular layer.
Phys.~Rev.~E  {\bf 72}, 051302 (2005).
\bibitem{Eshuis2005}
P. Eshuis, K. van der Weele, D. van der Meer, and D. Lohse,
Granular Leidenfrost effect: experiment and theory of floating particle clusters.
Phys.~Rev.~Lett. {\bf 95}, 258001 (2005).
\bibitem{CPS2007}
J.~A.~Carrillo, T. P\"{o}schel, and C. Saluena,
Granular hydrodynamics and pattern formation in vertically oscillated granular disk layers.
J.~Fluid Mech.  {\bf 597}, 119-144 (2007).
\bibitem{Kudrolli2004}
A. Kudrolli,
Size separation in vibrated granular materials,
Rep. Prog. Phys. {\bf 67}, 209-247 (2004).
\bibitem{AT2006}
I.~S.  Aranson and L.~S. Tsimring,
Patterns and collective behaviour in granular media: Theoretical concepts.
Rev.~Mod.~Phys. {\bf 78}, 641 (2006).
\bibitem{Shukla2014}
P. Shukla, I.~ Ansari, D. van der Meer, D. Lohse, and M. Alam,
Nonlinear instability and convection in a vertically vibrated granular bed.
J.~Fluid Mech.  {\bf 761}, 123-167 (2014).
\bibitem{Eshuis2007}
P. Eshuis, K. van der Weele, D. van der Meer, R. Bos, and D. Lohse,
Phase diagram of vertically shaken granular matter.
Phys.~Fluids  {\bf 19}, 123301 (2007).
\bibitem{Leidenfrost}
J.~G. Leidenfrost, 
De Aquae Communis Nonnullis Qualitatibus Tractatus 
(University of Duisburg, Duisburg, Germany, 1756).
\bibitem{AAlam2012}
I. Ansari and  M. Alam, 
Patterns, segregation and hysteresis in vertically vibrated granular mixtures.
Bull.~Am.~Phys.~Soc.~{\bf 57}, No. 17, p.~358 (2012).
\bibitem{AAlam2013}
I. Ansari and  M. Alam, 
Patterns and velocity field in vertically vibrated granular materials.
In {\it AIP Conf.~Proc.} (Ed. A.~Yu \etal), vol.~{\bf 1542}, pp.~775-778 (2013).
\bibitem{Eshuis2010}
P. Eshuis, K. van der Weele, M. Alam, H.~J. van Gerner, K. van der Weele, and D. Lohse,
Onset of convection in strongly  shaken granular matter.
Phys.~Rev.~Lett.  {\bf 104}, 038001  (2010).
\bibitem{Eshuis2013}
P. Eshuis, D. van der Meer, M. Alam, H.~J. van Gerner, M. van der Hoef, H. Kuipers, S. Luding, D. van der Meer, and D. Lohse,
Buoyancy driven convection in vertically shaken  granular matter: experiments, numerics and theory.
Granul.~Matt.  {\bf 15}, 893  (2013).
\bibitem{CG1996}
J.~C.~Crocker and D.~Grier,
Methods of digital video microscopy for colloidal studies.
J.~Coll.~Interface Sci., {\bf 179}, 298-310 (1996).
\bibitem{SK2005}
I.~F.~Sbalzarini and P.~Koumoutsakos,
Feature point tracking and trajectory analysis for video imaging in cell imaging.
J.~Struc.~Biology, {\bf 151}(2), 182-195 (2005).
\bibitem{Dantec}
Dantec Dynamics A/S
Product Information on adaptive correlation in dynamic studio.
http://www.dantecdynamics.com (2012).
\bibitem{SR2000} 
F. Scarano and R.~L.~Riethmuller,
Advances in iterative multigrid PIV image processing.
Exp.~Fluids {\bf 29}, S51-S60 (2000).
\bibitem{RLT2013}
N. Rivas, S. Luding, and A.~R.~Thornton, 
Low-frequency oscillations in narrow vibrated granular systems.
New J.~Phys.  {\bf 15}, 113043 (2013).
\bibitem{AT1989}
M.~P.~Allen and D.~J.~Tildesley,
{\it Computer Simulations of Liquids}.
(Clarendon, Oxford, 1989).
\bibitem{Bogdan2006}
T.~V. Bogdan,
Atom-atomic potentials and the correlation distribution functions for modelling liquid Benzene by molecular dynamics methods.
Russian J. Phys. Chem. {\bf 80}, 14-20 (2006).
\bibitem{BPP2012}
J. Bougie, V. Policht, and J.~K. Pearce,
Time-dependence and density-inversion in simulations of vertically oscillated granular layers.
Phys.~Rev.~E  {\bf 86}, 020302(R) (2012).
\bibitem{RTLM2015}
N. Rivas, A.~R.~Thornton, S. Luding, and D. van der Meer,
From the granular Leidenfrost state to buoyancy-driven convection.
Phys.~Rev.~E  {\bf 91}, 042202 (2015).
\bibitem{AS2013}
M. Alam and P. Shukla,
Nonlinear stability, bifurcation and vortical patterns in three-dimensional granular plane Couette flow.
J.~Fluid Mech.  {\bf 716}, 349-413 (2013).
\bibitem{SA2009}
P. Shukla  and M. Alam,
Landau-type Order Parameter Equation for Shear Banding in Granular Couette Flow. 
Phys.~Rev.~Lett.  {\bf 100}, 068001 (2009).
\end{thebibliography}
\end{document}